\documentclass[iop]{emulateapj}
\usepackage{color}
\usepackage{amssymb, amsmath}

\usepackage[utf8]{inputenc}
\usepackage{lipsum}
\usepackage{pifont}
\usepackage{bm}
\usepackage{ulem}
\usepackage{balance}
\usepackage{multirow}
\usepackage{mathtools}
\usepackage{enumitem}
\usepackage{booktabs}
\usepackage{appendix}
\usepackage{listings}

\shorttitle{The CAMELS project}
\shortauthors{Villaescusa-Navarro, Angl\'es-Alc\'azar, Genel, et al.}

\usepackage[usenames,dvipsnames]{xcolor}
\usepackage{hyperref}
\hypersetup{
    colorlinks = true,
    citecolor = {MidnightBlue},
    linkcolor = {BrickRed},
    urlcolor = {blue}
}

\newcommand{\Msun}{M$_{\odot}$}

\newcommand{\kms}{km\,s$^{-1}$}

\newcommand{\Msunh}{M$_{\odot}h^{-1}$}

\newcommand{\be}{\begin{equation}}
\newcommand{\ee}{\end{equation}}
\newcommand{\ba}{\begin{eqnarray}}
\newcommand{\ea}{\end{eqnarray}}

\usepackage{soul}
\definecolor{nicegreen}{HTML}{2CA02C}
\definecolor{orange}{rgb}{1,0.5,0}

\definecolor{darkgreen}{RGB}{0,120,0}

\begin{document}

\title{The CAMELS project: Cosmology and Astrophysics with MachinE Learning Simulations}

\author{Francisco Villaescusa-Navarro$^{1,2, \dagger}$, Daniel Angl\'es-Alc\'azar$^{3,2, \ddagger}$, Shy Genel$^{2,4,\mathsection}$, David N. Spergel$^{2,1}$, Rachel S. Somerville$^{2,5}$, Romeel Dave$^{6}$, Annalisa Pillepich$^{7}$,  Lars Hernquist$^{8}$, Dylan Nelson$^{9}$, Paul Torrey$^{10}$, Desika Narayanan$^{10,11}$, Yin Li$^{2}$,  Oliver Philcox$^{1}$,  Valentina La Torre$^{2,12}$, Ana Maria Delgado$^{2,8}$, Shirley Ho$^{2,1,13}$, Sultan Hassan$^{2,14}$,  Blakesley Burkhart$^{5,2}$, Digvijay Wadekar$^{15}$, Nicholas Battaglia$^{16}$, Gabriella Contardo$^{2}$, Greg L. Bryan$^{17,2}$}

\affil{$^1$ Department of Astrophysical Sciences, Princeton University, Peyton Hall, Princeton, NJ, 08544, USA}
\affil{$^2$ Center for Computational Astrophysics, Flatiron Institute, 162 5th Avenue, New York, NY, 10010, USA}
\affil{$^3$ Department of Physics, University of Connecticut, 196 Auditorium Road, U-3046, Storrs, CT, 06269, USA}
\affil{$^4$ Columbia Astrophysics Laboratory, Columbia University, 550 West 120th Street, New York, NY, 10027, USA}
\affil{$^5$ Department of Physics and Astronomy, Rutgers University, 136 Frelinghuysen Rd, Piscataway, NJ 08854, USA}
\affil{$^{6}$ Institute for Astronomy, University of Edinburgh, Royal Observatory, Edinburgh EH9 3HJ, United Kingdom}
\affil{$^{7}$ Max-Planck-Institut f\"{u}r Astronomie, K\"{o}nigstuhl 17, 69117 Heidelberg, Germany}
\affil{$^{8}$Center for Astrophysics | Harvard \& Smithsonian, 60 Garden St, Cambridge, MA 02138, USA}
\affil{$^{9}$Universit\"{a}t Heidelberg, Zentrum f\"{u}r Astronomie, Institut f\"{u}r theoretische Astrophysik, Albert-Ueberle-Str. 2, 69120 Heidelberg, Germany}
\affil{$^{10}$ Department of Astronomy, University of Florida, 211 Bryant Space Sciences Center, Gainesville, FL, USA}
\affil{$^{11}$ University of Florida Informatics Institute, 432 Newell Drive, CISE Bldg E251, Gainesville, FL, USA}
\affil{$^{12}$ Physics and Astronomy Department, Tufts University, Medford, MA 02155}
\affil{$^{13}$ Department of Physics, Carnegie Mellon University, Pittsburgh, PA 15213, USA}
\affil{$^{14}$ Department of Physics and Astronomy, University of the Western Cape Cape Town 7535, South Africa}
\affil{$^{15}$ Center for Cosmology and Particle Physics, Department of Physics, New York University, New York, NY 10003, USA}
\affil{$^{16}$ Department of Astronomy, Cornell University, Ithaca, NY 14853, USA}
\affil{$^{17}$ Department of Astronomy, Columbia University, 550 West 120th Street, New York, NY 10027, USA}

\altaffiltext{$\dagger$}{villaescusa.francisco@gmail.com}
\altaffiltext{$\ddagger$}{angles-alcazar@uconn.edu}
\altaffiltext{$\mathsection$}{sgenel@flatironinstitute.org}

\begin{abstract}
We present the Cosmology and Astrophysics with MachinE Learning Simulations --CAMELS-- project. CAMELS is a suite of 4,233 cosmological simulations of $(25~h^{-1}{\rm Mpc})^3$ volume each: 2,184 state-of-the-art (magneto-)hydrodynamic simulations run with the AREPO and GIZMO codes, employing the same baryonic subgrid physics as the IllustrisTNG and SIMBA simulations, and 2,049 N-body simulations. The goal of the CAMELS project is to provide theory predictions for different observables as a function of cosmology and astrophysics, and it is the largest suite of cosmological (magneto-)hydrodynamic simulations designed to train machine learning algorithms. CAMELS contains thousands of different cosmological and astrophysical models by way of varying $\Omega_m$, $\sigma_8$, and four parameters controlling stellar and AGN feedback, following the evolution of more than 100 billion particles and fluid elements over a combined volume of $(400~h^{-1}{\rm Mpc})^3$. We describe the simulations in detail and characterize the large range of conditions represented in terms of the matter power spectrum, cosmic star-formation rate density, galaxy stellar mass function, halo baryon fractions, and several galaxy scaling relations.  We show that the IllustrisTNG and SIMBA suites produce roughly similar distributions of galaxy properties over the full parameter space but significantly different halo baryon fractions and baryonic effects on the matter power spectrum.  This emphasizes the need for marginalizing over baryonic effects to extract the maximum amount of information from cosmological surveys. We illustrate the unique potential of CAMELS using several machine learning applications, including non-linear interpolation, parameter estimation, symbolic regression, data generation with Generative Adversarial Networks (GANs), dimensionality reduction, and anomaly detection.\\

\end{abstract}


\section{Introduction}
\label{sec:introduction}

We are in the epoch of precision cosmology. The large amount of data, and the accuracy and precision of theory, have allowed us to constrain the values of the cosmological parameters with exquisite precision \citep{Planck_2018}. An illustrative example of this is the quest to constrain neutrino masses: it seems feasible to achieve a $\sim5\sigma$ constraint on the sum of the masses of the lightest particles in nature using data from upcoming cosmological missions \citep{Massara_2020, Chang_19, Thejs_2019, Cora_19}.

In the coming years, missions such as CMB-S4\footnote{https://cmb-s4.org}, DESI\footnote{https://www.desi.lbl.gov},
eROSITA\footnote{https://www.mpe.mpg.de/eROSITA},
Euclid\footnote{https://www.euclid-ec.org}, Roman Observatory\footnote{https://wfirst.gsfc.nasa.gov/index.html}, Rubin Observatory\footnote{https://www.lsst.org}, SO\footnote{https://simonsobservatory.org}, PFS\footnote{https://pfs.ipmu.jp/index.html}, and SKA\footnote{https://www.skatelescope.org} will survey large cosmological volumes collecting data that will allow us to constrain neutrino masses and the nature of dark energy, among many other fundamental quantities. One of the ultimate goals of these surveys is to constrain the value of the cosmological parameters with the smallest errors, in order to improve our understanding of fundamental physics and of the origin and fate of our Universe. To achieve this, it is necessary to extract the maximum amount of information from the data.

The standard methodology to constrain cosmological parameters is as follows. First, summary statistics are computed from the observed data. Second, predictions from theory are made for the said statistics. Third, parameter bounds are derived by matching theory to observations. A very deep and important question naturally arises: what summary statistic, or statistics, should be computed from the data to achieve the most accurate and precise parameter constraints?

The answer depends on the properties of the observed field. In the case of Gaussian density fields, their properties are fully described by the 2-point correlation function, or its Fourier transform, the power spectrum. On the other hand, for non-Gaussian density fields, the answer is unknown. Furthermore, the problem of finding that summary statistic is mathematically intractable. 

Unfortunately, most cosmological surveys observe non-Gaussian density fields, e.g.~galaxy redshift surveys, weak lensing surveys, \textit{et cetera.} Currently, we place constraints on the parameters by using primarily the power spectrum \citep{Philcox_2020, Ivanov_2019, eBOSS, Guido_2020}. How much information could we gain by using other summary statistics? The answer to that question is difficult, as theory predictions in the mildly non-linear to non-linear regime are needed. Recently, the \textsc{Quijote} project \citep{Quijote} has run a very large number of N-body simulations to quantify the information content on generic cosmological statistics. Several studies have shown that statistics other than the power spectrum contain a huge amount of cosmological information when we consider smaller scales \citep{Chang_19, Massara_2020, Cora_19, Friedrich_2019, Allys_2020, Arka_Tom_2020, Gualdi_2020, Jay_2019}. 

An alternative approach to extract information is through neural networks, that are trained to search through all possible summary statistics and use the best to extract the cosmological information \citep{Siamak_16, Schmelzle_17, Gupta_18, Ribli_19, Fluri_19, Ntampaka_19, Sultan_2019, Jose_2020, Niall_2020}. These methods have also shown that much tighter constraints on the value of the cosmological parameters can be placed than those obtained from the power spectrum. 

All these findings show that small scales embed a wealth of cosmological information which, if extracted, could allow us to better constrain the value of the cosmological parameters and therefore improve our understanding of fundamental physics. However, uncertainties in the complex physics of galaxy formation limits its use. Astrophysical effects such as feedback from Active Galactic Nuclei (AGN) can expel gas to very large distances, which will also affect the distribution of dark matter via gravitational interactions. These effects are expected to alter the spatial distribution of the underlying matter field on small scale. We will be referring to these complex and poorly understood processes as \textit{baryonic effects}. 

Significant progress has been made in the field of galaxy formation and evolution using large volume cosmological hydrodynamic simulations \citep{SIMBA, IllustrisTNG_public,HorizonAGN,Eagle,Illustris, Lee_2020, Mark_2020, Bahamas, Dianoga, BlueTides, Magneticum}, where a comprehensive modeling of baryonic processes is required. These simulations have improved to the point where they are able to produce galaxies that match observations in many aspects, e.g.~mass functions, morphologies, and colors.
However, key astrophysical processes such as star formation, massive black hole growth, stellar feedback, and AGN feedback remain very uncertain and are modeled via phenomenological subgrid models that require extensive tuning of free parameters to match observations \citep{SomervilleDave2015}. Thus, while large volume cosmological hydrodynamical simulations have achieved many important milestones, uncertainties in subgrid physics and degeneracies of free parameters limit their applicability for cosmological analyses in the era of precision cosmology.
Cosmological ``zoom-in'' simulations can achieve much higher resolution and reduce uncertainties in subgrid physics \citep[e.g.][]{Agertz2016,Hopkins2018,Angles-Alcazar2020, Dubois_2020} but at the expense of modeling significantly smaller volumes. 

A Bayesian approach to address this problem is to marginalize over baryonic effects. While the predictions of a given simulation may not be accurate enough to match observations, reality may lie within the range of predictions from simulations spanning a sufficiently broad range of astrophysical models. By using simulations with different values of astrophysical parameters and different subgrid physics implementations, we can in principle quantify the errors associated with our poor understanding of baryonic physics.

Thus, it would be desired to extract cosmological information from small scales from either summary statistics, or from the field itself, while marginalizing over baryonic effects. For this purpose, theory predictions for those summary statistics, or at the field level, are needed as a function of cosmology, astrophysics, and potentially the initial conditions. Obtaining these theory predictions by running hydrodynamic simulations for each point in parameter space needed to either evaluate the likelihood or perform likelihood-free inference may be too computationally demanding. However, it may be possible to run a large, but finite number of state-of-the-art (magneto-)hydrodynamic simulations that cover a large volume in parameter space to provide theory predictions in those points, while using machine learning techniques to fill the gaps not covered by the simulations.

We illustrate the above idea using two examples. First, one can imagine building an emulator for the matter power spectrum through Gaussian random processes or bayesian neural networks, by using the measurements of the matter power spectrum from a set of simulations that expand a large volume in the cosmological and astrophysical parameter space. Second, convolutional networks can be used to find the mapping between the 3D spatial distribution of dark matter and cosmic neutral hydrogen \citep{Jay_2020}. Finding that mapping as a function of cosmology and astrophysics will enable the possibility of having theory predictions at the field level for 21cm surveys. For the latter, hydrodynamic simulations with different cosmologies and astrophysics are needed to calibrate the mapping learned by the network.

In this work, we introduce the Cosmology and Astrophysics with MachinE Learning Simulations (CAMELS) project\footnote{\url{https://www.camel-simulations.org}}. The main goal of CAMELS is to provide theoretical predictions, as a function of cosmology and astrophysics, for key observables in cosmology and galaxy formation by combining the direct output of (magneto-)hydrodynamic simulations with machine learning techniques. CAMELS contains thousands of simulations, spanning thousands of different cosmological and astrophysical models that, combined together, host more than 100 billion dark matter particles, gas elements, stars, and black holes over a combined volume larger than $(400~h^{-1}{\rm Mpc})^3$. However, we note that the presence of extreme objects (e.g. large voids or massive clusters) is essentially limited in the combined volume because each simulation only samples a comoving volume of $(25 h^{-1} {\rm Mpc})^3$. We further discuss the limitations of CAMELS in section \ref{subsec:limitations}.

This paper is organized as follows. We outline the main goals of the CAMELS project in Section~\ref{sec:goals}. We describe in detail the CAMEL simulations in Section~\ref{sec:Simulations}. We then present some key cosmological and astrophysical properties of the different simulation suites in Section~\ref{sec:Properties}. Next, we show several machine learning applications of CAMELS to illustrate its unique potential in Section~\ref{sec:ML_applications}. Finally, we summarize the main aspects of this work and conclude in Section~\ref{sec:Conclusions}. 
CAMELS has been developed as part of the Simulating Multiscale Astrophysics to Understand Galaxies (SMAUG) collaboration,\footnote{\url{https://www.simonsfoundation.org/flatiron/center-for-computational-astrophysics/smaug}} which aims to improve our understanding of the key baryonic processes driving galaxy formation and evolution in order to maximize the scientific return of upcoming cosmological missions.

\section{Project goals}
\label{sec:goals}

In order to determine the value of the cosmological parameters with the highest accuracy we need to extract cosmological information from mildly non-linear to non-linear scales, where poorly understood baryonic effects take place. 
The main goal of the CAMELS project is to provide theory predictions for a given observable, or field, as a function of cosmology and astrophysics:
\begin{equation}
    S(z) = f(\vec{\theta}_{\rm c}, \vec{\theta}_{\rm a},z),
\end{equation}
where $S$ denotes a generic statistic (e.g.~power spectrum) at redshift $z$ and $\vec{\theta}_{\rm c}$ and $\vec{\theta}_{\rm a}$ represent the value of the cosmological and astrophysical parameters, respectively. The above equation holds for a summary statistic, but it can be generalized to 2D or 3D density fields $F(\vec{x})$:
\begin{equation}
    F(\vec{x},z)=g(\vec{\theta}_{\rm c}, \vec{\theta}_{\rm a}, \delta(\vec{x}), \delta(\vec{y})),
    \label{Eq:master_equation}
\end{equation}
where $\delta(\vec{x})$ and $\delta(\vec{y})$ denote the initial conditions at position $\vec{x}$ and its environment $\vec{y}$ ($\vec{y}\ne\vec{x}$). Finding the mapping between the initial conditions and a given field as a function of cosmology and astrophysics (i.e.~Eq.~\ref{Eq:master_equation}) may require several sub-steps. For instance, one can first go from initial conditions to the matter density field of N-body simulations \citep{Siyu_2019}, and then use CAMELS to find the mapping between N-body and full hydrodynamic simulations \citep{Leander_2020,Jay_2020,Jacky_2019,Zhang_2019}. Changes in the cosmology or astrophysics model can be done at the N-body or hydrodynamic level \citep{Giusarma_19}.

Determining the function in Eq.~\ref{Eq:master_equation} will enable the generation of fast mock observations, as only the initial density field and the value of the cosmological and astrophysical parameters is needed. Those mocks can then be used to extract the maximum amount of cosmological information while marginalizing over baryonic effects \citep{Paco_2020b, Pablo_2020}. For instance, one can train neural networks to predict the value of the cosmological parameters on simulations with different feedback strengths to force neural networks to marginalize over them. A more stringent test will be to train neural networks on simulations run with a given code and subgrid physics model, and then test if the trained network is able to recover the correct cosmology from simulations run with a different code and subgrid physics \citep{Paco_2020d}.  For that purpose, CAMELS includes independent simulation suites performed with two different galaxy formation codes (see Section~\ref{sec:Simulations}). 

On the other hand, quantifying systematically the dependence of a given galaxy or circumgalactic medium observable on the value of the cosmological and astrophysical parameters will allow us to improve our understanding of galaxy formation and evolution. Furthermore, one can use machine learning techniques to find the mapping between input parameters (e.g.~strength of feedback mechanisms) and output observables (e.g.~the galaxy stellar mass function) to improve the calibration of subgrid parameters in the next generation of cosmological large volume simulations.

CAMELS scientific goals can thus be summarized as follows:
\begin{itemize}
    \item Provide theory predictions for summary statistics and full 3D fields as a function of cosmology and astrophysics.
    \item Train neural networks to extract cosmological information while marginalizing over baryonic effects.
    \item Develop machine learning techniques to find the mapping between N-body simulations and hydrodynamic simulations with full baryonic physics.
    \item Quantify the dependence of galaxy formation and evolution on astrophysical and cosmological parameters.
    \item Use machine learning to efficiently calibrate subgrid parameters in cosmological hydrodynamic simulations to match a set of observations.
\end{itemize}

\section{Simulations}
\label{sec:Simulations}

CAMELS is a suite of 4,233 numerical simulations. Approximately half of them are N-body, while the rest are state-of-the-art (magneto-)hydrodynamic simulations implementing the subgrid physics models of IllustrisTNG \citep{IllustrisTNG_public} and SIMBA \citep{SIMBA}. From $z = 127$ to today, each simulation follows the evolution of $256^3$ dark matter particles of mass $6.49\times10^7~(\Omega_{\rm m}-\Omega_{\rm b})/0.251~h^{-1}M_\odot$ and (in the case of the hydrodynamic simulations) $256^3$ gas resolution elements with an initial mass of $1.27\times10^7~h^{-1}M_\odot$ in a periodic box of comoving volume equal to $(25~h^{-1}{\rm Mpc})^3$. This volume and particle number has been chosen as a trade-off between (1) the large number of simulations required to cover the parameter space as densely as possible (at least 1,000 variations for machine learning applications) and (2) the increased computational cost of larger volume simulations at the resolution required to model complex galaxy formation processes.  Importantly, CAMELS implements the same resolution as the original SIMBA simulation and similar to that of the original IllustrisTNG300-1 simulation.  While this choice of volume and resolution enables a broad range of science, we discuss some of the limitations of CAMELS as well as future extensions in Section~\ref{sec:Conclusions}.  

The initial conditions are generated at $z=127$ using second order Lagrangian perturbation theory (2LPT). For simplicity, we assume that the initial power spectra of dark matter and gas in the hydrodynamic simulations are the same, and equal to that of total matter. Note that in that case both the growth factor and the growth rate are scale-independent, which allow us to use standard rescaling codes. 

We save snapshots at redshifts $z = 6, 5, 4, 3.5$, and then at 30 additional redshifts logarithmically spaced from $(1+z)=4$ to $(1+z)=1$. Dark matter halos and subhalos/galaxies are identified through friends-of-friends (FoF; \citealt{FoF_82, FoF}) and SUBFIND\footnote{These codes are run on-the-fly for the IllustrisTNG suite and during post-processing for the SIMBA suite.} \citep{Subfind}; 
additional (sub)halo catalogs are produced with the Amiga Halo Finder \citep[AHF;][]{Knollmann2009_AHF}, which are not used here but are available for future work. In total, more than 140,000 snapshots are generated, comprising of more than 100 billion resolution elements at each redshift over a combined volume\footnote{See section \ref{subsec:limitations} for more details on how to interpret this number.} of $\sim (400~h^{-1}{\rm Mpc})^3$.

The value of th e following cosmological parameters is fixed in all simulations: $\Omega_{\rm b}=0.049$, $h=0.6711$, $n_s=0.9624$, $M_\nu=0.0$ eV, $w=-1$, $\Omega_K=0$. The value of $\Omega_{\rm m}$ and $\sigma_8$ is varied across simulations. We have used a very wide range for these parameters in order to mitigate prior effects on the output of neural networks: $\Omega_{\rm m}\in[0.1,0.5]$, $\sigma_8\in[0.6,1.0]$. In the N-body simulations these are the only two parameters that change, besides the value of the initial random seed that determines the initial Gaussian density field. 

In the hydrodynamic simulations, besides $\Omega_{\rm m}$, $\sigma_8$, and the initial random seed, we also vary the value of four astrophysical parameters related to stellar and AGN feedback. We refer to these parameters generically as $A_{\rm SN1}$, $A_{\rm SN2}$, $A_{\rm AGN1}$ and $A_{\rm AGN2}$, which we introduce in CAMELS as simple normalization factors to generate variations relative to the original IllustrisTNG and SIMBA feedback models.  While we do not consider parameter variations for other key baryonic processes, we explore variations over a broad range of efficiencies for some of the most important and uncertain feedback processes in galaxy formation: galactic winds driven by supernovae (SNe) and kinetic feedback from massive black holes.
We emphasize that CAMELS implements identical baryonic feedback models as IllustrisTNG and SIMBA, aside from the varying numerical values of $A_{\rm SN1}$, $A_{\rm SN2}$, $A_{\rm AGN1}$ and $A_{\rm AGN2}$, which are summarized in Table~\ref{table:sims2} and described in detail below. The range of variation of these parameters is identical in IllustrisTNG and SIMBA but their actual definition and overall effect can be quite different, reflecting the rather different feedback implementations in each model.

\subsection{IllustrisTNG}
\label{subsec:IllustrisTNG}
The IllustrisTNG galaxy formation model is fully described in \citet{WeinbergerR_16a,PillepichA_16a}. Building on its predecessor Illustris model \citep{VogelsbergerM_13a, Torrey_2014}, IllustrisTNG utilizes the Arepo code\footnote{\url{https://arepo-code.org}} \citep{Arepo,Arepo_public} to solve the coupled equations of gravity (using an N-body tree-particle-mesh approach; TreePM) and magneto-hydrodynamics (MHD; using a Voronoi moving-mesh approach), in addition to models for sub-grid physics. These models cover radiative cooling and heating, star-formation, stellar evolution, feedback from galactic winds, the formation and growth of supermassive black holes (SMBH), and feedback from AGN.

Radiative cooling from both an on-the-fly network for the primordial elements and tabulated metal cooling are modeled following \citet{KatzN_96b,WiersmaR_09a} assuming a spatially-uniform ionising background \citep{FaucherGiguereC_09a} and hydrogen self-shielding \citep{Rahmati2013}. Stars form from gas above a hydrogen number density threshold of $n_{\rm H}\sim0.13{\rm\thinspace cm}^{-3}$ on timescales that become shorter at increasing densities, following \citet{SpringelV_03a}. Star-forming gas is also pressurised using an equation of state that derives from a sub-grid multi-phase model for the interstellar medium \citep{SpringelV_03a}. IllustrisTNG tracks chemical enrichment from Type II SNe, Type Ia SNe, Asymptotic Giant Branch (AGB) stars and neutron star-neutron star mergers, tracking nine individual elements (H, He, C, N, O, Ne, Mg, Si, and Fe).

Galactic winds driven by stellar feedback are implemented kinetically via temporarily hydrodynamically-decoupled particles that are stochastically and isotropically ejected from the star-forming gas, in a scheme based on \citet{SpringelV_03a} with the addition of a sub-dominant ($10\%$) component of thermal energy. The prescribed wind speed $v_{\rm w}$ and total energy injection rate (power) per unit star-formation $e_w$ depend on local conditions of the gas (metallicity, dark matter velocity dispersion), redshift, and two global normalization parameters. The latter are modulated by the parameters $A_{\rm SN1}$ and $A_{\rm SN2}$ that we vary between CAMELS simulations, as follows:
\begin{eqnarray}
e_w &=& A_{\rm SN1} \times \overline{\vphantom{t} e}_{w}~ 
                   \left[ f_{w,Z} + \frac{1-f_{w,Z}}{1 + (Z/Z_{w, \rm ref})^{\gamma_{w,Z}}} \right] \nonumber \\[2ex]
            && \times ~ N_{\rm SNII} ~E_{\rm SNII,51} ~10^{51} ~\rm{erg}~ \hbox{$\rm\thinspace M_{\odot}$}^{-1}
\end{eqnarray}
and
\begin{equation}
 v_{\rm w} = A_{\rm SN2} \times {\rm max} \left[ \kappa_w ~\sigma_{\rm DM} \left( \frac{H_0}{H(z)} \right)^{1/3} , ~ v_{w, \rm min} \right]
\end{equation}
(see Eqs.~3 and 1 from \citet{PillepichA_16a}, respectively), where $Z$ is the gas metallicity, $\sigma_{\rm DM}$ is the local dark matter velocity dispersion as calculated by SUBFIND, and $H$ is the Hubble constant, and details on the parameters $\overline{\vphantom{t} e}_{w}$, $f_{w,Z}$, $Z_{w, \rm ref}$, $\gamma_{w,Z}$, $N_{\rm SNII}$, $E_{\rm SNII,51}$, $\kappa_w$, and $v_{w, \rm min}$ can be found in \citet{PillepichA_16a} (see their Table 1). The resulting wind mass loading factor equals $\eta_{\rm w} \equiv \dot{M}_{\rm wind}$/SFR$ = 1.8v_{\rm w}^{-2}e_w$.

The SMBH models in IllustrisTNG are described in \citet{WeinbergerR_16a} and builds upon the earlier models of \citet{SpringelV_05d,SijackiD_07a,VogelsbergerM_13a}. SMBH particles with initial mass $M_{\rm seed} = 8\times10^5$\,\Msunh~are seeded in halos with mass $M_{\rm FoF}> 5\times10^{10}$\,\Msunh. Throughout their evolution, SMBH particles are repositioned to the location of the potential minimum within their kernel. SMBH particles located within each other's `feedback spheres' (see below) are instantaneously merged. SMBH growth by gas accretion follows the spherical \citet{Bondi1952} parameterization, limited by the Eddington rate.

SMBH feedback follows a three-mode approach: thermal, kinetic, and radiative. The latter modifies the cooling and heating of gas in and around the dark matter halo host of the SMBH by adding its radiation flux to the cosmic ionizing background, and is always active, though its effect is limited to very bright AGN and is relatively minor to the overall heating and cooling rates of the halo atmospheres. The thermal and kinetic modes operate separately and are delineated by the Eddington ratio of the instantaneous accretion rate onto the SMBH. This Eddington ratio is in itself dependent on the SMBH mass, such that the transition from the high-mode to the low-mode occurs typically at $M_{\rm SMBH}\sim10^8\hbox{$\rm\thinspace M_{\odot}$}$. The high accretion rate thermal mode operates by injection of thermal energy into a `feedback sphere' that is defined as containing a prescribed (fixed) amount of mass around the SMBH. The injection rate is directly proportional to the mass accretion rate with an overall mass-to-energy conversion efficiency of 0.02.

The low accretion rate kinetic SMBH feedback mode is the one that is modulated in CAMELS by the $A_{\rm AGN1}$ and $A_{\rm AGN2}$ parameters, which, as in the case of the stellar feedback, correspond to an overall energy and speed normalizations. The power injected in the kinetic mode scales with the accretion rate $\dot{M}_{\rm BH}$ as follows:
\begin{equation}
 \dot{E}_{\rm low} = A_{\rm AGN1} \times {\rm min} \left[ \frac{\rho}{0.05\rho_{\rm SFthresh}} , ~ 0.2 \right] \times \dot{M}_{\rm BH}c^2
\end{equation}
(see Eqs.~8 and 9 from \citealp{WeinbergerR_16a}), where $\rho$ is the gas density around the SMBH, $\rho_{\rm SFthresh}$ is the density threshold for star-formation, and $c$ is the speed of light. This energy is accumulated over time and not spent except for in discrete events, whence 
it is injected as kinetic energy in a random direction into the `feedback sphere'. These discrete events occur every time the SMBH has accumulated a total amount of energy since the last feedback event; 
\begin{equation}
 E_{\rm inj,min} = A_{\rm AGN2} \times f_{\rm re}\frac{1}{2}\sigma^2_{\rm DM}m_{\rm enc}
\end{equation}
(see Eq.~13 from \citealp{WeinbergerR_16a}), where $\sigma^2_{\rm DM}$ is the one-dimensional dark matter velocity dispersion around the SMBH, $m_{\rm enc}$ is the gas mass in the feedback sphere, and $f_{\rm re}=20$ is a constant of the fiducial TNG model. Hence, $A_{\rm AGN2}$ controls the burstiness and speed of the low-mode SMBH feedback: the larger $A_{\rm AGN2}$, the rarer yet more energetic the individual feedback events are.

Finally, it is noted that at the mass resolution used in CAMELS, the IllustrisTNG model is set to have a spatial resolution (gravitational softening length of the dark matter) of approximately $2{\rm\thinspace kpc}$ comoving. The mass and spatial resolution of these simulations are comparable to the original IllustrisTNG300-1 \citep{IllustrisTNG_public}.

\begin{table*}
\begin{center}
\renewcommand{\arraystretch}{0.6}
\resizebox{1.0\textwidth}{!}{\begin{tabular}{| c || c | c | c | c |}
\hline
\multirow{2}{*}{Simulation} & \multirow{2}{*}{$A_{\rm SN1}$} & \multirow{2}{*}{$A_{\rm SN2}$} & \multirow{2}{*}{$A_{\rm AGN1}$} & \multirow{2}{*}{$A_{\rm AGN2}$}\\
& & & & \\
\hline \hline
\multirow{3}{*}{IllustrisTNG} & Galactic winds: & Galactic winds: & Kinetic mode BH feedback: & Kinetic mode BH feedback: \\[0.5ex] 
& energy per unit SFR & wind speed & energy per unit BH accretion rate  & ejection speed / burstiness \\[0.5ex]
\hline
\multirow{3}{*}{SIMBA} & Galactic winds: & Galactic winds: & QSO \& jet-mode BH feedback: & Jet-mode BH feedback:\\[0.5ex] 
& mass loading & wind speed & momentum flux & jet speed \\[0.5ex]
\hline \hline
Variation range & [0.25 - 4.00] & [0.50 - 2.00] & [0.25 - 4.00] & [0.50 - 2.00]\\[0.5ex]
\hline
\end{tabular}}
\end{center}
\caption{We vary two cosmological ($\Omega_{\rm m}$, $\sigma_8$) and four astrophysical ($A_{\rm SN1}$, $A_{\rm SN2}$, $A_{\rm AGN1}$, $A_{\rm AGN2}$) parameters in CAMELS. This table summarizes their physical meaning for both the IllustrisTNG and SIMBA suites. The last row shows the range of variation of each parameter; for the cosmological parameters the range of variation is $\Omega_{\rm m}\in$[0.1 - 0.5], $\sigma_8\in$ [0.6 - 1.0]. We emphasize that simulations with $A_{\rm SN1}=A_{\rm SN2}=A_{\rm AGN1}=A_{\rm AGN2}=1$ correspond to the original IllustrisTNG and SIMBA models at the resolution considered here. Note that the N-body simulations only have $\Omega_{\rm m}$ and $\sigma_8$ as free parameters.}
\label{table:sims2}
\end{table*}

\subsection{SIMBA}
\label{subsec:SIMBA}

The SIMBA galaxy formation model is fully described in \citet{SIMBA}.  Building on its predecessor Mufasa \citep{Dave2016}, SIMBA utilizes the $N$-body+hydrodynamics code GIZMO\footnote{\url{http://www.tapir.caltech.edu/~phopkins/Site/GIZMO.html}} \citep{Hopkins2015_Gizmo} in its ``Meshless Finite Mass'' (MFM) hydrodynamics mode. 
Gravitational forces are computed using a modified version of the TreePM algorithm of the GADGET-III code \citep{Springel2005_Gadget}, including adaptive gravitational softenings for the gas, stellar, and dark matter components. 

Radiative cooling and photoionization heating are modeled using the {\sc Grackle-3.1} library~\citep{Smith2017_Grackle}, including non-equilibrium evolution of primordial elements, metal cooling, a spatially-uniform ionising background \citep{HaardtMadau2012}, and hydrogen self-shielding \citep{Rahmati2013}.
Stars form from molecular gas at a rate given by the H$_2$ density divided by the dynamical time: SFR$=\epsilon_*\rho_{H2}/t_{\rm dyn}$, where $\epsilon_*=0.02$~\citep{Kennicutt1998} and the H$_2$ fraction is computed based on the metallicity and local column density following \citet{Krumholz2011}.   A minimum level of pressurization is included as required to resolve the Jeans mass in star-forming gas \citep{Dave2016}.
SIMBA tracks chemical enrichment from Type II SNe, Type Ia SNe, and AGB stars, following eleven individual elements (H, He, C, N, O, Ne, Mg, Si, S, Ca, and Fe), and includes a dust physics module that tracks the formation of dust in stellar ejecta, dust growth by accretion of metals, and destruction by thermal sputtering and SNe \citep{Li2019_SimbaDust}.

\begin{table*}
\begin{center}
\renewcommand{\arraystretch}{0.9}
\resizebox{1.0\textwidth}{!}{\begin{tabular}{| c || c | c | c | c | c |}
\hline
Name & Type & Code & Simulations & Set & Varying parameters\\[0.7ex]
\hline \hline
\multirow{6}{*}{IllustrisTNG} & \multirow{4}{*}{Magneto-} & \multirow{6}{*}{AREPO} & 1,000 & LH & $\Omega_{\rm m}$, $\sigma_8$, $A_{\rm SN1}$, $A_{\rm SN2}$, $A_{\rm AGN1}$, $A_{\rm AGN2}$, $S$ \\[0.5ex] 
\cline{4-6}
&  \multirow{4}{*}{hydrodynamic} & & 61 & 1P & $\Omega_{\rm m}$, $\sigma_8$, $A_{\rm SN1}$, $A_{\rm SN2}$, $A_{\rm AGN1}$, $A_{\rm AGN2}$ \\[0.5ex]
\cline{4-6}
& & & 27 & CV & $S$ \\[0.5ex]
\cline{4-6}
& & & 4 & EX & $A_{\rm SN1}$, $A_{\rm SN2}$, $A_{\rm AGN1}$, $A_{\rm AGN2}$ \\[0.5ex]
\hline
\multirow{6}{*}{SIMBA} & \multirow{6}{*}{Hydrodynamic} & \multirow{6}{*}{GIZMO} & 1,000 & LH & $\Omega_{\rm m}$, $\sigma_8$, $A_{\rm SN1}$, $A_{\rm SN2}$, $A_{\rm AGN1}$, $A_{\rm AGN2}$, $S$ \\[0.5ex] 
\cline{4-6}
& & & 61 & 1P & $\Omega_{\rm m}$, $\sigma_8$, $A_{\rm SN1}$, $A_{\rm SN2}$, $A_{\rm AGN1}$, $A_{\rm AGN2}$ \\[0.5ex]
\cline{4-6}
& & & 27 & CV & $S$ \\[0.5ex]
\cline{4-6}
& & & 4 & EX & $A_{\rm SN1}$, $A_{\rm SN2}$, $A_{\rm AGN1}$, $A_{\rm AGN2}$ \\[0.5ex]
\hline
\multirow{6}{*}{Dark Matter} & \multirow{6}{*}{N-body} & \multirow{6}{*}{GADGET-III} & 2,000 & LH & $\Omega_{\rm m}$, $\sigma_8$, $S$ \\[0.5ex] 
\cline{4-6}
& & & 21 & 1P & $\Omega_{\rm m}$, $\sigma_8$ \\[0.5ex]
\cline{4-6}
& & & 27 & CV & $S$ \\[0.5ex]
\cline{4-6}
& & & 1 & EX & - \\[0.5ex]
\hline
\hline
Total & & & 4,233 & & \\[0.5ex]
\hline
\end{tabular}}
\end{center}
\caption{Characteristics of the CAMELS suite. $A_{\rm SN1}$, $A_{\rm SN2}$, $A_{\rm AGN1}$, and $A_{\rm AGN2}$ represent the value of subgrid physics parameters controlling stellar and AGN feedback (see text and table \ref{table:sims2}). $S$ is the initial random seed of a simulation. The LH set is a latin hypercube where the values of $\Omega_{\rm m}$, $\sigma_8$, $A_{\rm SN1}$, $A_{\rm SN2}$, $A_{\rm AGN1}$, $A_{\rm AGN2}$, and $S$ are varied simultaneously. Note that the latin hypercubes of the IllustrisTNG and SIMBA simulations are different. Simulations in the 1P set have the same initial random seed and only the value of one parameter is varied at a time. The CV set is made of simulations with fixed cosmology and astrophysics (at fiducial values) but different initial random seeds. Finally, simulations in the EX set have the same cosmology and random seed but even more extreme values in the feedback parameters than represented in the other simulation sets. See text for further details.}
\label{table:sims}
\end{table*}

Galactic winds driven by stellar feedback are implemented kinetically via hydrodynamically-decoupled, two-phase, metal-enriched winds with 30\% of wind particles heated to a temperature set by the SNe energy minus the wind kinetic energy.
Star-forming gas elements are stochastically ejected according to prescribed values of mass loading factor, $\eta \equiv \dot{M}_{\rm wind}$/SFR, and wind velocity, $v_{\rm w}$.  The mass loading factor is based on the FIRE ``zoom-in'' simulations \citep{Hopkins2014_FIRE} and scales with galaxy stellar mass $M_\star$ following \citet{Angles-Alcazar2017_BaryonCycle}:
\begin{equation}
\eta(M_\star) = A_{\rm SN1} \times
\begin{cases}
    9\Big(\frac{M_\star}{M_0}\Big)^{-0.317},& \text{if } M_\star<M_0\\
    9\Big(\frac{M_\star}{M_0}\Big)^{-0.761},& \text{if } M_\star>M_0\\
\end{cases}
\end{equation}
where $M_0=5.2\times 10^9 M_\odot$ and we have introduced $A_{\rm SN1}$ to control the overall normalization of the mass loading factor in CAMELS.
The wind velocity parameterization is also based on FIRE and scales with galaxy circular velocity $v_{\rm circ}$ following \citet{Muratov2015}:
\begin{equation}
v_{\rm w} = A_{\rm SN2} \times 1.6 \Big(\frac{v_{\rm circ}}{200\,{\rm km\,s}^{-1}}\Big)^{0.12} v_{\rm circ} + \Delta v(0.25R_{\rm vir})
\end{equation}
where $\Delta v(0.25R_{\rm vir})$ is the velocity corresponding to the potential difference between the launch point and 0.25\,$R_{\rm vir}$ (where \citealt{Muratov2015} measured the wind velocity; see \citealt{Dave2016}) and $A_{\rm SN2}$ controls the wind velocity normalization for the parameter variations in CAMELS.
An on-the-fly approximate friends-of-friends (FoF) finder is applied to stars and dense gas to compute $M_\star$ and then $v_{\rm circ}$ is estimated based on the baryonic Tully-Fisher relation. SIMBA limits the wind kinetic energy to the available SNe energy by attenuating the wind velocity when needed but in CAMELS we remove this constraint to have full control of the galactic wind efficiency.  All other aspects of stellar feedback, including Type Ia SNe and AGB wind heating, are identical to the original SIMBA simulations.

The black hole module in SIMBA builds on the gravitational torque accretion and kinetic feedback implementation in GIZMO presented in \citet{Angles-Alcazar2017_BHfeedback}. Black holes with initial mass $M_{\rm seed} = 10^4$\,\Msunh~are seeded in galaxies with $M_{\star} \gtrsim 10^{9.5}$\,\Msun~by means of the on-the-fly FoF, with the $M_{\star}$ threshold for seeding motivated by FIRE simulations showing that stellar feedback strongly suppresses black hole growth in lower mass galaxies \citep{Angles-Alcazar2017_BHsOnFIRE,Catmabacak2020}. Black hole particles are repositioned to the location of the potential minimum within the FoF host group if it is found within a distance $<4\times R_0$, where $R_{0}$ is the size of the black hole kernel enclosing the nearest 256 gas elements. Black holes located within $R_{0}$ of each other are instantaneously merged if their relative velocity is smaller than three times their mutual escape velocity. Black hole growth follows a two-phase model where cold gas is accreted at a rate given by the transport of angular momentum by gravitational torques from the stars \citep{Hopkins2011_Analytic} and hot gas accretion proceeds at a rate given by the spherical \citet{Bondi1952} parameterization. 

\begin{figure*}
\centering
\includegraphics[width=0.95\textwidth]{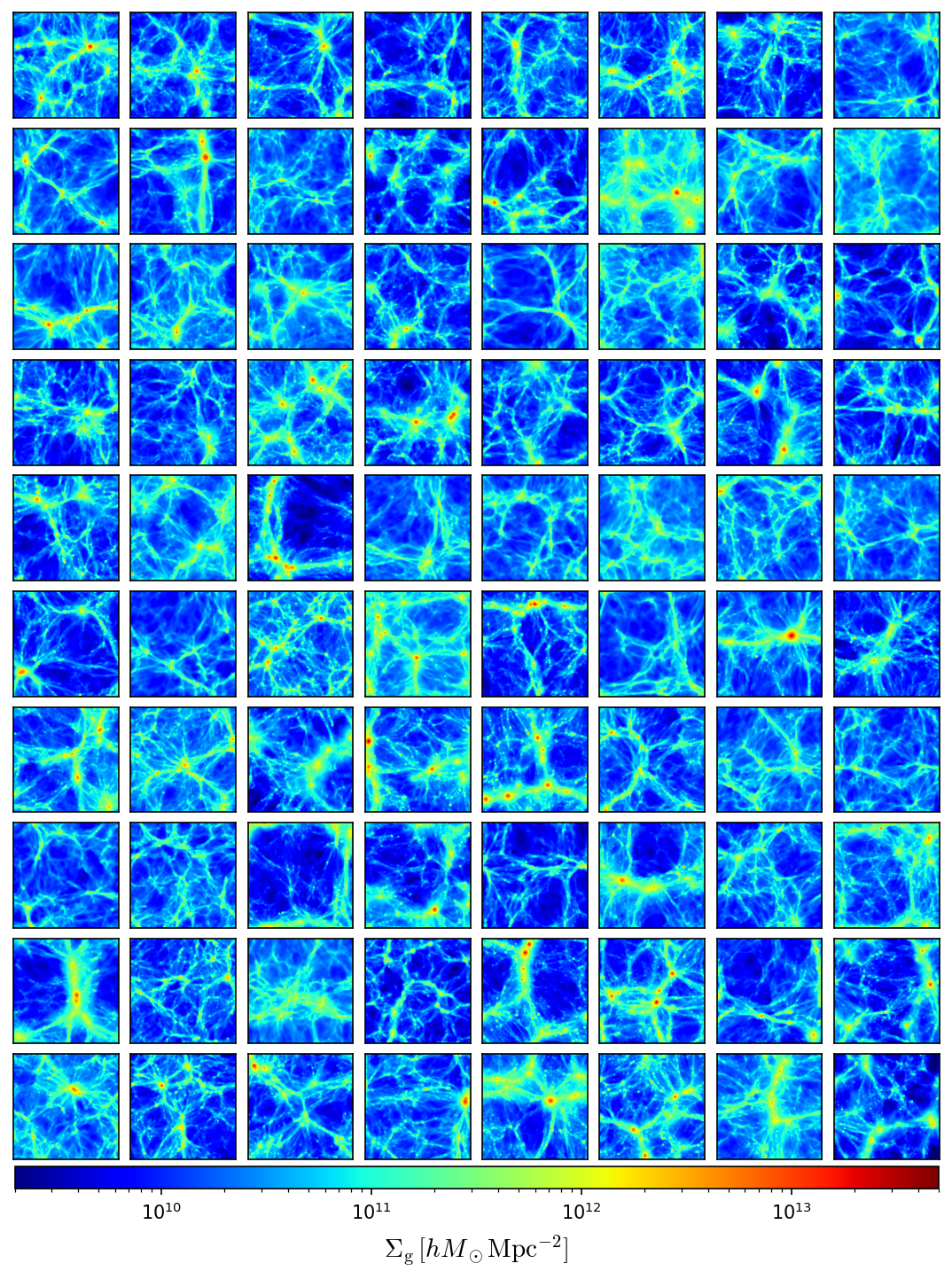}
\caption{This figure shows projected gas density maps from 80 different simulations of the SIMBA LH set at $z=0$. Each panel represents a region of $25\times25\times5~(h^{-1}{\rm Mpc})^3$. The LH set covers a large variety of environments and physical conditions, simultaneously varying the initial random field, cosmology, and feedback parameters.}
\label{fig:image_LH}
\end{figure*}

\begin{figure*}
\centering
\includegraphics[width=0.79\textwidth]{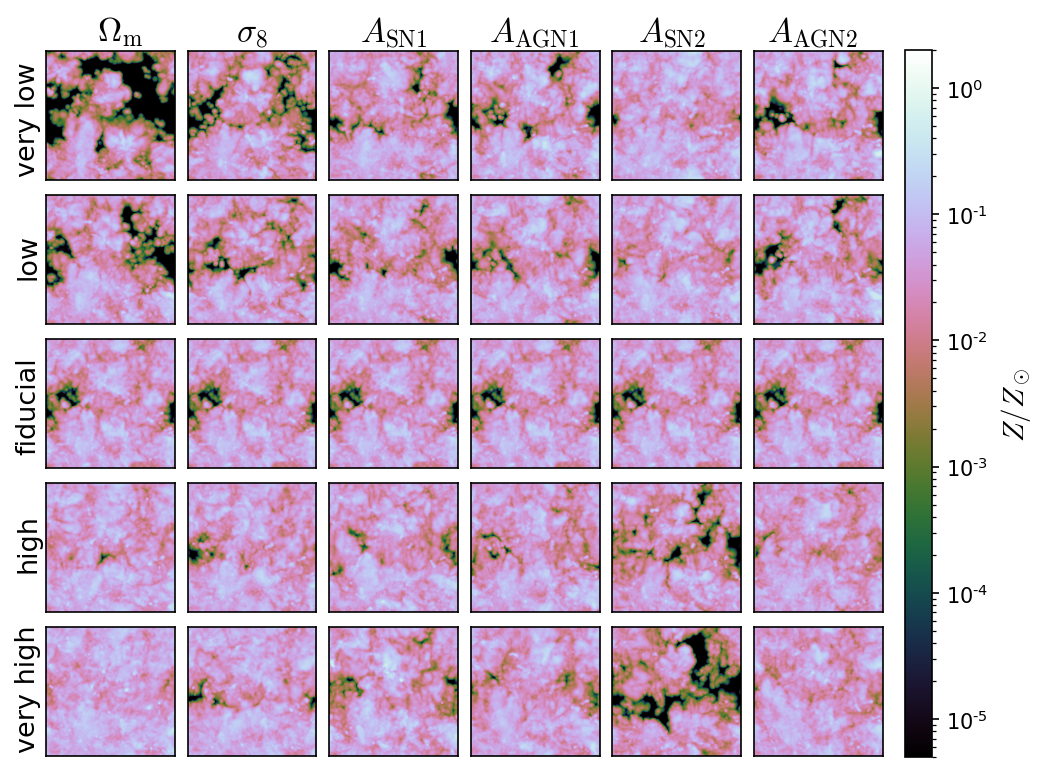}
\caption{Each panel shows the projected gas metallicity on a region of $25\times25\times5~(h^{-1}{\rm Mpc})^3$ at $z=0$ from the SIMBA 1P set. In these simulations we only vary the value of one parameter at a time; $\Omega_{\rm m}$, $\sigma_8$, $A_{\rm SN1}$, $A_{\rm SN2}$, $A_{\rm AGN1}$, and $A_{\rm AGN2}$, from left to right. The different rows show the results when the value of the changed parameter is very low (first row), low (second row), its fiducial value (third row), high (fourth row), and very high (fifth row). As can be seen, variations on each parameter affect the metallicity of the gas in a different way.}
\label{fig:image_1P}
\end{figure*}

Black hole feedback also follows a two-mode approach motivated by observations \citep{Heckman2014}, with high mass loading outflows in the radiative ``QSO'' mode and lower mass loading but faster outflows at low Eddington ratios in the jet mode.  In both cases, gas elements are ejected in a purely bipolar fashion, following a direction parallel and anti-parallel to the angular momentum vector of gas within the black hole kernel, and the total momentum flux satisfies:
\begin{equation}\label{eq:momflux}
\dot{P}_{\rm out} \, \equiv \, \dot{M}_{\rm out} \, v_{\rm out} \, = A_{\rm AGN1} \times 20 \,L_{\rm bol} / c,
\end{equation}
where $L_{\rm bol} = \epsilon_{\rm r} \, \dot{M}_{\rm BH} \, c^2$ is the bolometric luminosity, $ \epsilon_{\rm r}=0.1$ is the radiative efficiency, $c$ is the speed of light, and $A_{\rm AGN1}$ is introduced in CAMELS to vary the total momentum flux.  The outflow velocity in the radiative mode is based on observations of X-ray detected AGN from SDSS \citep{Perna2017}, parameterized in terms of $M_{\rm BH}$ (in \Msun) as $v_{\rm rad} = 500+500({\rm log}_{10}(M_{\rm BH})-6)/3$\;\kms.  Black holes with mass $M_{\rm BH}>10^{7.5}$\,\Msun~accreting at low Eddington ratio ($\lambda_{\rm Edd}<0.2$) are transitioned into a jet mode, which we model by including an additional velocity kick $v_{\rm jet} = 7000 \times {\rm min}[1,{\rm log}_{10}(0.2/\lambda_{\rm Edd})]$\,\kms, where $\lambda_{\rm Edd}\equiv \dot{M}_{\rm BH}/\dot{M}_{\rm Edd}$ and $\dot{M}_{\rm Edd}$ is the Eddington accretion rate.  In this way, full speed jets are only achieved in massive black holes with $\lambda_{\rm edd}\lesssim0.02$, as motivated by observations \citep{Fabian2012,Barisic2017}.  In summary, the velocity of black hole-driven outflows is given by:  
\begin{equation}
v_{\rm out} = 
\begin{cases}
    v_{\rm rad} + A_{\rm AGN2} \times v_{\rm jet}& \text{if } \begin{cases}
    \lambda_{\rm Edd}<0.2\\
    M_{\rm BH}>10^{7.5}\,{\rm M}_{\odot}\\
    \end{cases}\\[3ex]
    v_{\rm rad}& \text{otherwise,} \\
\end{cases}
\label{eq:vjet}
\end{equation}
where we have introduced $A_{\rm AGN2}$ to control the maximum jet velocity in CAMELS.  Once $v_{\rm out}$ is determined, the momentum flux specified in Eq.~\ref{eq:momflux} defines the mass outflow rate relative to the black hole accretion rate.  
Black holes with full-speed jets in gas-poor galaxies ($M_{\rm gas}/M_{\star} < 0.2$) can also inject energy into the surrounding gas from X-rays following \citet{Choi2012_BHmodel}.
All other aspects of the black hole model are identical to the original SIMBA simulations \citep{SIMBA}.

\subsection{Simulation sets}
\label{subsec:sim_sets}
Each of the ``IllustrisTNG'' and ``SIMBA'' suites in CAMELS contains 1,092 simulations from four different sets: 1) a set of 1,000 simulations with different initial random seeds varying all parameters using sampling from a latin hypercube (LH), 2) a set of 61 simulations with the same initial random seed varying only one parameter at a time (1P), 3) a set of 27 simulations with fixed cosmology and astrophysics that sample cosmic variance using different initial random seeds (CV), and 4) a set of four simulations representing extreme feedback models (EX) with fixed initial random seed. We now describe in detail each different set.

\subsubsection{LH set}
\label{subsec:LH}
LH stands for latin hypercube. This is a set of 1,000 simulations where the value of the cosmological and astrophysical parameters is arranged in a latin hypercube with $\Omega_{\rm m}\in[0.1,0.5]$, $\sigma_8\in[0.6, 1.0]$, $A_{\rm SN1}\in[0.25,4.0]$, $A_{\rm AGN1}\in[0.25,4.0]$, $A_{\rm SN2}\in[0.5,2.0]$ and $A_{\rm AGN2}\in[0.5,2.0]$. The initial random seed is different for each simulation. The latin hypercubes of the IllustrisTNG and SIMBA suites are different. Fig.~\ref{fig:image_LH} shows 2D projections of the gas density from 80 simulations of the SIMBA LH set, from which the large variety covered by the CAMEL simulations can be appreciated. These simulations are designed to make predictions as a function of cosmology and astrophysics that take into account the effect of cosmic variance. Thus, they can also be used to train neural networks to marginalize over baryonic effects.

\begin{figure*}
\centering
\includegraphics[width=0.65\textwidth]{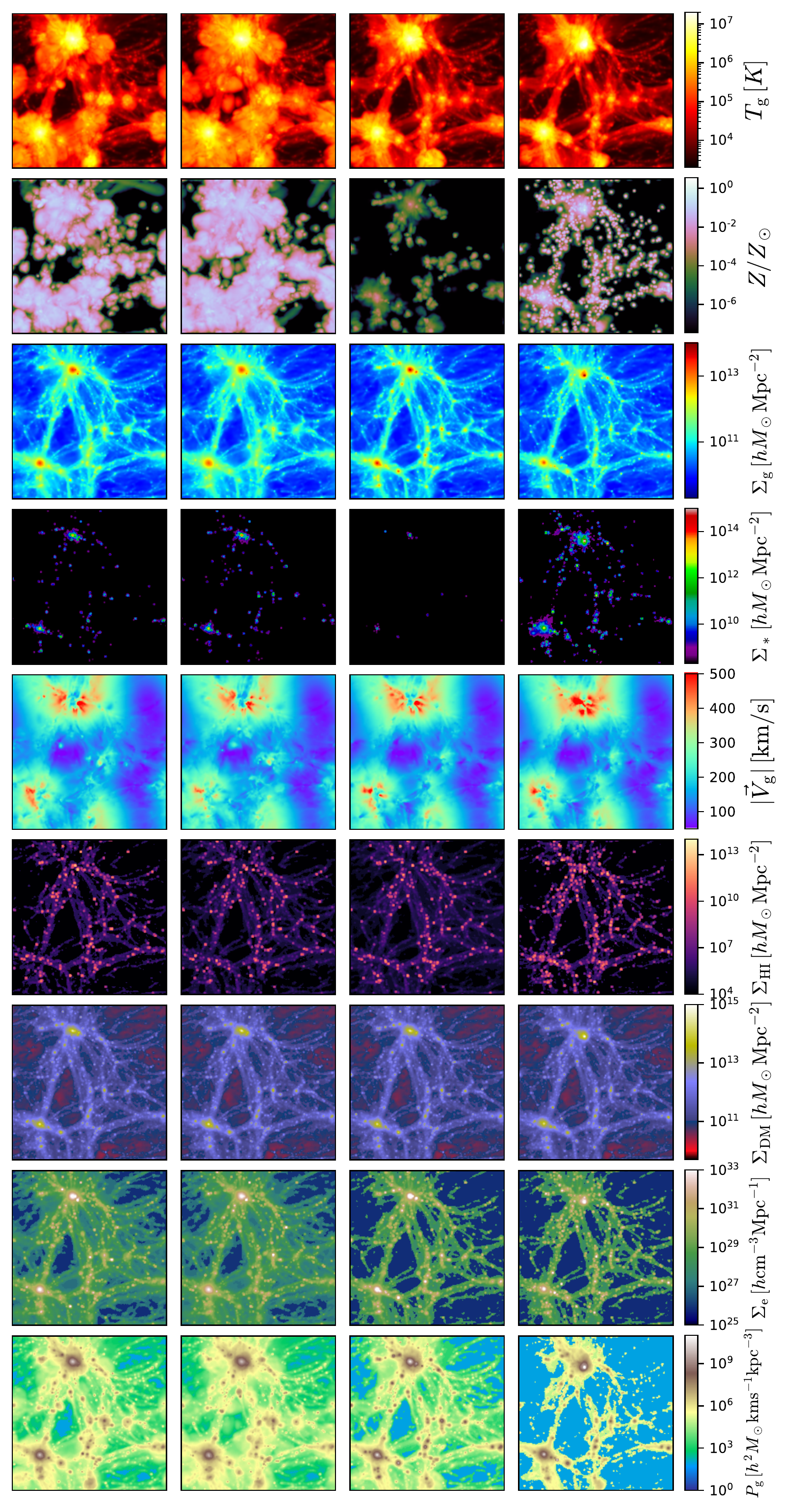}
\caption{This figure shows 2D projections of the gas temperature, gas metallicity, gas density, stellar mass, gas velocity, neutral hydrogen, dark matter, electron number density, and gas pressure (from top to bottom) over a region with $25\times25\times5~(h^{-1}{\rm Mpc})^3$ from the four different IllustrisTNG EX simulations at $z=0$. The different columns show results for the simulations with 1) fiducial astrophysics (left), 2) extreme AGN feedback (center-left), 3) extreme supernova feedback (center-right), and 4) no feedback (right). It can be seen how feedback affects different fields in a different manner.}
\label{fig:image_EX}
\end{figure*}

\subsubsection{1P set}
\label{subsec:1P}
1P stands for 1-parameter. This set contains 61 simulations sharing the value of the initial random seed. In these simulations we vary the values of all parameters ($\Omega_{\rm m}$, $\sigma_8$, $A_{\rm SN1}$, $A_{\rm SN2}$, $A_{\rm AGN1}$ and $A_{\rm AGN2}$), but only one at a time. The parameters are varied in the same range as the LH set: $0.1\leq\Omega_{\rm m}\leq0.5$, $0.6\leq\sigma_8\leq1.0$, $0.25\leq (A_{\rm SN1}, A_{\rm AGN1})\leq4.00$, and $0.5\leq (A_{\rm SN2}, A_{\rm AGN2})\leq2.0$. 
For $\Omega_{\rm m}$ and $\sigma_8$ the spacing is linear while for the astrophysical parameters it is in log scale. Fig.~\ref{fig:image_1P} shows the $z=0$ projected gas metallicity of 25 of these simulations from the SIMBA set, illustrating that the nature of the response of this field to parameter variations differs from parameter to parameter.

The purpose of this set is to understand responses of an observable or quantity to individual changes of the parameters, without being contaminated by cosmic variance or changes of other parameters. The IllustrisTNG and SIMBA 1P sets share the same initial random seed; therefore, these simulations can also be used to study changes across simulation schemes.

\subsubsection{CV set}
\label{subsec:CV}
CV stands for cosmic variance. There are 27 simulations in this set. The value of the cosmological and astrophysical parameters is fixed at $\Omega_{\rm m}=0.3$, $\sigma_8=0.8$, $A_{\rm SN1}=A_{\rm SN2}=A_{\rm AGN1}=A_{\rm AGN2}=1$. The only difference among the simulations is the value of the initial random seed, i.e.~the initial conditions. These simulations have the same value of the feedback parameters as the original IllustrisTNG and SIMBA simulations. We will refer to simulations on this set, and in general, to simulations with $\Omega_{\rm m}=0.3$, $\sigma_8=0.8$, $A_{\rm SN1}=A_{\rm SN2}=A_{\rm AGN1}=A_{\rm AGN2}=1$ as the fiducial model.

These simulations are designed to quantify the effect of cosmic variance on a given observable.

\subsubsection{EX set}
\label{subsec:Set4}
EX stands for extreme. This set only contains 4 simulations which share the value of the initial random seed and the cosmological parameters: $\Omega_{\rm m}=0.3$ and $\sigma_8=0.8$. One simulation is the fiducial model with $A_{\rm SN1}=A_{\rm SN2}=A_{\rm AGN1}=A_{\rm AGN2}=1$. The other three simulations are designed to show extreme cases; one simulation has extreme AGN feedback ($A_{\rm AGN1}=100$), another has extreme supernova feedback ($A_{\rm SN1}=100$), while the last one does not have any feedback at all ($A_{\rm AGN1}=A_{\rm SN1}=0$, as well as analogously setting to zero the energy input from the other AGN feedback modes). The value of the initial random seed of these simulations is different from the one in the 1P set, but it is the same between the IllustrisTNG and SIMBA EX sets. Fig.~\ref{fig:image_EX} shows different fields from these 4 simulations from the IllustrisTNG set.

The purpose of these simulations is to investigate the response of observables or quantities to extreme feedback scenarios. 

\subsection{Dark Matter}
\label{subsec:Nbody}

For each (magneto-)hydrodynamic simulation, we have additionally run its N-body counterpart. These simulations have the same value of $\Omega_{\rm m}$, $\sigma_8$ and initial random seed, but are run using GADGET-III \citep{Springel2005_Gadget}. The softening length is set to 0.5 $h^{-1}{\rm kpc}$. Table \ref{table:sims} summarizes the different sets of simulations in CAMELS.

\section{Cosmological and astrophysical properties}
\label{sec:Properties}

In this section we show some of the cosmological and astrophysical properties of the simulations. We also quantify the range of variation due to changes in cosmology and astrophysics, together with the scatter arising from cosmic variance. 

We first show the results when varying cosmology, astrophysics and the initial random seed by using the LH set. We quantify what fraction of the range of variation is due to cosmic variance versus changes in cosmology and astrophysics by using the CV set. We then focus our attention in the variations due to individual changes in cosmology and astrophysics by using the 1P set.

\subsection{Median, range of variation, and scatter}

\begin{figure*}
\centering
\includegraphics[width=0.99\textwidth]{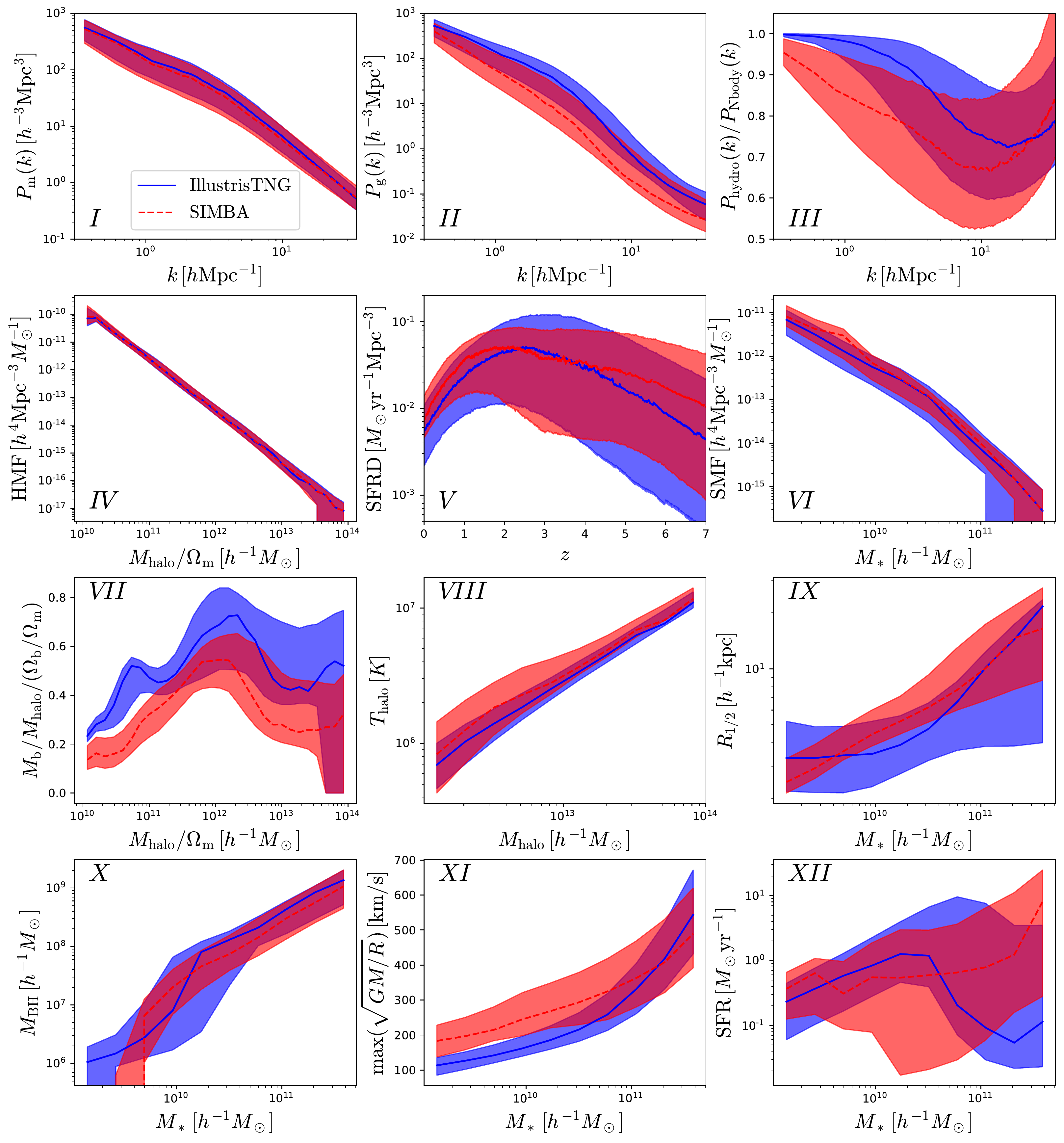}
\caption{Comparison between different cosmological and astrophysical properties of the IllustrisTNG (blue) and SIMBA (red) suites. Results are shown at $z=0$, with the exception of panel $V$, that shows the cosmic star formation rate density as a function of redshift. For each property we show the median of the CV set (27 simulations with fixed cosmology and astrophysics and different initial random seed, which correspond to the original IllustrisTNG and SIMBA models) and 16-84 percentiles from the LH set (1,000 simulations with different cosmologies, astrophysics, and initial random seeds). From top left to bottom right: matter power spectrum ($I$), gas power spectrum ($II$), ratio of matter power spectrum in hydrodynamic simulations to that of the corresponding N-body simulations ($III$), halo mass function ($IV$), star formation rate density ($V$), stellar mass function ($VI$), halo baryon fraction ($VII$), mass-weighted halo temperature ($VIII$), galaxy size versus stellar mass ($IX$), black hole mass versus host galaxy stellar mass ($X$), galaxy circular velocity versus stellar mass ($XI$), and star formation rate versus stellar mass for galaxies with non-zero star formation ($XII$). The wide range of parameters explored translates into a large spread in almost all quantities analyzed. The IllustrisTNG and SIMBA sets overlap in most cases, with some systematic differences in e.g.~gas power spectrum and baryon fractions.}
\label{fig:Properties}
\end{figure*}

We consider twelve different cosmological and astrophysical quantities and calculate their statistical properties. For this, we made use of the 1,000 simulations of the LH set, for both the IllustrisTNG and SIMBA suites. For each simulation suite, we consider a given quantity and compute its mean, median, standard deviation, and 16-84 percentiles over all simulations. Fig. \ref{fig:Properties} shows the results at $z=0$.

\begin{figure*}[ht!]
\centering
\includegraphics[width=0.995\textwidth]{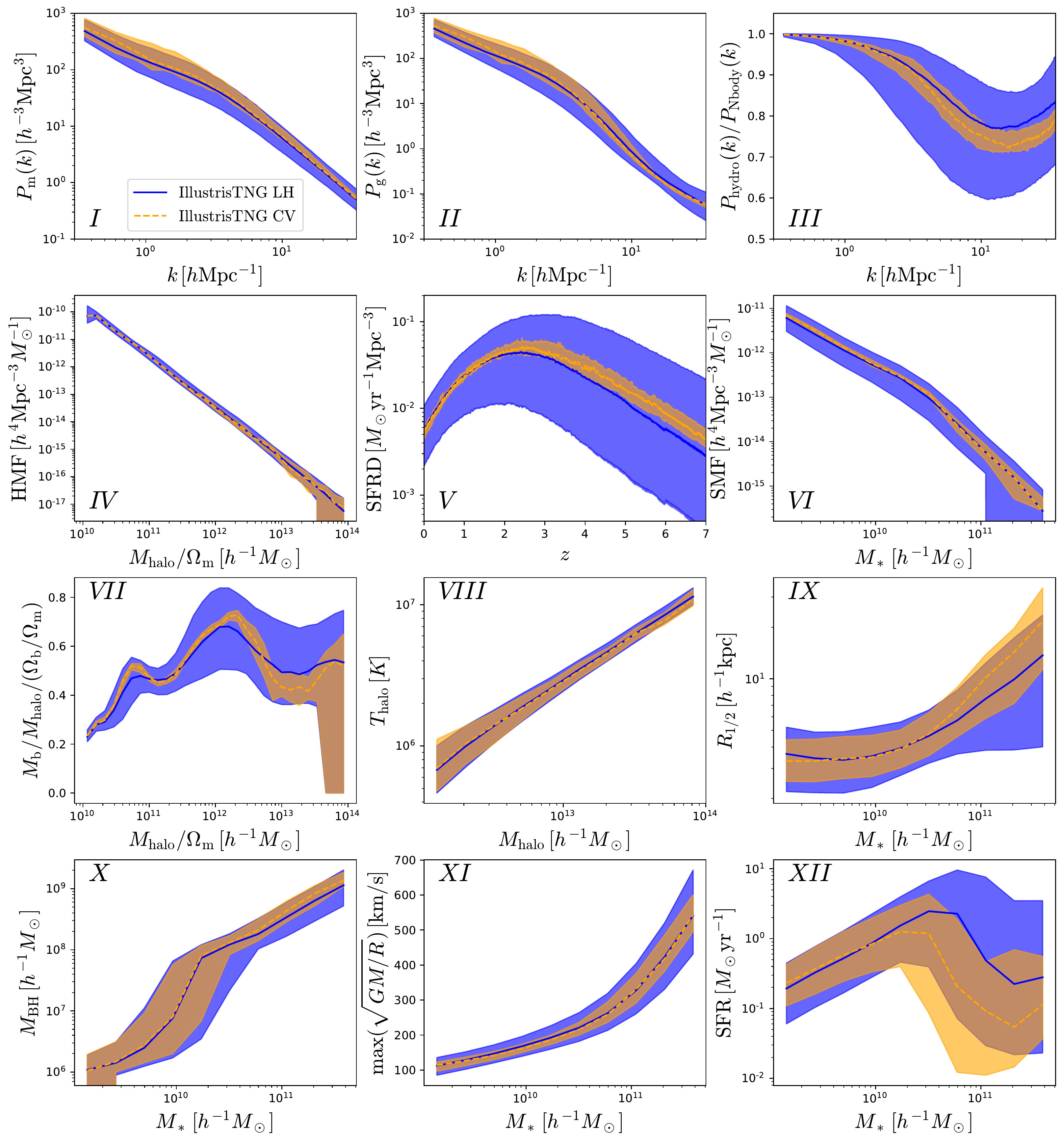}
\caption{This figure shows different cosmological and astrophysical properties of the simulations at $z=0$, as in Fig. \ref{fig:Properties}. For each property, we show the median and 16-84 percentiles from all simulations in the IllustrisTNG LH (solid line and blue region) and IllustrisTNG CV (dashed line and orange region) sets. We remind the reader that in the LH simulations, we vary the cosmology, the astrophysics and the initial random seed, whereas in the CV we fix the cosmology and astrophysics and only vary the initial random seed. We can see that changes in cosmology and astrophysics dominate the range of variation on many of the considered properties, which the exception of the halo temperature, galaxy size, black hole mass, maximum circular velocity, and star formation rate, where the scatter arising from cosmic variance accounts for a significant fraction of the variability observed in the LH simulations.}
\label{fig:LH_vs_CV}
\end{figure*}

We also quantify what fraction of the range of variation of each quantity is due to cosmic variance versus changes in the cosmological and astrophysical parameters. For this, we compute the median and 16-84 percentiles of the IllustrisTNG LH and CV sets and show the results in Fig. \ref{fig:LH_vs_CV}. We obtain similar results if we use the SIMBA suite instead. From Fig.~\ref{fig:LH_vs_CV} we can also quantify whether the median of the LH set agrees with the one of the fiducial model, represented by the CV set.

As we shall see below, there are many quantities where variations due to changes in cosmology and astrophysics are much larger than those arising from cosmic variance. It is important to emphasize that in many cases, this is due to the fact that our changes in cosmology and astrophysics are rather extreme. For instance, $\Omega_{\rm m}$ and $\sigma_8$ are varied within $[0.1- 0.5]$ and $[0.6 - 1.0]$, respectively. Should we have varied $\Omega_{\rm m}$ and $\sigma_8$ within Planck constraints, differences will be much smaller. 

We now describe in more detail each considered quantity.

\subsubsection{Matter power spectrum}

We compute the total matter power spectrum by assigning particle masses to a regular grid with $512^3$ voxels. We consider all different particle types: gas, dark matter, stars, and black holes. We then Fourier transform the grid and compute the power spectrum by averaging over $k$-bins with a width equal to the fundamental frequency, $k_F=2\pi/L$, with $L=25~h^{-1}{\rm Mpc}$. 

The panel $I$ of Fig.~\ref{fig:Properties} shows the results at $z=0$. We find that both IllustrisTNG and SIMBA agree very well on both their medians and percentiles. The range of variation in the matter power spectrum goes from $40\%$ at $k\simeq30~h{\rm Mpc}^{-1}$ to $70\%$ at $k\simeq1~h{\rm Mpc}^{-1}$. Put in context, this is very large; for instance, fixing the values of initial scalar amplitude, $A_s$, and $\Omega_{\rm m}$, it is expected that neutrinos with masses $\sim m_\nu=0.6$ eV will produce a maximum change in the amplitude of the matter power spectrum equal to $\simeq45\%$ with respect to a model with massless neutrinos\footnote{Fixing $A_s$ and $\Omega_{\rm m}$ the maximum suppression in the matter power spectrum is expected to be $\delta P(k)/P(k)\simeq -10\Omega_\nu/\Omega_{\rm m}$ \citep{Brandbyge_2008}.}. We emphasize that the possibility that neutrinos have such large masses is ruled out at many sigmas with current cosmological data. This demonstrates the wide range in parameter space that our simulations cover. 

From Fig. \ref{fig:LH_vs_CV} we can see that the median of the LH set agrees well with the median of the CV set. We also find that the range of variation on large scales is dominated by cosmic variance, whereas on small scales changes in cosmology and astrophysics dominate.

\begin{figure}
\centering
\includegraphics[width=0.49\textwidth]{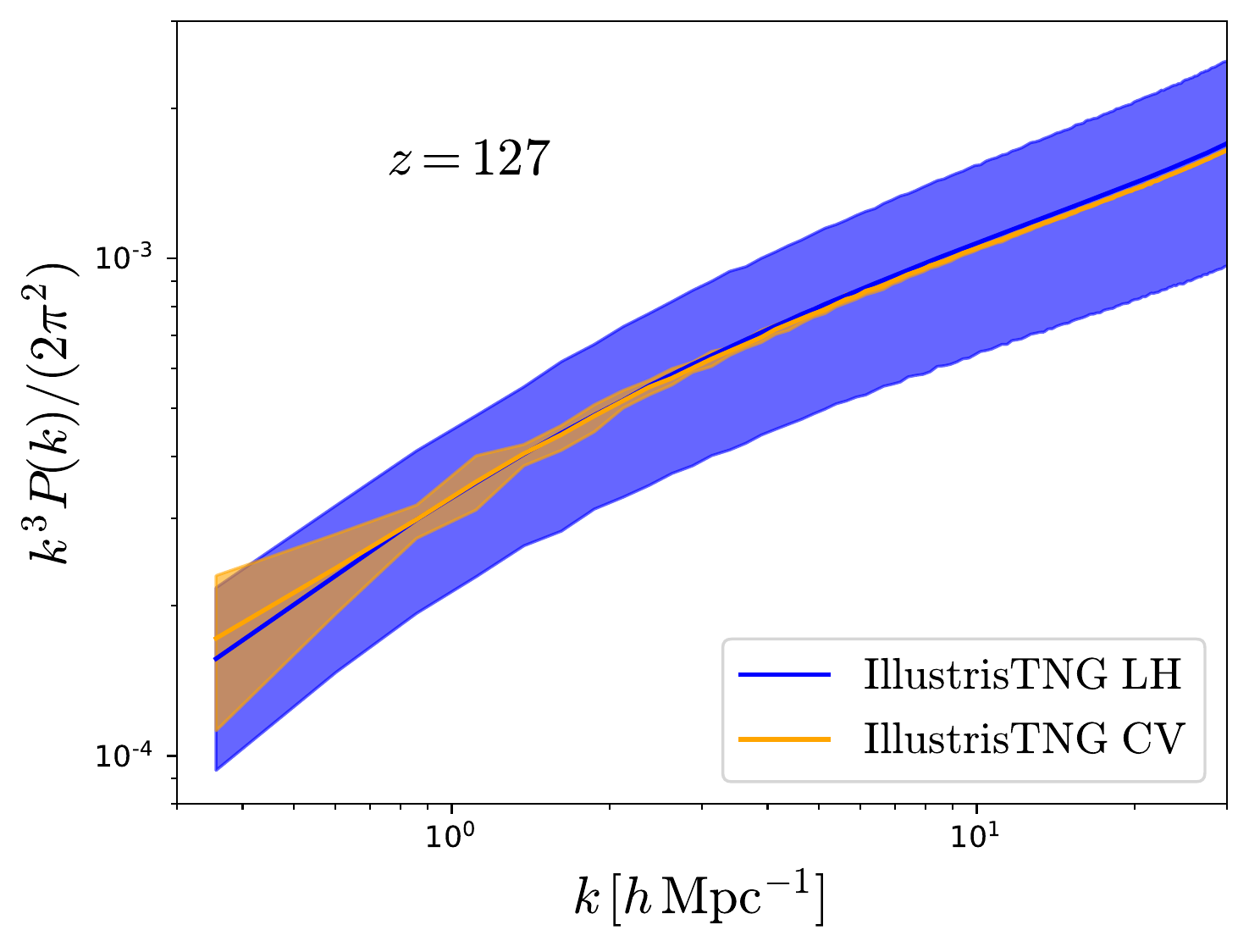}
\caption{The solid lines show the mean dimensionless matter power spectrum of the initial conditions ($z=127$) from the LH (blue) and CV (orange) sets for the IllustrisTNG suite, together with their scatter. We note that differences at this redshift are purely due to changes in cosmology and cosmic variance. While the scatter on large scales is comparable among the two sets, on small scales the scatter due to changes in cosmology is much larger than the one due to cosmic variance. Similar conclusions hold for the SIMBA suite.}
\label{fig:Pk_ICs}
\end{figure}

To further emphasize the scatter due to cosmic variance versus the one from changes in cosmology we plot in Fig. \ref{fig:Pk_ICs} the mean and 16-84 percentiles of the dimensionless matter power spectrum of the LH and CV sets from the IllustrisTNG suite at $z=127$, where changes in astrophysics are irrelevant. We find that while the scatter due to changes in cosmology and cosmic variance are similar on the largest scales probed by our simulations, differences on small scales are very large; changes due to cosmic variance are negligible in comparison with those due to cosmology. These changes couple with effects from astrophysics to produce the results shown in panel I of Fig. \ref{fig:LH_vs_CV}.

It is also interesting to note that the scatter in the CV set has grown significantly on small scales from $z=127$, where it was very small, down to $z=0$, where it is a fraction of the one from the LH set. The reason for this scatter enhanced can be due to both astrophysics effects but also due to the transfer of power from large to small scales \citep{Knebe_2003}.

\vspace{1cm}
\subsubsection{Gas power spectrum}

In the second panel of Fig.~\ref{fig:Properties} ($II$) we show the power spectrum of the gas component. We compute $P_{\rm g}(k)$ using the same procedure as above, but only using the gas resolution elements. In this case, we observe a larger difference between the IllustrisTNG and SIMBA suites, with the clustering of the latter being lower than the one of the former. Note however that results overlap in a wide region. This difference is due to the different subgrid feedback implementations and effective feedback strengths introduced by the parameter variations in the two simulation suites.  

In this case, the range of variation in the gas power spectrum is much larger than that of the total matter power spectrum: from $\sim50\%$ at $k=0.4~h{\rm Mpc}^{-1}$ to $\simeq130\%$ at $k=10~h{\rm Mpc}^{-1}$. The percentile range in the SIMBA suite is larger than in the IllustrisTNG  one, at all wavenumbers. This may point towards feedback being, on average, more effective at ejecting gas to large scales in SIMBA \citep{Borrow2020} than in IllustrisTNG.

From Fig. \ref{fig:LH_vs_CV} we can see that the medians of the LH and CV set agree well. The range of variation on this quantity is dominated by cosmic variance/changes in cosmology and astrophysics on large/small scales, similarly to the matter power spectrum. 

\subsubsection{$P_{\rm hydro}(k)/P_{\rm Nbody}(k)$}

We have computed the total matter power spectrum for each hydrodynamic simulation, $P_{\rm hydro}(k)$, and for its N-body counterpart, $P_{\rm Nbody}(k)$. We then compute the ratio $P_{\rm hydro}(k)/P_{\rm Nbody}(k)$ for each simulation. The third panel of Fig.~\ref{fig:Properties} ($III$) shows the median of the CV set and 16-84 percentiles of the LH set. We find that, on average, this ratio exhibits a characteristic spoon shape already noted and studied in the literature \citep{vanDaalen2011,Chisari_2019,Schneider_2019,vanDaalen2020}.

\begin{figure*}
\centering
\includegraphics[width=0.99\textwidth]{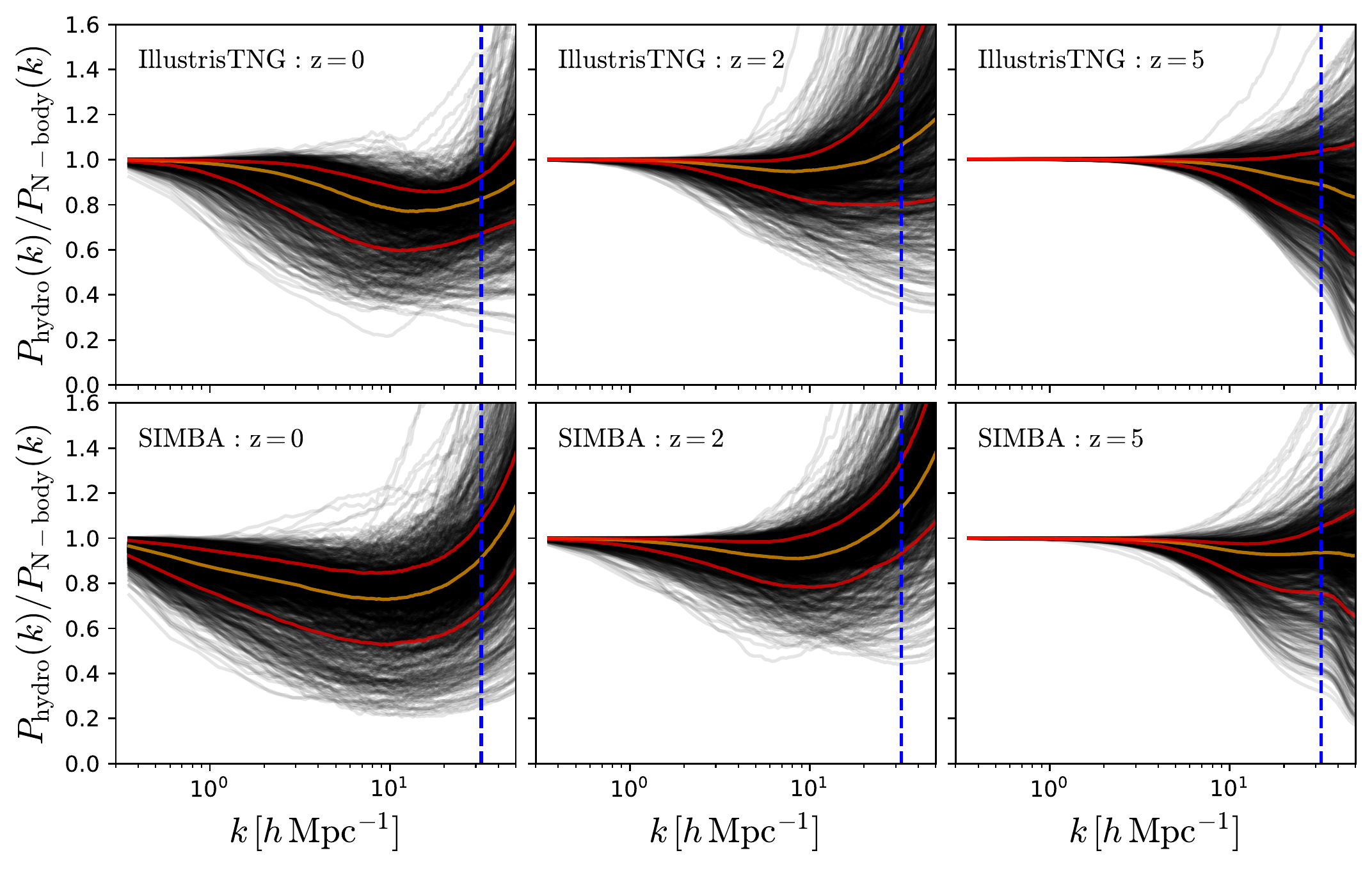}
\caption{Ratio between the matter power spectrum from hydrodynamic simulations to that of their N-body counterparts. Each panel shows the results for either the IllustrisTNG LH set (top row) or the SIMBA LH set (bottom row) at redshifts $z=0$ (left column), $z=2$ (middle column), and $z=5$ (right column). The orange line shows the median while the red lines display the 16-84 percentiles. The strength of feedback processes is varied over a wide range in CAMELS, giving rise to a large variation in the matter power spectrum ratio. It can be seen how astrophysical effects start on small scales at high redshift and propagate to larger scales at lower redshifts.}
\label{fig:Pk_ratio}
\end{figure*}

We note that the power spectrum ratio can reach $\sim 50\%$ at $k \sim 10\,h{\rm Mpc}^{-1}$ in strong feedback cases. This demonstrates the importance of including baryonic effects in the epoch of precision cosmology aimed at extracting information from small scales. Overall, the results of both simulations overlap but the range of variation is smaller in IllustrisTNG, where the $P_{\rm hydro}(k)/P_{\rm Nbody}(k)$ ratio tends to be closer to 1. This points toward the same conclusion as above: feedback appears to be less efficient in the IllustrisTNG than in SIMBA model.

It is expected that the range of variation of this quantity will be dominated by changes in the cosmological and astrophysical parameters. This is because it is computed as the ratio between two simulations that have the same initial random seed. We can see the contribution of cosmic variance to the range of variation in the panel $III$ of Fig. \ref{fig:LH_vs_CV}. Indeed, we find that cosmic variance does not dominated the scatter, although its contribution is not negligible.

In Fig.~\ref{fig:Pk_ratio} we show $P_{\rm hydro}(k)/P_{\rm Nbody}(k)$ again but at different redshifts for the LH sets of both IllustrisTNG and SIMBA. Each individual line represents the result of one simulation. It can be seen that, at high redshift, baryonic effects only appear at $k\simeq3~h{\rm Mpc}^{-1}$, while as time proceeds, baryonic effects propagate to larger scales. At $z=0$, percent-level differences appear on the largest scales that our simulations probe: $k\simeq0.3~h{\rm Mpc}^{-1}$.

\subsubsection{Halo mass function}

\begin{figure}
\centering
\includegraphics[width=0.49\textwidth]{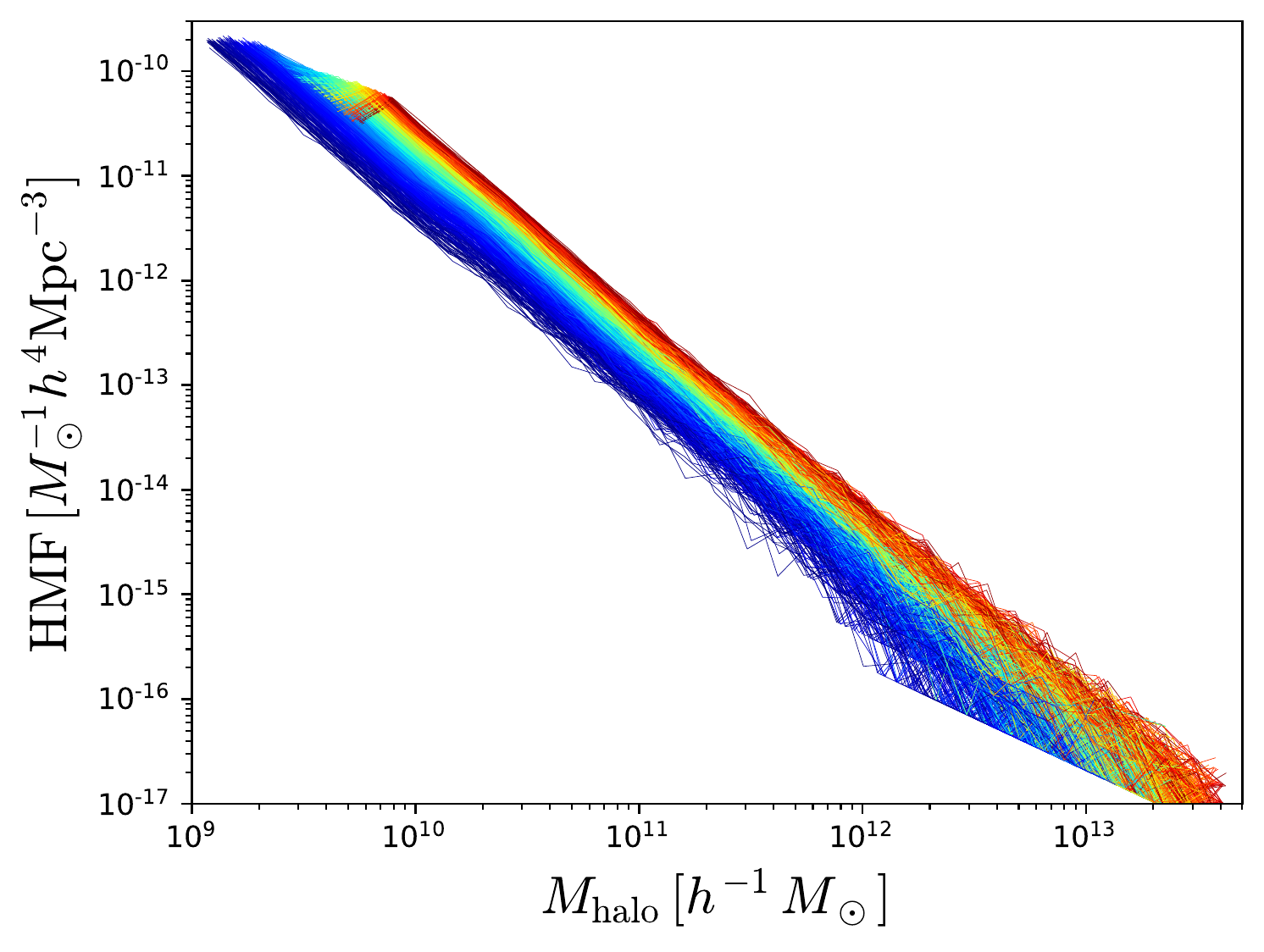}
\caption{Halo mass functions for the 1000 IllustrisTNG LH simulations at $z=0$. Each line is color coded according to the value of $\Omega_{\rm m}$, from 0.1 (blue) to 0.5 (red). At fixed halo mass, there is a very large range of variation ($>$1 dex) that strongly correlates with $\Omega_{\rm m}$. This variation is greatly reduced if the halo mass function is plotted against the reduced halo mass, $M_{\rm halo}/\Omega_{\rm m}$, instead of halo mass $M_{\rm halo}$ (see panel $IV$ of Fig.~\ref{fig:Properties}).}
\label{fig:HMF}
\end{figure}

For each simulation, we compute the halo mass function as the number density of FoF halos per unit of mass. The mass of our FoF halos includes dark matter, gas, stars and black holes. To minimize numerical effects, we only consider halos with more than 50 dark matter particles. The halo mass function for the 1,000 IllustrisTNG simulations of the LH set is shown in Fig.~\ref{fig:HMF}. Each line is colored according to the value of $\Omega_{\rm m}$. For a given halo mass, we observe a large range of variation in the halo mass function that correlates very well with $\Omega_{\rm m}$: the halo mass function increases with $\Omega_{\rm m}$, as expected. 

The forth panel ($IV$) of Fig.~\ref{fig:Properties} shows the halo mass function against the \textit{reduced halo mass}, defined as $M_{\rm halo}/\Omega_{\rm m}$, where $M_{\rm halo}$ is the FoF halo mass. We observe a much smaller range of variation when using the reduced halo mass: from $\simeq 25\%$ for reduced halo masses smaller than $10^{11}~h^{-1}M_\odot$ to $100\%$ at $10^{14}~h^{-1}M_\odot$. It is interesting to observe such small range of variation in the low mass end, considering the wide range of feedback strengths used in our simulations. The agreement between the results from the SIMBA and IllustrisTNG suites is excellent for all reduced halo masses, for both the median values and the range of variation. 

From the panel $IV$ of Fig. \ref{fig:LH_vs_CV} we can see that the small range of variation on this quantity is dominated by changes in cosmology and astrophysics, with the exception of the high-mass end, where cosmic variance dominates. We also find that the median from the LH and CV sets agree very well.

\subsubsection{Star formation rate density}

The fifth panel ($V$) of Fig.~\ref{fig:Properties} shows the results for the cosmic star formation rate density (SFRD). For each simulation, we have computed the global star formation rate per unit comoving volume. We interpolate the measurements in 10,000 bins from $z=0$ to $z=10$, then compute the median and 16-84 percentiles for each simulation set. 

We find that the SFRDs of the IllustrisTNG and SIMBA sets overlap at all redshifts, with some interesting differences.
At low redshift, the median SFRD of the SIMBA simulations is $\simeq30\%$ higher than that of the IllustrisTNG simulations, though well within the range of variation. In this case, the range of variation in the IllustrisTNG simulations is larger, at all redshifts, than the one in SIMBA.

The panel $V$ of Fig. \ref{fig:LH_vs_CV} shows that cosmic variance only contributes a small fraction to the range of variation of the SFRD, at any redshift. While the medians of the LH and CV sets agree well at low redshift, they significantly differ at high-redshift. This indicates that some extreme models may induce long tails in the distribution of SFRD at high-redshifts.

\subsubsection{Stellar mass function}

We have computed the stellar mass function as the number density of galaxies per unit stellar mass. We consider as a galaxy any subhalo with non-zero stellar mass identified by SUBFIND. For each simulation, we compute the stellar mass function using 10 bins equally spaced in log from $10^9$ to $5\times10^{11}~h^{-1}M_\odot$ and show the median and 16-84 percentiles in the sixth panel ($VI$) of Fig.~\ref{fig:Properties}.

The means of both sets agree well, with the exception of the low mass end, where the SIMBA set is slightly higher. As with the SFRD, the IllustrisTNG simulations exhibit a larger range of variation for all stellar masses than the SIMBA simulations. In both cases, the typical range of variations is $\sim0.5-1$ dex, which is similar to that seen in a previous study that explored different types of variations around the original IllustrisTNG model \citep{Pillepich_2018}.

From the panel $VI$ of Fig. \ref{fig:LH_vs_CV} we can see that the medians of the LH and CV sets agree very well. The range of variation of the stellar mass function is dominated by changes in cosmology and astrophysics, although cosmic variance plays an important role in the high-mass end.

\subsubsection{Baryon fractions}

For each FoF halo in the simulations, we compute its baryon fraction as $f_{\rm b}=M_{\rm b}/M_{\rm halo}$, where $M_{\rm b}$ and $M_{\rm halo}$ are the baryonic and total FoF halo mass, respectively; the baryonic mass includes gas, stars, and black holes whilst the total halo mass additionally includes dark matter.In this case, we only consider halos with more than 50 dark matter particles.

We compute the median and 16-84 percentile range of the baryon fraction of all halos in bins of halo mass. As with the halo mass function, we find more convenient to work with the reduced FoF halo mass, $M_{\rm halo}/\Omega_{\rm m}$. We show the results on the seventh panel ($VII$) of Fig.~\ref{fig:Properties}. Note that we have normalized the baryon fraction in each simulation by the corresponding cosmic mean, $\Omega_{\rm b}/\Omega_{\rm m}$.

This quantity exhibits the largest difference between the two simulation sets. For reduced halo masses below $\sim10^{11}~h^{-1}M_\odot$, the results of the LH simulation sets do not overlap. For the full range of reduced halo mass considered, the baryon fractions of the SIMBA simulations are always below those of the IllustrisTNG simulations. These differences arise because of the different feedback implementations and parameter variations. The range of variation of the SIMBA set is comparable to that of the IllustrisTNG set for almost all reduced halo masses. However, in both simulation sets we observe the same trend with halo mass: low baryon fractions for low reduced mass halos ($\sim10^{10}~h^{-1}M_\odot$), increasing to reach peak values around $\sim10^{12}~h^{-1}M_\odot$, and decreasing for higher mass halos.

Since baryon fractions are lower in the SIMBA set, we expect gas to be distributed in a more uniform manner compared to IllustrisTNG set, which should correspond to a lower amplitude of the gas power spectrum. This is in agreement with the aforementioned results for the gas power spectrum. The panel $VII$ of Fig. \ref{fig:LH_vs_CV} shows that the medians of the LH and CV sets of IllustrisTNG are in good agreement within the range of variation. The contribution of cosmic variance to the range of variation is very small for all halo masses, with the exception of the high-mass end.

\subsubsection{Halo temperature}

We compute the average halo temperature for each halo in all the simulations. In this case, we use Spherical Overdensity (SO) halos, where the halo radius is defined such that the mean enclosed density is equal to $\bigtriangleup_c$ times the critical density of the Universe. We adopt $\bigtriangleup_c$ from spherical collapse calculations as in \cite{Bryan_Norman_97}. We compute the average gas temperature of a halo as  
\begin{equation}
\bar{T}_{\rm halo}=\frac{\sum_{i\in{\rm halo}}T_im_i}{\sum_{i\in{\rm halo}}m_i}~,
\end{equation}
where $m_i$ and $T_i$ are the mass and temperature of gas particle $i$ belonging to the halo. We then consider 10 bins logarithmically spaced from $10^{12}$ to $10^{14}~h^{-1}M_\odot$ and compute the median and 16-84 percentiles of the temperature of halos in each mass bin. The results are shown in the panel $VIII$ of Fig.~\ref{fig:Properties}.

As expected, we find that the median halo temperature increases with halo mass, for both the IllustrisTNG and SIMBA sets. We find that halo temperatures, at fixed mass, are systematically higher in SIMBA in comparison with IllustrisTNG. The range of variation in the SIMBA simulation set is also larger than in IllustrisTNG, particularly for low mass halos. However, within the range of variation, both simulation sets agree very well.

The panel $VIII$ of Fig. \ref{fig:LH_vs_CV} shows good agreement between the medians of the LH and CV sets of IllustrisTNG. We find that cosmic variance is responsible for most of the range of variation observed at all halo masses.

\vspace{0.5cm}
\subsubsection{Galaxy sizes}

We now consider different galaxy properties, starting with galaxy sizes. We define the size of a galaxy as the radius at which a subhalo contains half of its stellar mass. We take 10 bins equally spaced in log from $10^9$ to $5\times10^{11}~h^{-1}M_\odot$ in stellar mass and assign each galaxy radius to its stellar mass bin. We then compute the median and 16-84 percentiles. We show the results in the panel $IX$ of Fig.~\ref{fig:Properties}.

We find that SIMBA galaxies are, on average, larger than those in IllustrisTNG, with the exception of galaxies with low stellar masses. The range of variation in the SIMBA simulation set is, in general, smaller than in IllustrisTNG. At almost all stellar masses, the full 16-84 percentile range of the two simulation sets overlap, while for galaxies with stellar masses around $4\times10^{10}~h^{-1}M_\odot$ the mean in SIMBA is almost at the 84\% tail of the IllustrisTNG simulations.

From the panel $IX$ of Fig. \ref{fig:LH_vs_CV}, we can see that the difference between the medians of the LH and CV sets in IllustrisTNG is within the range of variation. Also for this quantity, cosmic variance itself can explain most of the range of variation seen when comparing to the LH set. We emphasize that the way we compute cosmic variance automatically incorporates the intrinsic scatter expected in this, and the subsequent galaxy scaling relations. Thus, while we expect that changes in cosmology and astrophysics may modify the mean and median of these scaling relations, the intrinsic scatter may be larger than these differences. We defer the study of the systematic changes in mean and medians as a function of cosmology and astrophysics to future work. 

\subsubsection{Galaxy black hole masses}

For each galaxy we compute the total mass contained in all black holes gravitationally bound to it. We take 10 bins equally spaced in log from $10^9$ to $5\times10^{11}~h^{-1}M_\odot$ in stellar mass and assign each galaxy total black hole mass to its stellar mass bin. We then compute the median and 16-84 percentiles, with the results shown in the panel $X$ of Fig.~\ref{fig:Properties}.

We find good agreement between the median black hole masses of both simulation suites; however those for the SIMBA set drop quickly for stellar masses below $\sim5\times10^9~h^{-1}M_\odot$, owing to the black hole seeding model \citep{SIMBA,Thomas2019}; this feature is not observed in the IllustrisTNG set. Both sets overlap at all stellar masses, with the exception of the lower stellar mass range. The range of variation in the IllustrisTNG set is larger than that of SIMBA at stellar masses $< 5\times10^{10}\,h^{-1}M_\odot$ but they are comparable at higher masses. See \citet{Habouzit2020} for a comprehensive comparison of black hole properties in the original IllustrisTNG and SIMBA as well as other recent cosmological simulations. 

The panel $X$ of Fig. \ref{fig:LH_vs_CV} shows that most of the range of variation observed in the LH simulations is due to cosmic variance (that incorporates the contribution from intrinsic scatter), in particular for low-mass halos. On the other hand, the agreement between the medians of the LH and CV sets is very good and within the scatter due to cosmic variance.

\subsubsection{Galaxy circular velocities}

Next, we consider the maximum value of the galaxy circular velocity, defined as
\begin{equation}
V_{\rm max}=\max\left( \sqrt{\frac{GM(r)}{r}}\right),
\end{equation}
where $M(r)$ is the total enclosed mass within a spherical radius $r$.
We take 10 bins equally spaced in log from $10^9$ to $5\times10^{11}~h^{-1}M_\odot$ in stellar mass and assign each galaxy maximum circular velocity to its stellar mass bin. We then compute the mean, median, standard deviation, and 16-84 percentiles, showing the results in the panel $XI$ of Fig.~\ref{fig:Properties}.

The median values of both sets agree very well in the high-mass end: $M_*>2\times10^{11}~h^{-1}M_\odot$. For lower stellar masses, SIMBA galaxies have, on average, larger maximum circular velocities than IllustrisTNG galaxies. Combined with the size-mass relation discussed above, this indicates complex differences in the mass distribution within galaxies between the two simulation suites. The range of variation in the SIMBA set is, for almost all stellar masses, larger than the one from the IllustrisTNG simulations. 
The panel $XI$ of Fig. \ref{fig:LH_vs_CV} shows that the medians of the LH and CV sets agree very well. Also for this property, cosmic variance is responsible for most of the range of variation. We remind the reader than our cosmic variance budget incorporates the intrinsic scatter for this relation.

\subsubsection{Galaxy star formation rates}

The last quantity we consider is the star formation rate of galaxies, which we compute as the sum of the instantaneous star formation rates of all the gas particles belonging to each galaxy. Here we only consider galaxies with non-zero star formation rate to avoid the distributions being heavily affected by fully quenched galaxies. We take 10 bins equally spaced in log from $10^9$ to $5\times10^{11}~h^{-1}M_\odot$ in stellar mass and assign each galaxy total star formation rate to its stellar mass bin. We then compute the mean, median, standard deviation, and 16-84 percentiles, showing the results in the last panel ($XII$) of Fig.~\ref{fig:Properties}. 

We find a large range of variation in this quantity, spanning up to three orders of magnitude at fixed stellar mass. The medians of both sets agree well for low to intermediate stellar masses but seem to diverge at the high mass end (although they are still consistent within the large range of variation).  
The discrepancy at the high mass end is due to a small number of galaxies with non-zero star formation rate, while most galaxies with stellar masses above $\sim2\times10^{10}~h^{-1}M_\odot$ and $\sim7\times10^{10}~h^{-1}M_\odot$ are fully quenched in IllustrisTNG and SIMBA, respectively, in agreement with previous work \citep[see e.g.][]{SIMBA,2020arXiv200800004D}. Overall, within the large range of variation, both simulations agree well.

From the last panel of Fig. \ref{fig:LH_vs_CV} we can see that the medians of the LH and CV sets agree very well for low mass galaxies, while some significant deviations take place in the high-mass end. We can also see that cosmic variance (that incorporates the contribution from the intrinsic scatter) is responsible for most of the range of variation that we find for this quantity, in the same way as the other galaxy properties examined in this work.

\begin{figure*}
\centering
\includegraphics[width=0.99\textwidth]{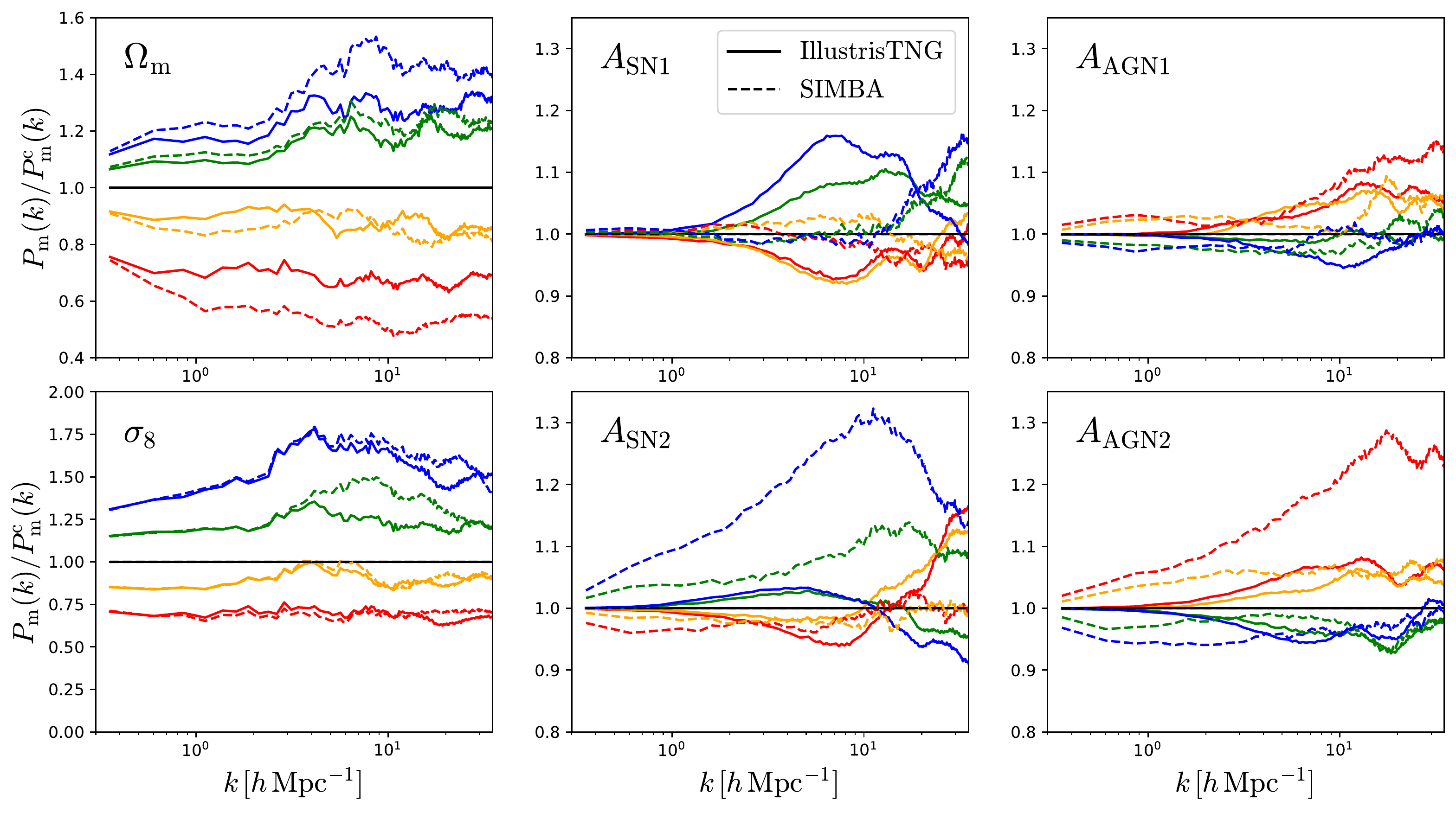}\\[5ex]
\includegraphics[width=0.99\textwidth]{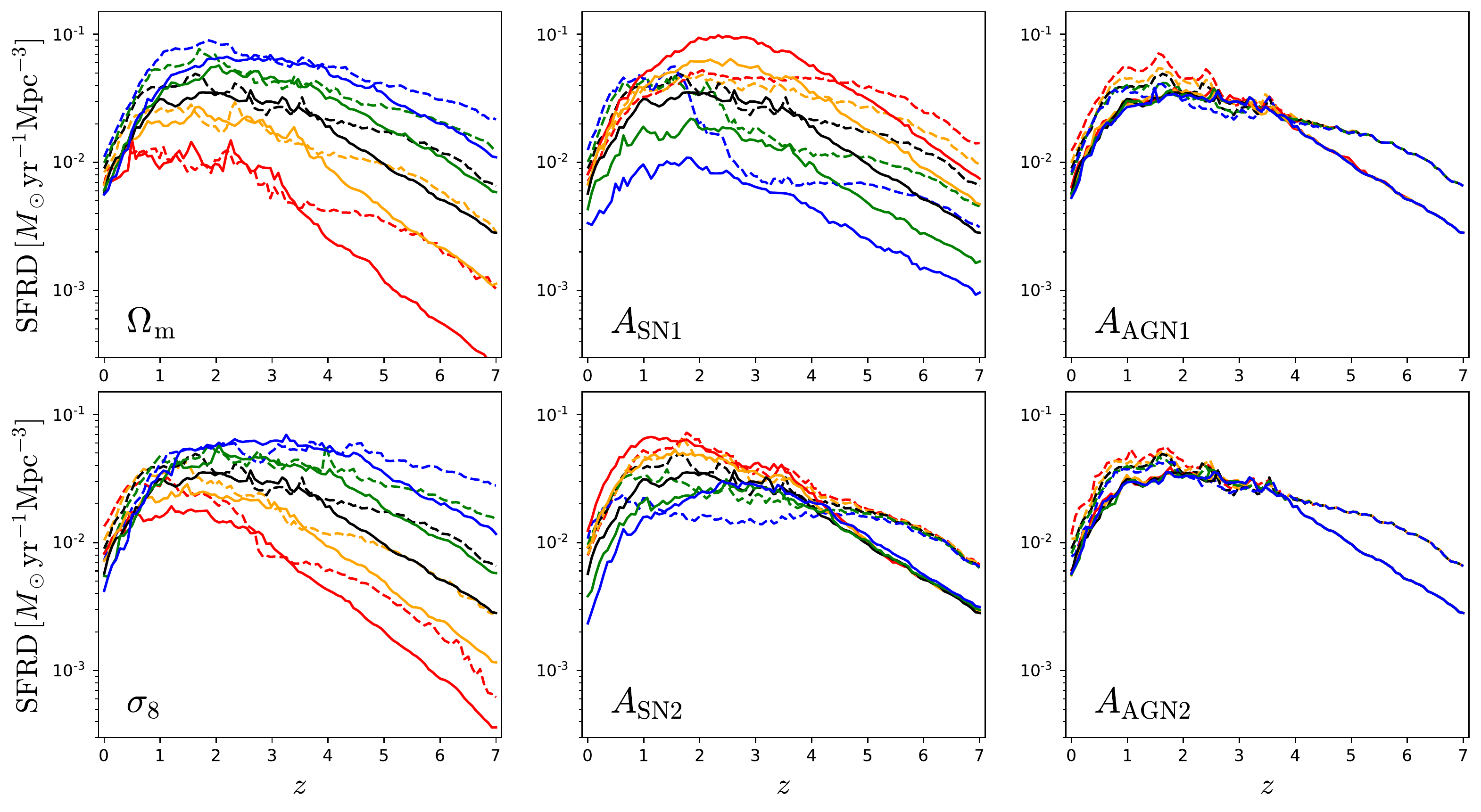}
\caption{\textbf{Top two rows:} Ratio between the matter power spectrum of the 1P simulations, where only one parameter is varied (while others are fixed), and the matter power spectrum of the central/fiducial model. Results are shown for $\Omega_{\rm m}$, $\sigma_8$, $A_{\rm SN1}$, $A_{\rm AGN1}$, $A_{\rm SN2}$, and $A_{\rm AGN2}$ as indicated. Solid and dashed lines correspond to the IllustrisTNG and SIMBA sets, respectively. The value of the parameters is varied within $\Omega_{\rm m}\in[0.1 - 0.5]$, $\sigma_8\in[0.6 - 1.0]$, ($A_{\rm SN1}, A_{\rm AGN1}) \in [0.25 - 4.00]$, ($A_{\rm SN2}, A_{\rm AGN2}) \in [0.5 - 2.0]$, and it is represented with different colors: from red (lowest) to blue (highest). The largest impact on the overall amplitude of the matter power spectrum is induced by $\sigma_8$, followed by $\Omega_{\rm m}$. The effect of the astrophysical parameters is mainly concentrated on small scales. Note that the IllustrisTNG and SIMBA sets respond differently to changes in the parameters. For instance, decreasing the value of $A_{\rm SN2}$ tends to increase the amplitude of the matter power spectrum on small scales in IllustrisTNG, while the opposite effect takes place for the SIMBA runs. \textbf{Bottom two rows:} Star formation rate density (SFRD) of the 1P sets with the same setup as above. $\Omega_{\rm m}$ and $\sigma_8$ have a large impact in the SFRD at $z>0.5$. $A_{\rm SN1}$ has a significant impact at all redshifts, while $A_{\rm SN2}$ affects significantly the SFRD at $z<4$.
The AGN feedback parameters $A_{\rm AGN1}$ and $A_{\rm AGN2}$ have a minor effect in IllustrisTNG, but a noticeable impact in the SFRD for SIMBA at $z<2$. }
\label{fig:Pk_individual}
\end{figure*}

\subsection{Variations due to cosmology and astrophysics}
\label{subsec:parameter_variation}

In the previous section we have computed the median and range of variation of 12 different properties when varying simultaneously the initial random field and the values of the cosmological and astrophysical parameters, using the LH sets for both suites. We have also used the CV sets to quantify what fraction of the range of variation is due to cosmic variance versus changes in cosmology and astrophysics. We now focus our attention on the response of two of these quantities, the matter power spectrum and the star formation rate density, to changes on the value of a single parameter, for which we make use of the 1P sets.

\subsubsection{Matter power spectrum}

In the top two panels of Fig.~\ref{fig:Pk_individual} we show the ratio between the power spectrum for each simulation belonging to the 1P set to that of the central model for each simulation suite (IllustrisTNG in solid lines and SIMBA in dashed). The central model represents a simulation run with fiducial parameters: $\{\Omega_{\rm m}, \sigma_8, A_{\rm SN1}, A_{\rm SN2}, A_{\rm AGN1}, A_{\rm AGN2} \}=\{ 0.3, 0.8, 1, 1, 1, 1\}$. We remind the reader that all the simulations in the 1P set share the same initial conditions, diminishing the effects of cosmic variance. The value of the cosmological and astrophysical parameters is varied in these simulations as in Table \ref{table:sims2}.

\begin{table*}
\begin{center}
\renewcommand{\arraystretch}{0.6}
\resizebox{1.0\textwidth}{!}{\begin{tabular}{| c || c | c | c |}
\hline
\multirow{2}{*}{Task} & \multirow{2}{*}{Model} & \multirow{2}{*}{Description} & \multirow{2}{*}{Section}\\
& & &\\
\hline \hline
\multirow{2}{*}{Emulator} & Fully connected layers & \multirow{2}{*}{Predict average SFRD from parameters} & \multirow{2}{*}{\ref{subsec:interpolation}}\\
& (supervised learning) & & \\
\hline
\multirow{2}{*}{Parameter regression} & Fully connected layers & \multirow{2}{*}{Constrain parameter values from measurements of the SFRD} & \multirow{2}{*}{\ref{subsec:LFI}}\\
& (supervised learning) & &\\
\hline
\multirow{2}{*}{Symbolic regression} & Genetic programming & \multirow{2}{*}{Approximate the mean SFRD with analytic expressions} & \multirow{2}{*}{\ref{subsec:symbolic_regression}}\\
& (supervised learning) & &\\
\hline
\multirow{2}{*}{Data generation} & Generative adversarial networks & \multirow{2}{*}{Generate 2D gas temperature maps} & \multirow{2}{*}{\ref{subsec:GANs}}\\
& (unsupervised learning) & &\\
\hline
\multirow{2}{*}{Dimensionality reduction} & Convolutional Auto-encoders & \multirow{2}{*}{Find lower dimensionality representation of 2D gas maps} & \multirow{2}{*}{\ref{subsec:dimensionality}}\\
& (unsupervised learning) & &\\
\hline
\multirow{2}{*}{Anomaly detection} & Convolutional Auto-encoders & \multirow{2}{*}{Find anomalies in 2D gas maps} & \multirow{2}{*}{\ref{subsec:dimensionality}}\\
& (unsupervised learning) & &\\
\hline
\end{tabular}}
\end{center}
\caption{This table summarizes the different machine learning applications presented in this paper, the model employed, and the section where they are presented.}
\label{table:ML_methods}
\end{table*}

Some expected trends are recovered, e.g.~the amplitude of the matter power spectrum on large scales increases with $\sigma_8$. On large scales, and for changes on the value of the cosmological parameters ($\Omega_{\rm m}$, and $\sigma_8$), the agreement between IllustrisTNG and SIMBA is very good. On smaller scales, however, the matter power spectrum responds differently in the SIMBA and IllustrisTNG suites even for changes in the value of the cosmological parameters.

We find that variations in any astrophysical parameter affect the amplitude and shape of the matter power spectrum. $A_{\rm AGN1}$ exhibits the weakest influence and $A_{\rm AGN2}$ also shows moderate changes, in particular for the IllustrisTNG simulation set. For the astrophysical parameters, we find very large differences between the output of the IllustrisTNG and SIMBA sets. For instance, decreasing the value of $A_{\rm SN2}$ tends to increase the amplitude of the matter power spectrum on small scales in IllustrisTNG, while the opposite trend takes place for the SIMBA runs. It is very interesting to observe that the velocities of both galactic winds and AGN jets (connected to $A_{\rm SN2}$ and $A_{\rm AGN2}$) can imprint percent level changes in the amplitude of the matter power spectrum on the largest scales that we probe. However, this only occurs in the SIMBA set while weaker effects are seen for the IllustrisTNG set. 

The largest variations on the matter power spectrum are induced by changes in cosmological parameters rather than astrophysics. This is expected, as the total matter power spectrum is dominated by the clustering of dark matter, which is expected to respond more weakly to astrophysics (via backreaction) than to changes in cosmology. However, note that this is in part due to the very large range of parameter variations for $\Omega_{\rm m}$ and $\sigma_8$, where most of the values considered are ruled out by current cosmological constraints.

\subsubsection{Star formation rate density}

We now turn our attention to the cosmic star formation rate density. In the two bottom panels of Fig.~\ref{fig:Pk_individual}, we show the changes in the SFRD due to individual variations in the value of the different parameters. In this case, we find that the SFRD is significantly affected by changes in $\Omega_{\rm m}$, $\sigma_8$, and $A_{\rm SN1}$ at most redshifts and by $A_{\rm SN2}$ at $z<4$ in both IllustrisTNG and SIMBA. On the other hand, $A_{\rm AGN1}$ and $A_{\rm AGN2}$ barely have any noticeable effect on the SFRD for the IllustrisTNG set while they have a clear effect in SIMBA at $z<2$. We note that these conclusions are significantly affected by our box size, which limits the abundance of massive galaxies and therefore likely leads to an underestimate of the effects of AGN feedback \citep{Joshua_2015, 2017MNRAS.472..949B}.

The strong dependence we observe with cosmology at high redshift may be attributed to the fact that changing the value of these parameters, in the large range we consider, will significantly affect the abundance of halos at high redshifts. This in turn is expected to affect the SFRD, as we find.

The SFRD also exhibits different behavior between the IllustrisTNG and SIMBA sets. For instance, while at $z>3$ the value of the SFRD decreases with increasing $A_{\rm SN1}$ for both simulations, at lower redshifts the trends between the two simulations switch: in SIMBA, the SFRD at $z\lesssim1.5$ increases with $A_{\rm SN1}$, corresponding to larger mass loading of galactic winds, suggesting an enhanced supply of gas at late times owing to ``wind recycling'' \citep[e.g.][]{Oppenheimer2010,Angles-Alcazar2017_BaryonCycle}. We also observe some systematics differences between both sets. For instance, at high-redshifts the IllustrisTNG simulations tend to predict lower amplitudes than their SIMBA counterparts, for all parameters. The SFRD of the SIMBA simulations also exhibit more high-frequency structure, reflecting the very different way feedback is encoded in the subgrid models of the two suites.
These results highlight the complicated and rich physics involved in these simulations, with highly non-trivial interplay between different cosmological and feedback parameters. However, these complicated non-linear effects can be \textit{learned} by different machine learning algorithms, as we shall see below.

\section{Machine Learning Applications}
\label{sec:ML_applications}

In this section we show a few applications of machine learning using the CAMEL simulations, that we summarize in Table \ref{table:ML_methods}. These applications are meant to showcase the potential of CAMELS, illustrating some of its scientific goals. We emphasize that all machine learning methods employed in this paper use traditional deep learning architectures; that is why we do not describe them in detail but refer the reader to other works in the literature.

\subsection{Interpolation with neural networks}
\label{subsec:interpolation}

We begin with a simple, yet powerful application: using neural networks \citep[see][for a review]{NN_review} as non-linear interpolators.

We illustrate this by training a neural network to predict the star formation rate density (SFRD) of a simulation just taking as input the value of its cosmological and astrophysical parameters
\begin{equation}
{\rm SFRD}(z)=f(\Omega_{\rm m}, \sigma_8, A_{\rm SN1}, A_{\rm SN2}, A_{\rm AGN1}, A_{\rm AGN2})~.
\label{eq:params_2_SFRD}
\end{equation}
Note that the SFRD of a given simulation is expected to depend not only on the underlying cosmological and astrophysical model, but also on the particular realization considered\footnote{We also expect stochastic noise from the nature of the simulation itself.} (i.e.~on the initial random seed). In the fifth panel of Fig. \ref{fig:LH_vs_CV} we show the median and 16-84 percentiles of the SFRD of 1) 27 simulations that have the same cosmology and astrophysics (the CV set), and 2) simulations with different cosmologies and astrophysics (the LH set); in both cases for the IllustrisTNG model (similar results hold for the SIMBA model). We find that the scatter due to cosmic variance is much smaller than the range of variation due to changes in cosmology and astrophysics. Thus, as a first order approximation, we neglect the dependence of the SFRD on the random seed. Note that it is thus expected that the network will learn to predict the SFRD for a `typical' cosmological volume.

\begin{figure}
\centering
\includegraphics[width=0.49\textwidth]{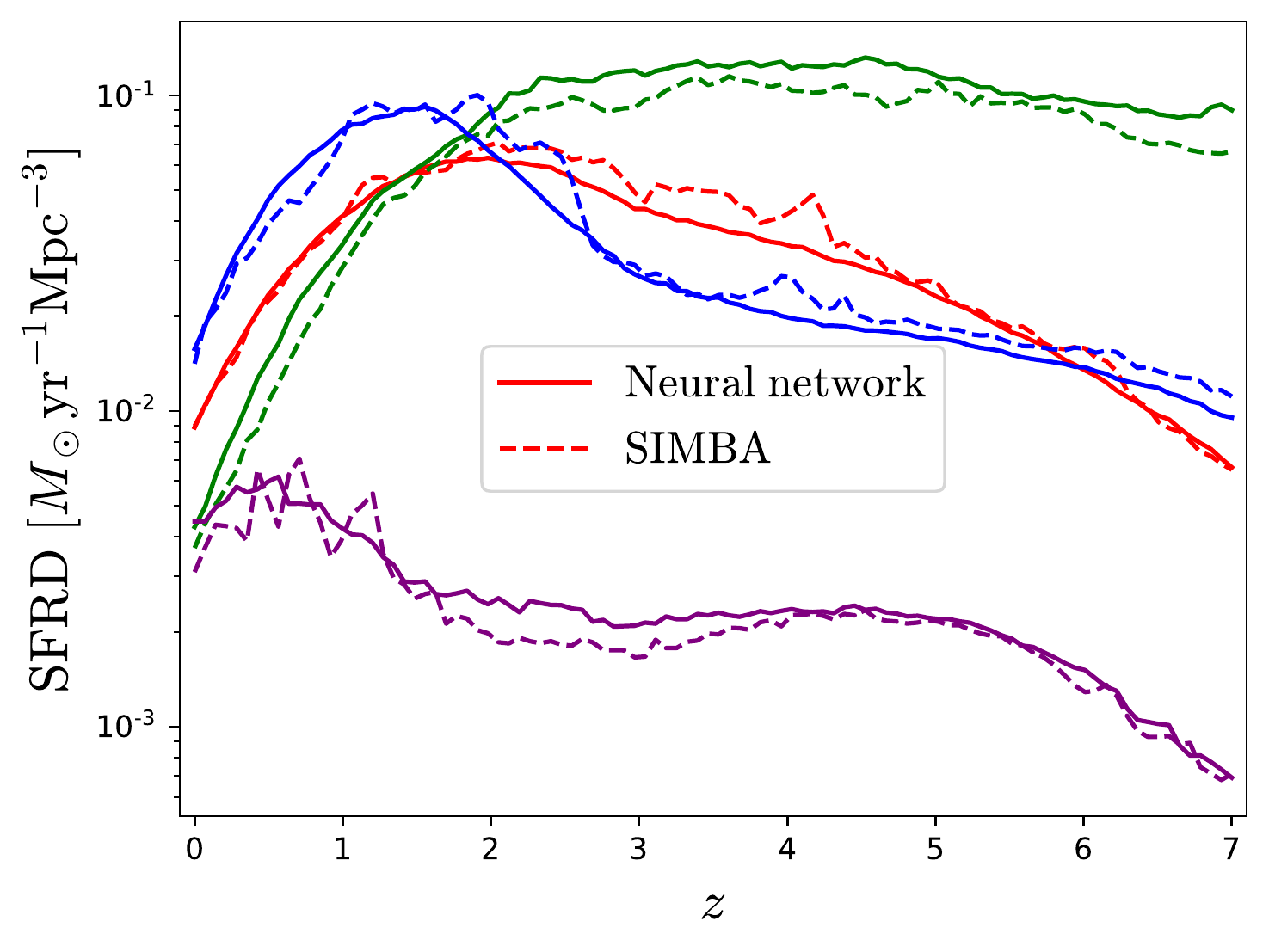}
\caption{We have trained a neural network to predict the star-formation rate density of the SIMBA LH simulations, taking as input only the value of $\Omega_{\rm m}$, $\sigma_8$, $A_{\rm SN1}$, $A_{\rm SN2}$, $A_{\rm AGN1}$, and $A_{\rm AGN2}$. The plot shows the star-formation rate density for four different simulations that were not included in the training/validation set (dashed lines) together with the predictions of the neural network (solid lines). Our network is able to predict the star-formation rate density with a precision of $0.12$dex on average. Note that each simulation in the data set has a different random initial seed.}
\label{fig:params_2_SFRD}
\end{figure}

\begin{figure*}
\centering
\includegraphics[width=0.99\textwidth]{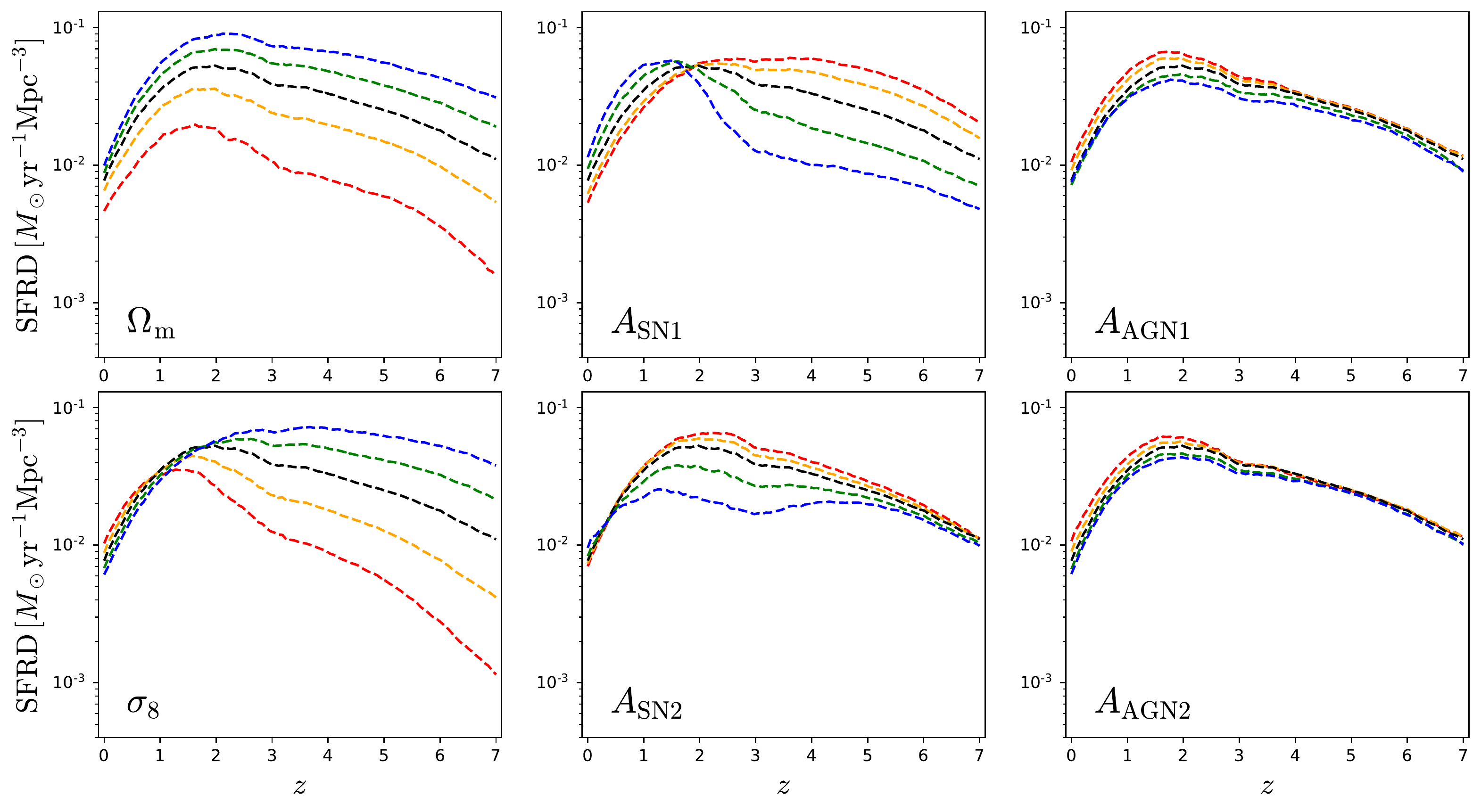}
\caption{We take the trained neural network and use it to predict how the SFRD responds to changes to a single parameter. The input to the network is the fiducial value of all parameters with the exception of one, that varies from almost the minimum (red) to almost the maximum (blue). The black line shows the prediction for the fiducial model. These curves can be directly compared with the dashed lines of the bottom panels of Fig.~\ref{fig:Pk_individual}. We find that the neural network, even if trained on data where all the parameters are varied at the same time, is able to capture the response of the SFRD to individual parameters as in the full hydrodynamic simulations.}
\label{fig:params_2_SFRD_1param}
\end{figure*}

We start by taking the SFRD of the 1,000 simulations of the SIMBA LH set. We then sample those SFRDs at 100 redshifts equally spaced between $z=0$ and $z=7$. We split that dataset into training (700), validation (150), and test (150) sets, and train a simple neural network to approximate the function $f$ of Eq.~\ref{eq:params_2_SFRD}. 

We have tried different architectures and performed a relatively wide hyper-parameter search, finding that a model with a single hidden fully connected layer performs better than models with more layers. The best-model is chosen as the one with the lowest validation error. We have tried with models having different number of neurons per layer. Already with 40 neurons we achieve a very good validation loss. However, that model provides a test score $\sim10\%$ worse than our best model, which contains 500 neurons in the hidden layer. The architecture of our model is a simple perceptron \citep{perceptron} as follows:
\begin{enumerate}
\item Input: Cosmo+Astro parameters; 6 numbers
\item Fully connected layer; 500 neurons
\item Leaky ReLU activation
\item dropout (0.2)
\item Output: SFRD from z=0 to z=7; 100 numbers
\end{enumerate}
where ReLU stands for Rectified Linear Unit. We have used the Adam optimizer with values of the $\beta$ parameters equal to $\{0.5, 0.999\}$. We also use weight decay with a value of $10^{-7}$. Learning rate is set at $2\times10^{-4}$, and we train the model for 15,000 epochs.

The network achieves an error $\delta=0.12$dex in predicting the SFRD just from the value of the cosmological and astrophysical parameters, where the error is defined as
\begin{equation}
\delta=\sqrt{\left\langle\left[\log_{10}({\rm SFRD}_{\rm predicted}) - \log_{10}({\rm SFRD}_{\rm true}) \right]^2 \right\rangle},
\label{Eq:SFRD_accuracy}
\end{equation}
where the average runs over realizations and redshifts. We have repeated the same exercise with the simulations from the IllustrisTNG LH set and, in that case, we obtain a slightly better error: $\delta=0.106$dex. This is because the SFRD of the SIMBA simulations are \textit{noisier} than those from IllustrisTNG, namely they are less smooth \citep{2020MNRAS.tmp.2258I}.

We show an example of the performance of the network in Fig.~\ref{fig:params_2_SFRD}, where we display four random SFRDs from the test set, together with the prediction from the network. As can be seen, the network is able to capture the general trend very well in all cases. The individual SFRDs exhibit significant high-frequency variability in some redshifts, which the network is not able to capture as that arises mainly due to cosmic variance. 

The $\simeq30\%$ error reached by our network should be compared with the scatter due to cosmic variance, of around $20\%$ (see panel $V$ of Fig. \ref{fig:LH_vs_CV}). That $\sim20\%$ scatter represents the minimum error our network can achieve, since it is trained without any variable that can account for the effects of cosmic variance such as the initial density field. This shows that our network achieves a good accuracy. More training, hyperparameter tuning, and in particular more data, can improve the network 
error even further.

Finally, once the network has been trained, it can be used for understanding the dependence of the SFRD on the parameters. For instance, we can use the network to predict the SFRD for the fiducial value of the cosmological and astrophysical parameters, and then, when changing the value of a single parameter while keeping fixed the values of the others. This is illustrated in 
Fig.~\ref{fig:params_2_SFRD_1param}, which  
can be directly compared with the dashed lines in the bottom rows of Fig.~\ref{fig:Pk_individual}, where the actual measurements from the 1P SIMBA set are shown. We find that the neural network has learned the correct dependence on the parameters, e.g.~whether the amplitude of the SFRD increases or decreases when varying a given parameter. We note that the comparison between the actual measurement from the SIMBA 1P set and the prediction of the network may not be perfect; this happens because the network has been trained to predict the mean value of the SFRD, while the actual measurements from the simulations are affected by cosmic variance.

In the future, this approach can enable a fast and accurate exploration of the parameter space that can be used for selecting the cosmological and astrophysical parameters needed to reproduce a given observable. Other possibility is to use active learning to select points in parameter space that reproduce a set of observations, such as the stellar mass function and star-formation rate density. We emphasize that we are not attempting to do a proper and rigours analysis here, where things like binning, observational uncertainties...etc, need to be taken into account when building an emulator. Instead, we are just illustrating the potential of the method with data that is equally spaced in redshift, while most physical quantities operate in physical time.

\begin{figure*}
\centering
\includegraphics[width=0.99\textwidth]{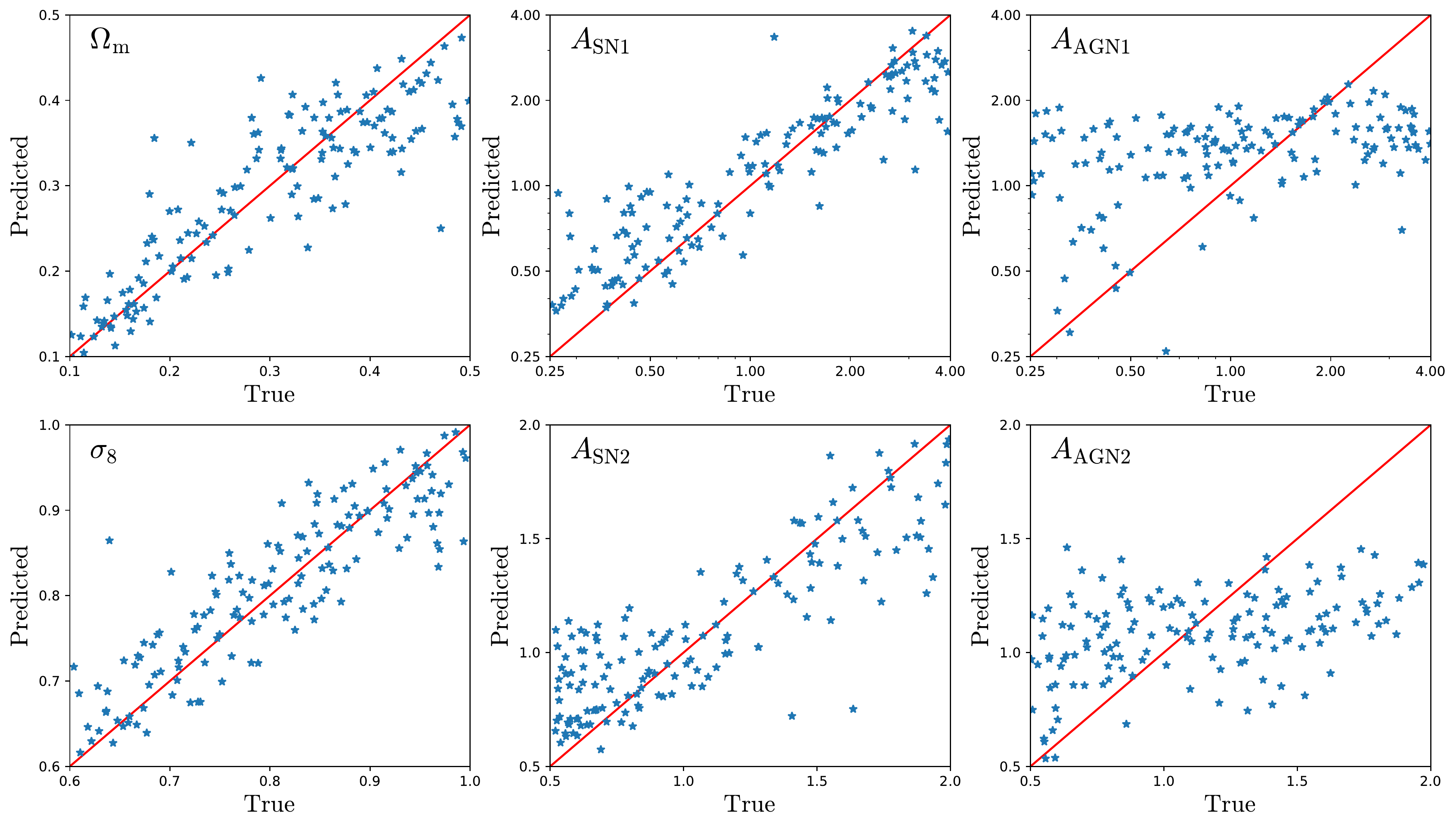}
\caption{We train a neural network to predict the value of the cosmological and astrophysical parameters from measurements of the star formation rate density (SFRD) from $z=0$ to $z=7$, taken from the SIMBA LH set. Each point represents the predicted value of the parameter from the neural network against its true value. The red solid line represents predicted=true; the closer the points to the line the more accurate the prediction. We find that with measurements of the SFRD we are able to predict the value of $\Omega_{\rm m}$, $\sigma_8$, $A_{\rm SN1}$ and $A_{\rm SN2}$, but not $A_{\rm AGN1}$ and $A_{\rm AGN2}$.}
\label{fig:SFRD_2_params}
\end{figure*}

\subsection{Constraining parameters}
\label{subsec:LFI}

We now present what can be seen as the \textit{inverse} application of the previous case: constraining the value of the cosmological and astrophysical parameters from measurements of an observable. In this case, we take the star formation rate density as our observable, and our goal is to use neural networks to approximate the function
\begin{equation}
\vec{\theta} = f({\rm SFRD}(z)),
\end{equation}
where $\vec{\theta}=\{\Omega_{\rm m}, \sigma_8, A_{\rm SN1}, A_{\rm SN2}, A_{\rm AGN1}, A_{\rm AGN2} \}$. We note that our goal is not to get the full posterior on the value of the parameters, but simply find the value of the parameters that maximize the likelihood function.

The data we use is the same as in section \ref{subsec:interpolation}: 100 values of the SFRD equally spaced from $z=0$ to $z=7$ from the SIMBA LH set. The manner in which we normalize the value of the parameters and the SFRD is also the same as in the previous section. The architecture of our network is also a simple multilayer perceptron \citep{perceptron}:

\begin{enumerate}
\item Input: SFRD; 100 values equally spaced from $z=0$ to $z=7$ 
\item Fully connected layer: 500 neurons
\item Leaky ReLU activation; 0.01
\item Dropout
\item Fully connected layer: 500 neurons
\item Leaky ReLU activation; 0.01
\item Dropout
\item Fully connected layer: 500 neurons
\item Leaky ReLU activation; 0.01
\item Dropout
\item Fully connected layer: 6 neurons
\item Output: 6 numbers; value of the cosmo+astro params
\end{enumerate}

We have tried with different values for the dropout rate, but we found that setting it to zero yields good validation scores. We have however tuned the value of the weight decay to achieve the best validation scores; the optimal value was found to be $2\times10^{-3}$. We use a learning rate equal to $10^{-4}$ and we use the Adam optimizer with $\vec{\beta}=\{0.5,0.999 \}$. 

The dataset was split into training (700 simulations), validation (150 simulations), and test (150 simulations) sets. The training is carried out by minimizing the mean squared error (MSE), which guarantees that the prediction of the network represents the mean of the posterior \cite[see appendix A of][]{Paco_2020b}. After training, we use the network to predict the value of the cosmological and astrophysical parameters for each of the 150 SFRD of the test set. We show the results in Fig.~\ref{fig:SFRD_2_params}. 

The network is able to constrain the value of $\Omega_{\rm m}$, $\sigma_8$, $A_{\rm SN1}$ and $A_{\rm SN2}$, while it cannot determine the value of $A_{\rm AGN1}$ and $A_{\rm AGN2}$. This is expected as we saw in the bottom panels of Fig.~\ref{fig:Pk_individual} that $A_{\rm AGN1}$ and $A_{\rm AGN2}$ have a very weak effect on the SFRD in the SIMBA suite (the effect is also very weak on the IllustrisTNG suite). We find that the network can constrain the value of $\Omega_{\rm m}$, $\sigma_8$, $A_{\rm SN1}$ and $A_{\rm SN2}$ with an error equal to $0.055$, $0.051$, $0.55$ and $0.25$, respectively, where the error $\delta^i$ on the parameter $\theta^i$ is defined as 
\begin{equation}
\delta^i = \sqrt{\frac{1}{N}\sum_j\left( \theta^i_{{\rm True},j}-\theta^i_{{\rm NN},j}\right)^2}~,
\label{Eq:regression_error}
\end{equation}
where $j$ runs over all realizations of the test set ($N$). We note that the constraints are not particularly tight, but it is important to keep in mind that our data arises from simulations with a very small volume of $(25~h^{-1}{\rm Mpc})^3$ and with a different initial random seed for each realization, giving rise to significant scatter from cosmic variance. 

We note that the error on a parameter may depend on its actual value. Thus, by using Eq. \ref{Eq:regression_error} we are computing an average error. Ideally, for each point in parameter space we would like to compute the error by calculating the scatter on the network prediction from multiple SFRD with the same cosmology and astrophysics. In cases where this option is not available (e.g. simulations are too expensive), a different possibility is to estimate the errors using Bayesian neural networks \citep{Bayesian_NN1,Bayesian_NN2, Bayesian_NN3, Bayesian_NN}.

This method represents a powerful way to constrain the value of the parameters in cases where it is not obvious how to write down a likelihood function and where a large number of simulations are needed to compute some of its ingredients, e.g.~the covariance matrix. Combining different observables, e.g.~ the star formation rate density with the matter power spectrum and the stellar mass function, is trivial using this method, while it may be very difficult to write down the likelihood function for that observable vector. We emphasize that more sophisticated methods, that also output the posterior distribution, have been recently developed \citep[see for instance][]{Justin_2019}.

\subsection{Symbolic regression}
\label{subsec:symbolic_regression}

Neural networks can approximate a very general set of functions. Their strength relies on the fact that they work without the need to know the underlying function; only samples from that function are needed. Once a neural network has been trained, it will produce an output for a given input. Unfortunately, knowing what exactly the neural network is doing may be complicated, and in some situations almost impossible. 

In some cases, it is better to have an analytic function that describes, or at least approximates, the relationships of interest within data \citep[see e.g.][]{Miles_2020}. Furthermore, in general, analytic expressions extrapolate better than neural networks\footnote{It is known that in general, neural networks do not generalize well, and can learn undesired things like priors on the data they are trained on.}. There are different techniques that can be used to derive analytic expressions that capture the structure or relations in a given data. In this subsection we make use of genetic programming to find an analytic function that approximates the relation between the SFRD in IllustrisTNG suite and the value of the parameters and redshift:
\begin{equation}
{\rm SFRD}=f(z,\Omega_{\rm m}, \sigma_8, A_{\rm SN1}, A_{\rm SN2})~.
\end{equation}
Note that we have dropped the dependence with $A_{\rm AGN1}$ and $A_{\rm AGN2}$, as we have shown that for the IllustrisTNG suite it is very weak. 

The way genetic programming works is as follows. First, the user chooses a set of different functions to use, e.g.~sum (+), multiplication ($\cdot$) or sinus ($\sin$). Next, a set of random combinations of functions, variables, and constants is created, e.g.~$\sin(3\cdot x + 2.789)$. Those functions are then evaluated on the dataset and the functions that perform better are more likely to go to the next generation. If a function is chosen to go into the next generation, it can experience different possibilities. It can cross-over with other function, it can randomly mutate entirely or a part of it, or it can go unchanged. All the functions in the new generation are evaluated as before and some will go into a new generation. This process is repeated for several generations until the underlying function, or an approximation of it, is found. 

We have used \textit{Eureqa} code\footnote{\url{https://www.nutonian.com/products/eureqa/}} to carry out the calculation\footnote{The publicly available package PySR, \url{https://github.com/MilesCranmer/PySR}, achieves a very similar performance on the same task.}. For the function set we have chosen: $+, -, \times, \div, \log, \exp, a^b$. For the operators $+, -, \times, \div$ we use a complexity\footnote{Complexity is an integer number that is associated to each operator. It is used to penalize more complex operations, e.g.~$\sin$, over more standard ones, e.g.~$+$. It is a free parameter that the user needs to specify; being its value very problem-specific.} equal to 1, while for the other operators we set it to 2. We trained for approximately 250,000 generators, and stopped our search once the stability and maturity achieved a level above $50\%$. We tried the previous procedure several times, to test the code on different initial random states.

For each simulation of the IllustrisTNG LH set we have measured the SFRD. We have then split the data into two sets, a training set containing 700 realizations and a test set with 300 measurements. We do not make use of a validation set as we do not tune the value of the hyper-parameters. Our input to the code is the value of the redshift, $\Omega_{\rm m}$, $\sigma_8$, $A_{\rm SN1}$ and $A_{\rm SN2}$, while the output is the $\log_{10}$ of the SFRD.

We have found three expressions that are short enough to be useful while at the same time achieve a good accuracy on both the training and test sets:
\begin{widetext}
\begin{eqnarray}
\label{eq:sg1}
\log_{10}({\rm SFRD}) &=&  - \frac{1.777 + \sigma_8A_{\rm SN2}}{1+z} - (1+z)\left(\frac{0.365}{\sigma_8} - 0.559\Omega_{\rm m} + 3.57\times10^{-3}\frac{A_{\rm SN1}}{\Omega_{\rm m}\sigma_8}\right)
 + 0.39^{A_{\rm SN1}}\\[3ex]
\label{eq:sg2}
\log_{10}({\rm SFRD}) &=&  \frac{0.696}{(1+z)A_{\rm SN2}} -(1+z)\left( \frac{0.0389}{\Omega_{\rm m}} + \frac{0.379}{\sigma_8}\right) - 1.333 + 2.317\log(1+z) - A_{\rm SN1}^{0.391}\\[3ex]
\label{eq:sg3}
\log_{10}({\rm SFRD}) &=&   \frac{0.692A_{\rm SN1}^{0.458}}{(1+z)A_{\rm SN2}} -(1+z)\left(\frac{0.038}{\Omega_{\rm m}} + \frac{0.425}{\sigma_8}\right) + \frac{0.36}{\sigma_8}- 1.631 + 2.534\log(1+z) - 1.21A_{\rm SN1}^{0.406}
\end{eqnarray}
\end{widetext}

These expressions achieve a 0.19, 0.18 and 0.16 
error, respectively, on both the training and test sets, where the error is defined as in Eq.~\ref{Eq:SFRD_accuracy};
$\log_{10}({\rm SFRD})_{\rm prediction}$ is now the output of the symbolic regression equation. We emphasize that our criteria to chose these expressions over the others is that they achieve a low training error while, at the same time, being compact enough. We find that longer expressions are barely more accurate, while shorter expressions have a significant lower error.

\begin{figure}
\centering
\includegraphics[width=0.48\textwidth]{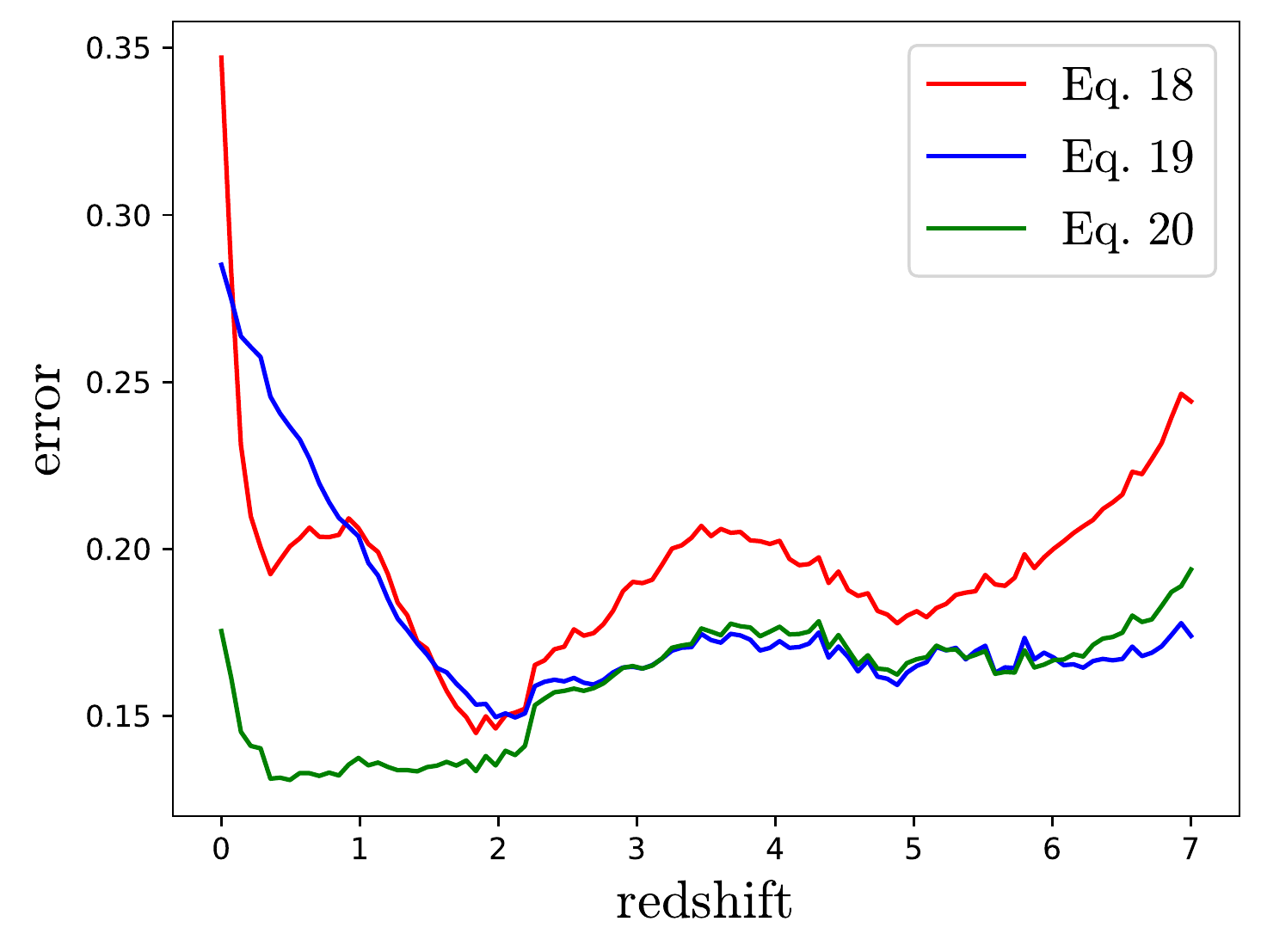}
\caption{We use genetic programming to find analytic expressions that relate the star-formation rate density from the IllustrisTNG simulations with the value of redshift, $\Omega_{\rm m}$, $\sigma_8$, $A_{\rm SN1}$ and $A_{\rm SN2}$. We show these expressions in Eqs.~\ref{eq:sg1}, \ref{eq:sg2}, and \ref{eq:sg3}. This plot shows the error at a given redshift, defined as $\delta(z)=\sqrt{\left\langle\left[\log_{10}({\rm SFRD}_{\rm predicted}(z)) - \log_{10}({\rm SFRD}_{\rm true}(z)) \right]^2 \right\rangle}$. Our analytic expressions achieve average accuracies between 0.16 and 0.195. Note that neural networks were able to get an error equal to 0.106.}
\label{fig:symbolic_regression}
\end{figure}

\begin{figure*}
\centering
\includegraphics[width=0.99\textwidth]{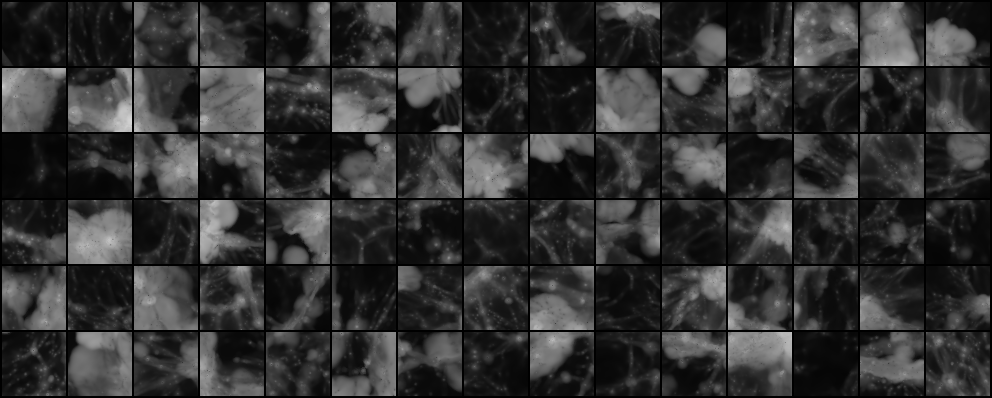}\\[1ex]
\includegraphics[width=0.99\textwidth]{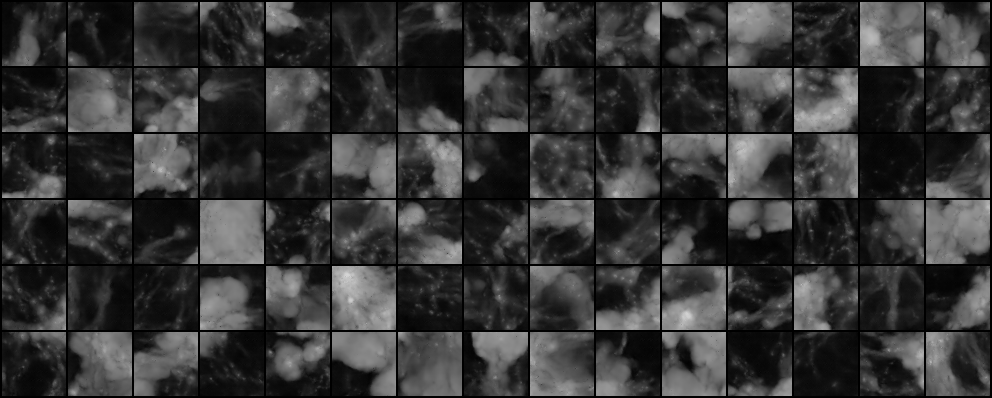}
\caption{\textbf{Top}: 2D gas temperature projections of regions of $6.4\times6.4\times5$ $(h^{-1}{\rm Mpc})^3$ from the IllustrisTNG LH set. \textbf{Bottom:} Fake  temperature images generated by our generative adversarial network (GAN). Our network not only produces images that visually resemble the true ones, but also they share very similar statistics,  in terms of probability distribution function and power spectrum, as the ones from the simulations. We show the images in black and white (and the same color scale), instead of the color palette employed in the top row of Fig. \ref{fig:image_EX}, since that allows us to better capture the dynamical range of the images. From now on, we will use black and white to show images.}
\label{fig:GAN_images}
\end{figure*}

We note that the neural network we trained in Sec.~\ref{subsec:interpolation} on the same simulations achieved a $\delta=0.106$ error. In Fig.~\ref{fig:symbolic_regression} we show the error each expression achieves as a function of redshift. As can be seen, the accuracy of the expressions depends on redshift. Eq.~\ref{eq:sg3} is the most accurate and the one that shows less dependence with redshift. The accuracy of Eq.~\ref{eq:sg1} exhibits the strongest dependence with redshift, and overall is the least accurate. 

Whilst these expressions perform worse than the neural network, their analytic form can be very useful for understanding the dependence of the SFRD on each parameter, and the interactions between the different parameters. For example, in Eq.~\ref{eq:sg2}, the role of $A_{\rm SN1}$ is purely that of a normalization constant, while in Eqs.~\ref{eq:sg1} and \ref{eq:sg3} it provides an additional redshift dependence (though very weak one in Eq.~\ref{eq:sg1}). This feature may be responsible for the overall better accuracy of Eq.~\ref{eq:sg3}, in particular at low redshift. 

We note that it is always possible to get more accurate expressions at the expense of having longer formulae, or to include more diverse operators (e.g.~trigonometric functions). However, longer and more complex expressions may not generalize as well as simpler ones. Our purpose here is to show that simple and short formulae can be derived that quite accurately capture the dependence on the different parameters. Symbolic regression can be used to derive accurate analytic expressions on physical processes where the dimensionality is relatively small. These equations can help us understanding the physics behind these complex processes.

\subsection{Generative models}
\label{subsec:GANs}

We now turn our attention to generative models. Up to now, we have been using supervised learning, i.e.~we had an input and an output, and we used neural networks to find the mapping between the two. Now, we consider unsupervised learning, where we have unlabeled data drawn from some underlying distribution, and our goal is to generate new samples from that distribution.

We work with temperature maps from the IllustrisTNG LH simulation set, created as follows. For each simulation, we take slices of width $5~h^{-1}{\rm Mpc}$ along the X-, Y- and Z- axes, extracting a total of 15 slices for each realization. For each slice we assign the following quantities to two 2D regular grids with $250\times250$ pixels each: $Tm$ and $m$, where $T$ and $m$ are the temperature and mass of each gas particle. Quantities are assigned to the grid by performing a single line integral perpendicular to the image plane for each pixel
\begin{equation}
    A_{\rm 2D}(\vec{x})=\int_l A_{\rm 3D}(\vec{x},l)dl
\end{equation}
where $A_{\rm 2D}$ and $A_{\rm 3D}$ are the considered quantities on the projected plane and in 3D, respectively. $\vec{x}$ is the position on the grid, while $l$ goes in the direction perpendicular to the plane. In calculating the integral, we assume that each gas cell is a uniform-density sphere with a radius equal to its distance to its 64$^{\rm th}$ closest dark matter particle. The temperature maps are then created by dividing the field with $Tm$ over the field with $m$. We then randomly take regions with $64\times64$ pixels over these maps. These temperature maps represent our training data. In total, our dataset consists in more than 135,000 images that we enlarge by using data augmentation.

We show examples of these maps in the upper part of Fig.~\ref{fig:GAN_images}. As can be seen, these maps cover a very rich variety of appearances, from very empty regions corresponding to voids, to filaments and rich galaxy groups. Note also the wide range in astrophysics parameters we cover: whilst in some cases gas is relatively smooth and homogeneous, in others large `bubbles' induced by feedback are also visible. 

Our goal is to identify a low-dimensional manifold in which this data lives, and by sampling from that manifold, create new images that have the same statistical properties as the original ones. To carry out this task we have used generative adversarial networks (GANs) \citep{Goodfellow_2014}. We emphasize that the generated images will be sampled from the underlying manifold that contains not only the different elements of the cosmic web, but also the different values of the cosmological and astrophysical parameters. In other words, a generated image does not have a label attached to it with the value of the cosmological or astrophysical parameters. 

We have two networks, the generator and the discriminator. The mission of the generator is to generate images with the same properties as the real ones. On the other hand, the discriminator's role is to distinguish real from fake images. As the discriminator improves at performing its task, it forces the generator to produce better images in order to fool it. 

The input to the generator is a random vector from a 100-dimension space (also known as latent space\footnote{The connection of this abstract space to the physical properties of the temperature maps is not trivial. It is however expected that the generator will use the coordinates of points in this space to create temperature fields with a given properties.}) that is sampled using a multivariate Gaussian distribution with unit covariance matrix. We outline the architecture of the generator and discriminator in the appendix \ref{sec:GAN_architecture}. We train both networks using the Adam optimizer with a learning rate of $2\times10^{-4}$. 

\begin{figure}
\centering
\includegraphics[width=0.48\textwidth]{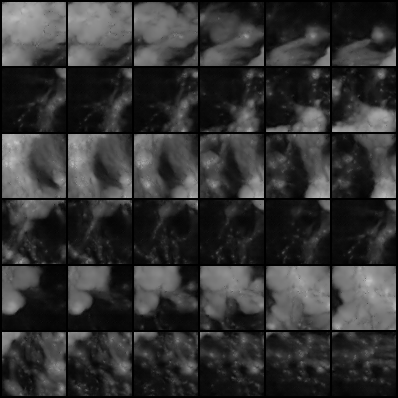}
\caption{We take different generated images from Fig.~\ref{fig:GAN_images} (first and last column) and interpolate them in latent space (each row). The figure shows the smooth transition between the two images, where each interpolated figure represents a temperature field with the proper statistical properties. This test reinforces our belief that our GAN does not suffer of mode collapse.}
\label{fig:GAN_interpolation}
\end{figure}

\begin{figure*}
\centering
\includegraphics[width=0.99\textwidth]{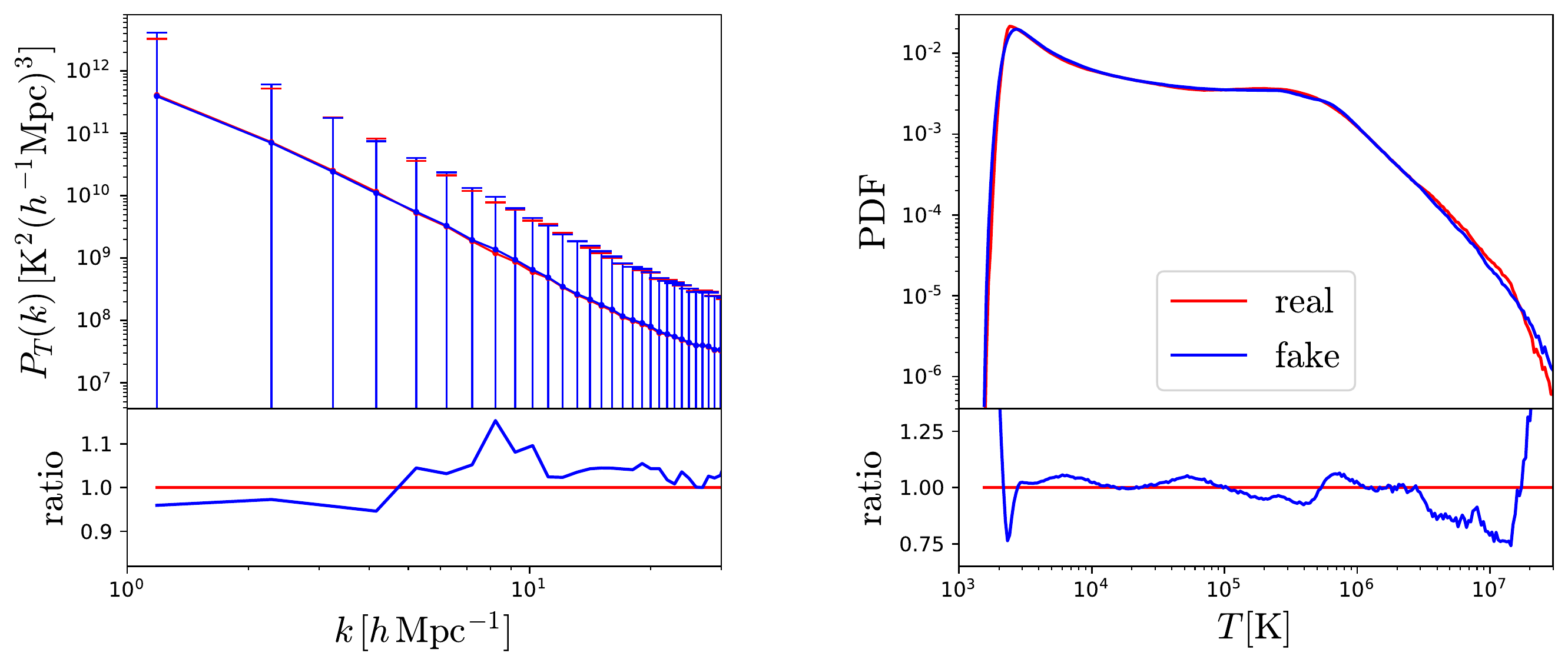}
\caption{We compute the power spectrum and the probability distribution function (PDF) of 15,000 real (red) and fake (blue) 2D temperature maps from the IllustrisTNG LH set. The left panel displays the mean and standard deviation of the power spectrum from individual maps while the right panel shows the PDF of all maps together. Bottom panels show the ratios. Our generative network is able to generate temperature maps whose statistical properties match very well those of the real maps.}
\label{fig:GAN_summary}
\end{figure*}

We show images generated by our GAN in the bottom part of Fig.~\ref{fig:GAN_images}. From their visual aspect, it is hard to distinguish the real from the fake images. Note that the GAN is able to produce images from a large variety of environments. It is also very interesting to note how the GAN produces images with such level of detail. For instance, black points can be seen in the real images showing the temperature around massive halos. Those black dots correspond to star-forming gas residing in galaxies. The fake images generated by the GAN also contain these black points, showing the quality of the generated images. 

One well-known problem with GANs is mode collapse; the generator learns to produce only a subset of images. In other words, the generator collapses the underlying distribution to a few peaks. This is clearly undesirable. One standard test to check whether our GAN suffers from this is to interpolate two points in latent space, and see what the generated images look like. If there is mode collapse, one would expect that the reconstructed images will not vary smoothly. We have performed this test and show the results in Fig.~\ref{fig:GAN_interpolation}. As can be seen, we do not find evidence for our latent space to have collapsed (at least for the considered trajectories); the interpolation between images is smooth and realistic. We have repeated the above exercise for many different images, and in all cases we find a smooth interpolation. This indicates that our GANs should not be heavily affected by mode collapse, if any at all. 

Whilst the images generated by the GAN look very realistic and almost indistinguishable from the real ones, it is important to quantify the agreement between both data sets using some summary statistics. Here we consider two: the temperature power spectrum and the temperature PDF. For each image in the training set we have computed the temperature power spectrum, as well as that from 15,000 fake images. We have then computed the mean and standard deviation of each set. We show the results in the left panel of Fig.~\ref{fig:GAN_summary}. We find an excellent agreement between the results from the real and fake images: results agree at the $\sim15\%$ level on scales from $k=1~h{\rm Mpc}^{-1}$ to $k=30~h{\rm Mpc}^{-1}$. Not only the mean values agree well, but also the scatter.

Next we evaluate the PDFs of the two sets. We have considered 300 bins equally spaced in $\log T$ between $1.6\times10^3$ and $4.9\times10^7$ K. We then compute the total number of pixels that fall inside each bin using all the images in a set (either real or fake), normalizing the numbers in each bin by the total number of pixels in all images. We show the results in the right panel of Fig.~\ref{fig:GAN_summary}. We also find an excellent agreement ($\sim25\%$) between both distributions over almost 4 orders of magnitude in temperature. 

We conclude that our GAN is able to produce temperature images that resemble very well those from the simulations, both visually and when using summary statistics. We emphasize that the GAN we have trained in this work, will produce temperature maps with an unknown cosmological and astrophysical model. One can think in more useful applications than the one illustrated in this paper. For instance, GANs can be used to generate temperature fields, and/or stellar mass fields, inside and around halos. In this case, one can use conditional GANs \citep{Conditional_GAN} to generate those fields for a given cosmology, astrophysics and halo mass. This will be a way to \textit{paint} physical fields on top of dark matter halos from cheap N-body simulations taking into account the scatter from the true distribution.

\subsection{Dimensionality reduction}
\label{subsec:dimensionality}

In this section we show one application of dimensionality reduction by training autoencoders on projected temperature fields. 

While traditional machine learning techniques such as Principal Component Analysis (PCA) can be used to reduce the dimensionality of generic data, its performance on complex data may not be satisfactory. Autoencoders, on the other hand, make use of neural networks that, while more difficult to train, can approximate generic non-linear functions; autoencoders can be seen as non-linear PCA methods.

The idea behind an autoencoder is to reconstruct the input, which can be an array, an image, a 3D field or any other tensor, by first lowering its dimensionality. If we look at the temperature fields from Fig.~\ref{fig:GAN_images}, we can see that there is an underlying structure in the data, in form of pixel correlations on different scales. Thus, while the number of pixels in those images is $64\times64=4096$, the number of independent numbers needed to reproduce (or approximate with high accuracy) a particular image is expected to be much lower. 

Autoencoders try to find a lower dimensionality representation of the data (the \textit{bottleneck}), from which the original image can be reconstructed as closely as possible. Some class of autoencoders are widely used as generative models (e.g.~variational autoencoders); once the lower dimensionality representation is found, new images can be generated by sampling it.

Here we are interested in reducing the dimensionality of the temperature maps and investigating whether the found lower dimensionality manifold also encodes maps from different cosmologies and astrophysics models. More specifically, we train a simple autoencoder on temperature fields from the IllustrisTNG CV set of 27 simulations at fixed cosmology and astrophysics. We then pass temperature maps from simulations with very different cosmologies and astrophysics to the autoencoder to see how well it can reconstruct these images. 

We show the architecture of our autoencoder and describe the details of the training procedure in the appendix \ref{sec:autoencoder_architecture}. The loss of our autoencoder is simply
\begin{equation}
\mathcal{L}=\frac{1}{N_{\rm pixels}}\sum_{i=1}^{N_{\rm pixels}}(\log_{10} T_i-\log_{10} \tilde{T}_i)^2~,
\label{eq:reconstruction_error}
\end{equation}
where $\log_{10}T_i$ and $\log_{10}\tilde{T}_i$ are the values of the normalized log 10 of the temperature field of the original and reconstructed $i$ pixel, respectively. We construct temperature fields from the 27 realizations of the IllustrisTNG CV set. In order to increase our dataset, we allow the maps to overlap in space (i.e.~the 3D slices we use to create the maps can partially overlap). We split that data into 3 sets: training, validation, and testing. We carry out data augmentation (i.e.~90, 180, and 270 degrees rotations and flipping of the images) to further enhance the dataset. 

\begin{figure}
\centering
\includegraphics[width=0.49\textwidth]{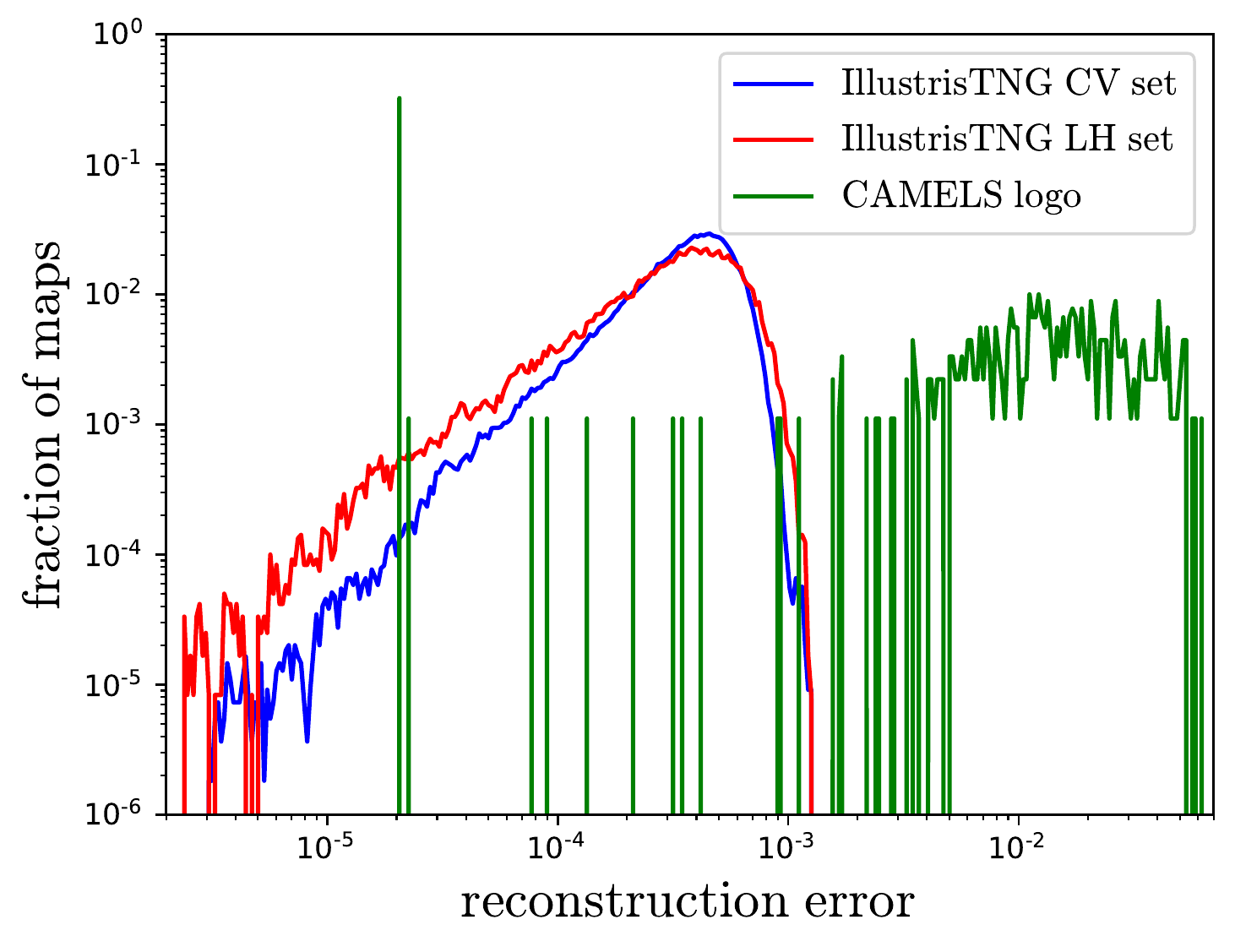}
\caption{Once the autoencoder is trained, we compute the reconstruction error (Eq.~\ref{eq:reconstruction_error}) for each image of the test set from the IllustrisTNG CV set. The blue line shows the PDF of that reconstruction error. We then use images from the IllustrisTNG LH set (i.e.~simulations with different cosmologies and astrophysics than those used to train the model). The PDF of the reconstructed error from those images is shown in red. As can be seen, the autoencoder reconstructs images with different cosmologies and astrophysics as well as those from the fiducial model. The green line shows the results when images from the CAMELS logo (see Fig. \ref{fig:CAMEL_logo}) are used. In this case, the reconstruction error is much larger, pointing out that these images are \textit{anomalous}.}
\label{fig:autoencoder_reconstruction_loss}
\end{figure}

Once the autoencoder has been trained, we use the maps from the test set\footnote{We find no differences in the results if we use maps from training, validation or all maps together.} and compute the reconstruction error as in Eq.~\ref{eq:reconstruction_error}. In Fig.~\ref{fig:autoencoder_reconstruction_loss} we show the distribution of the reconstruction error. We find that the maximum reconstructed error is around $1.3\times10^{-3}$, while the peak of the distribution takes place around $5\times10^{-4}$.

With the autoencoder trained, we feed it with temperature maps from the IllustrisTNG LH set, in which the values of the cosmological and astrophysical parameters are varied and hence are different from those of the IllustrisTNG CV set simulations the autoencoder was trained on. Since the autoencoder has been trained on temperature maps with fixed cosmology and astrophysics, one may expect that when using it to reconstruct temperature fields with different cosmologies and astrophysics models, it may not perform as well.

\begin{figure}
\centering
\includegraphics[width=0.47\textwidth]{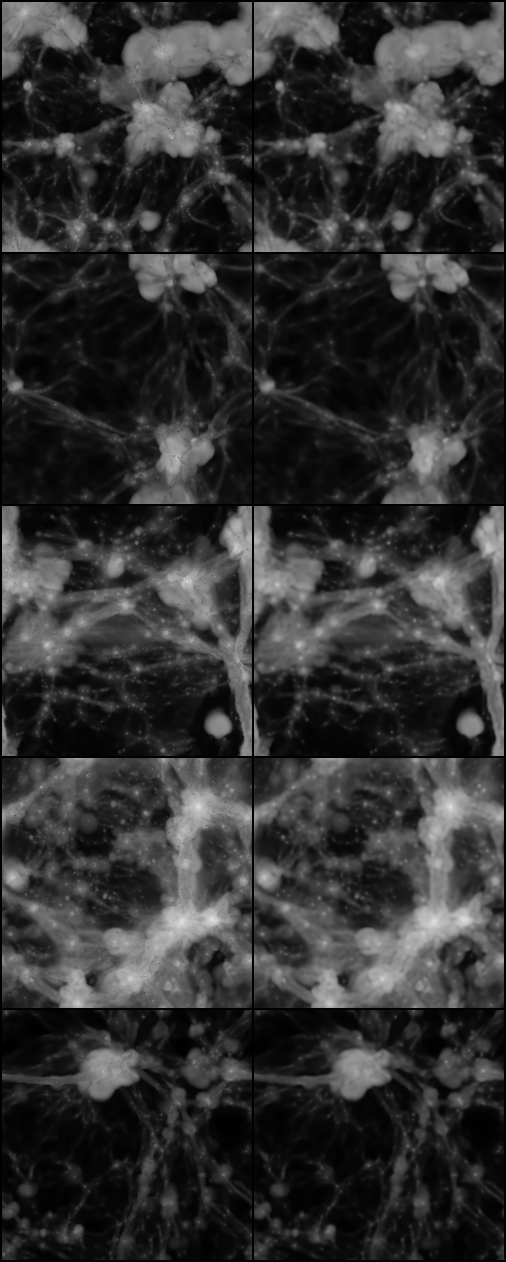}
\caption{We have trained an autoencoder on projected temperature fields from the IllustrisTNG CV set (fixed cosmology and astrophysics). We then use it on projected temperature fields from the IllustrisTNG LH set (different cosmologies and astrophysics). The above panels show the results; the left/right columns show the true/reconstructed maps. Our autoencoder is able to reconstruct temperature fields from different cosmologies and astrophysics with the same accuracy as the ones from the fiducial model it was trained on (see Fig. \ref{fig:autoencoder_reconstruction_loss}). See text for details.}
\label{fig:autoencoder}
\end{figure}

In Fig.~\ref{fig:autoencoder} we show examples of the input and reconstructed images. We note that our autoencoder is trained on $64\times64$ pixel images, while the images shown in Fig. \ref{fig:autoencoder} have $250\times250$ pixels; this is achieved by splitting the big images into small ones and input those into the autoencoder. As can be seen, from visual inspection the reconstruction is quite good. It is however noticeable how the reconstructed image is somewhat blurrier and it cannot get the very small scale structure right. This can be seen in the distribution of black points, which correspond to star forming regions, that are missing in the reconstructed images. 

Of greater interest is the distribution of the reconstructed errors for these images, which is shown with a red line in Fig.~\ref{fig:autoencoder_reconstruction_loss}. We find that the autoencoder can reconstruct these images with the same accuracy as those it was trained on. Whilst the distribution is slightly different, i.e.~the tails and peak differ in the two cases, the maximum reconstruction error is very similar in both cases. We have repeated the same exercise using temperature maps from the EX sets (i.e.~simulations with extreme values for the efficiency of the astrophysical processes), reaching the same conclusions as with the maps from the LH set. Furthermore, we have also use temperature maps from the SIMBA suite (all different sets), and we find that the autoencoder reconstruct these images as accurately as the IllustrisTNG suite.

This is somewhat surprising, as one may have na{\"i}vely expected that different astrophysical models, e.g.~very efficient AGN or supernova feedback, may have produced a different morphology of the temperature field that the autoencoder would not have been able to reconstruct. These results indicate that the simulations run with the fiducial model contain a set of images rich enough to be able to find a lower dimension manifold that is general enough to embed maps from other cosmological and astrophysical models. 

This statement depends sensitively on the size of the bottleneck. In our autoencoder, the size of the bottleneck is set to 500 neurons, i.e.~$\simeq10\%$ of the size of the original image. Larger bottlenecks will allow a better reconstruction of the maps, but data compression will be smaller.  We have trained the same autoencoder but with a bottleneck that contained only 100 neurons. In this case, the reconstructed images were much blurrier, but our conclusions did not change: even if trained only on images of the CV set, the autoencoder could reconstruct images from other models with the same accuracy. 

One possible explanation of this behavior is that our autoencoder has learned to compress general images, namely that it will be able to reconstruct any image, independently of its nature. This is not what we are after, as we are interested in finding a lower dimension manifold that captures the structure of our data. In order to test this hypothesis, we have fed the autoencoder with data whose nature is very different from the temperature maps it has been trained on: the CAMELS logo. 

\begin{figure*}
\centering
\includegraphics[width=0.49\textwidth]{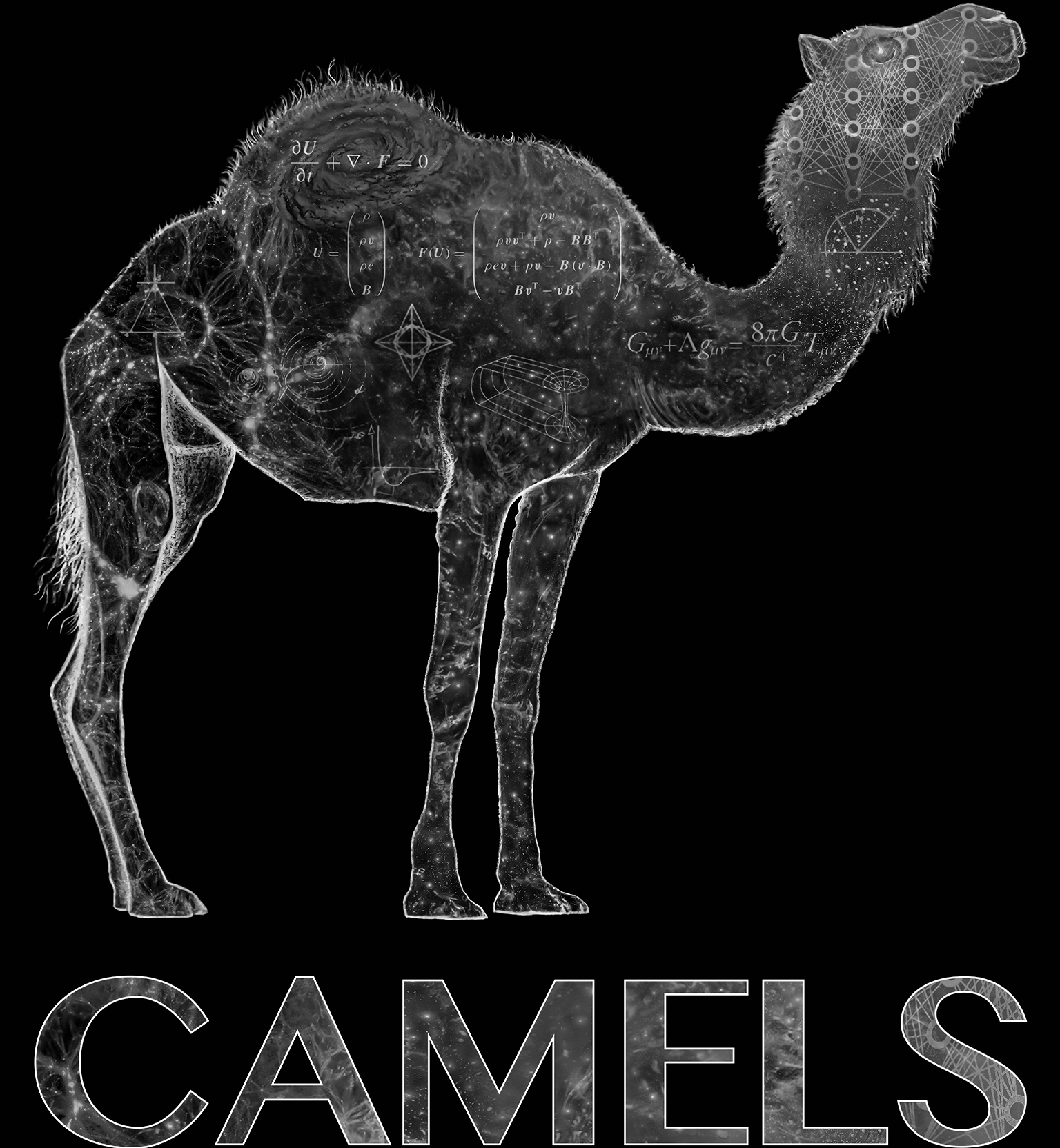}
\includegraphics[width=0.49\textwidth]{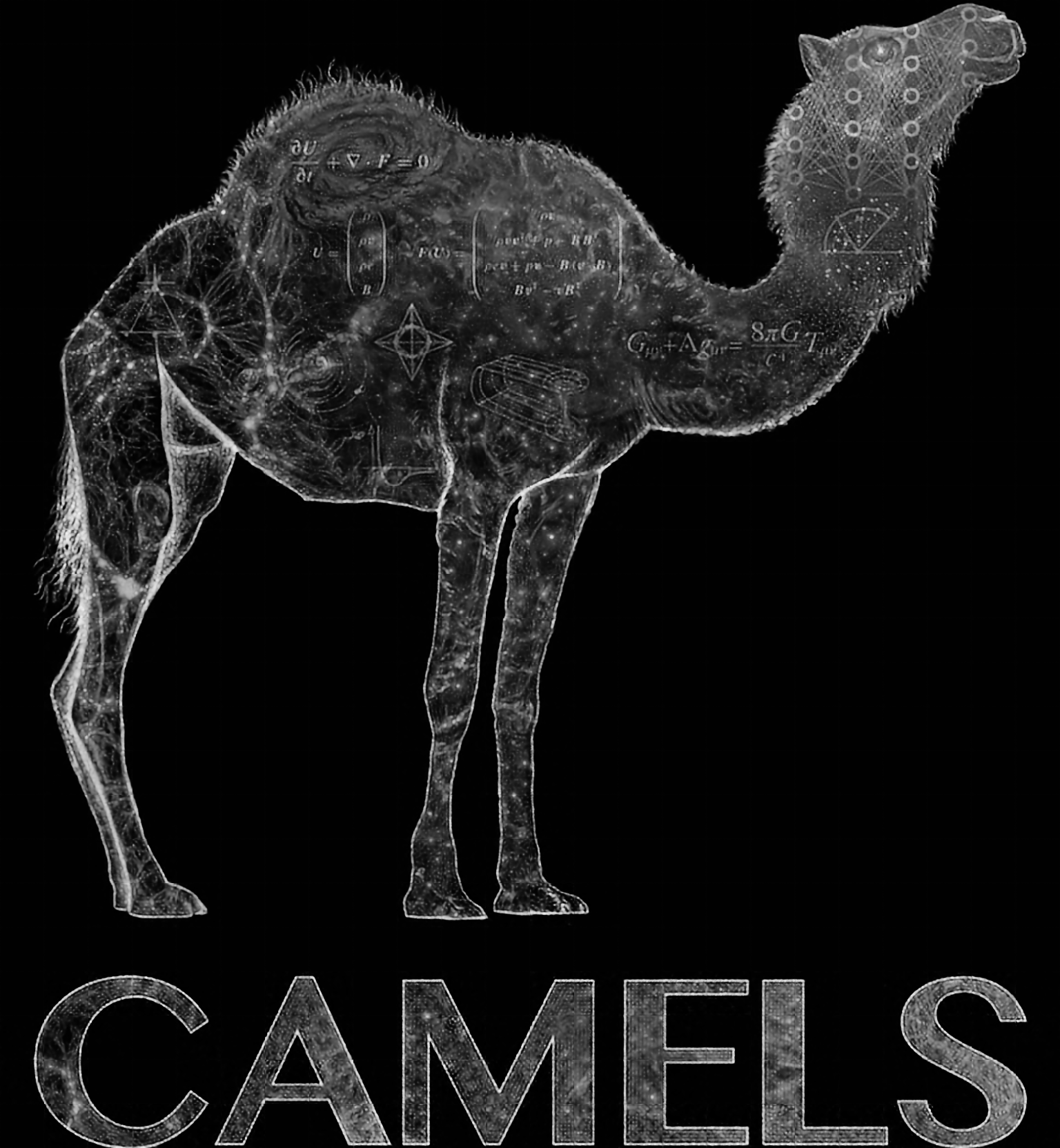}
\caption{We input the CAMELS logo (left panel) into our autoencoder trained on temperature maps. The reconstructed image is shown on the right panel. Whilst the autoencoder is able to reconstruct the logo very well, we find that the reconstruction error for any part that is not the background is larger than anything for the temperature fields (see Fig.~\ref{fig:autoencoder_reconstruction_loss}). In other words, the autoencoder identifies these regions as an anomaly/outlier. This shows that our autoencoder is not just learning to compress generic images, but is learning the manifold where the temperature fields live.}
\label{fig:CAMEL_logo}
\end{figure*}

In Fig.~\ref{fig:CAMEL_logo} we show the original logo on the left and its reconstruction on the right. Also in this case the split the big image into $64\times64$ pixels regions that we input into the autoencoder. Visually, the reconstructed logo looks attractively similar to the original, even though it has features, like camel hair, equations, diagrams of artificial neural nets, etc., which the autoencoder has never seen. More quantitatively, the green line of Fig.~\ref{fig:autoencoder_reconstruction_loss} shows the distribution of the reconstruction errors for this image. In this case, we find that around $30\%$ of the images have a good reconstruction loss, while $70\%$ of them have much larger reconstructed errors than anything produced for the temperature maps. 

Visual inspection of the maps shows that the lower reconstructed errors correspond to pixels in the black background of the image, while all images that lie in the body of the camel or the letters, exhibit a large reconstructed error. This demonstrates that the autoencoder has learned a manifold that captures the structure of the temperature maps, not just a general way to compress images. 

Representation learning is a branch of deep learning that aims, among other things, to understand the meaning of the neurons in the bottleneck. It will be interesting to apply it to our case, and try to understand the physical meaning of the numbers in the bottleneck, and the reason of its apparent universality. 

As we have shown with a rather extreme example, autoencoders can also be used to identify anomalies. Anomalies may be very hard to find, as they may show up as deviations in a high dimensional space. However, they may also carry with them a huge signal-to-noise ratio, that can allow to rule out an entire theory.

\section{Summary and discussion}
\label{sec:Conclusions}

In this paper, we have introduced the Cosmology and Astrophysics with MachinE Learning Simulations (\textsc{CAMELS}) project. We now summarize and discuss the main aspects of CAMELS and this presentation paper.

\subsection{Scientific goals}
The main scientific goals of CAMELS are:
\begin{itemize}
    \item Provide theory predictions for statistics, or fields, as a function of cosmology and astrophysics.
    \item Extract cosmological information while marginalizing over baryonic effects.
    \item Find the mapping between N-body and hydrodynamic simulations.
    \item Quantify the dependence of galaxy formation and evolution on astrophysics and cosmology.
    \item Use machine learning to efficiently calibrate subgrid parameters in cosmological hydrodynamic simulations to match a set of observations.
\end{itemize}

\subsection{Simulations}
CAMELS is 
a suite of 4,233 numerical simulations, including both N-body simulations run with \textsc{Gadget-III} \citep{Springel2005_Gadget} and state-of-the-art (magneto-)hydrodynamic simulations run with \textsc{AREPO} \citep{Arepo} and \textsc{GIZMO} \citep{Hopkins2015_Gizmo}. The subgrid models used in the hydrodynamic simulations are the same as those in the IllustrisTNG \citep{WeinbergerR_16a,PillepichA_16a} and SIMBA \citep{SIMBA} simulations, respectively.
Each simulation follows the evolution of $256^3$ dark matter particles and $256^3$ fluid elements (for the hydrodynamic simulations) in a periodic box of size $25~h^{-1}{\rm Mpc}$.

The simulations span thousands of different cosmological and astrophysical models, by varying the value of six parameters: $\Omega_{\rm m}$, $\sigma_8$, and four astrophysical parameters ($A_{\rm SN1}$, $A_{\rm SN2}$, $A_{\rm AGN1}$, and $A_{\rm AGN2}$) controlling the strength of stellar and AGN feedback. The (magneto-)hydrodynamic simulations are organized into four different sets: 
\begin{itemize}
    \item LH: A set of 1,000 simulations. Each simulation has a different value of the cosmological and astrophysical parameters. Each simulation has a different value of the initial random seed.
    \item 1P: A set of 61 simulations. In these simulations, the value of the cosmological and astrophysical parameters are varied only one at a time. The value of the random seed is the same in all simulations.
    \item CV: A set of 27 simulations. All the simulations share the value of the cosmological and astrophysical parameters, and they only differ in the value of their initial random seed.
    \item EX: A set of 4 simulations. All simulations share the value of the cosmological parameters, but differ in the value of the astrophysical parameters. One simulation has fiducial values; the other three represent extreme cases with 1) very efficient supernova feedback, 2) very efficient AGN feedback, and 3) no feedback. All simulations share the value of the initial random seed.
\end{itemize}
The above 1,092 simulations have been run with both AREPO/IllustrisTNG and GIZMO/SIMBA for a total of 2,184 (magneto-)hydrodynamic simulations. For each of these simulations, CAMELS provides their dark matter-only counterpart for a total of 2,049 N-body simulations.

\subsection{Cosmological and astrophysical properties}

We have considered twelve different cosmological and astrophysical quantities --- the matter and gas power spectrum, the ratio $P_{\rm hydro}(k)/P_{\rm Nbody}(k)$, the halo mass function, the star formation rate density, the stellar mass function, the halo baryon fraction, the average halo temperature, and the relation between stellar mass and galaxy radius, total black hole mass, circular velocity, and star formation rate --- and studied their properties and range of variation in CAMELS. 

We find that the results from the IllustrisTNG and SIMBA suites agree well within the range of variation for many of the considered quantities, producing roughly similar halo and stellar mass functions, star formation rate density, and galaxy/halo scaling relations over the full parameter space covered by the 1,000 simulations in each LH set.    
Using the simulations in the CV sets, we have shown that some of the overlap between IllustrisTNG and SIMBA can be explained by cosmic variance, which represents a significant contribution to the range of variation observed in the average halo temperature, galaxy scaling relations (that incorporates the contribution of intrinsic scatter), and the amplitude of the matter and gas power spectrum on large scales.
Nonetheless, IllustrisTNG and SIMBA make significantly different predictions for halo baryon fractions and the impact of baryonic effects on the matter power spectrum, emphasizing the need for marginalizing over uncertainties in baryonic effects to extract the maximum amount of information from cosmological surveys. 

Our results also highlight the complex non-linear interplay between different baryonic processes: systematic variations in feedback parameters may have different (even opposite) effects on observables depending on the model implementation, as seen in e.g. the response of the global star formation rate history to variations in parameter $A_{\rm SN1}$ for the IllustrisTNG and SIMBA models.
CAMELS significantly expands on earlier studies of model and parameter variations performed as part of the original IllustrisTNG and SIMBA simulations to understand the implications of different feedback mechanisms \citep[e.g.][]{WeinbergerR_16a,Pillepich_2018,PillepichA_16a,Christiansen2019,SIMBA,Appleby2020}. 
Previous studies have also emphasized the impact of baryonic effects on the matter power spectrum and performed comparisons between different baryonic physics implementations \citep[e.g.][]{Chisari_2019,vanDaalen2020}.
CAMELS represents an increase in the number of cosmological hydrodynamic simulations of two orders of magnitude compared to previous studies, allowing us to exploit the full potential of machine learning.

\subsection{Machine learning applications}

CAMELS has been designed to train machine learning algorithms by sampling the 6D parameter space with more than 1,000 simulations for each code/model. In this paper we have illustrated the potential of CAMELS with five simple applications: 
\begin{itemize}
\item We have trained a neural network to predict the cosmic star formation rate density of a simulation just from the value of its cosmological and astrophysical parameters. Our network achieves a $\simeq30\%$ accuracy on this task. We note that cosmic variance itself produces an average scatter of $\simeq20\%$ on the cosmic star formation rate density.
\item We have used neural networks to determine the value of the cosmological and astrophysical parameters from measurements of the cosmic star formation rate density, without knowing the likelihood function of the data. The network can constrain the value of $\Omega_{\rm m}$, $\sigma_8$, $A_{\rm SN1}$, and $A_{\rm SN2}$ with an average error equal to 0.055, 0.051, 0.55, and 0.25, respectively. These errors arise from the small cosmological volumes sampled by our simulations.
\item We have made use of symbolic regression to obtain simple analytic expressions that predict the star-formation rate density of a simulation as a function of the cosmological and astrophysical parameters. Our equations achieve $\simeq45\%$ accuracy, that should be compared with the $\simeq30\%$ of the neural network and the $\simeq20\%$ of intrinsic scatter due to cosmic variance.
\item We have trained Generative Adversarial Networks (GANs) using 2D temperature fields from the CAMELS suite. Our GAN learns to generate new projected temperature maps whose statistical properties agree very well with those from the simulations: the power spectra and PDF of the true and fake images agree within 15\% and 25\%, respectively.
\item We have trained autoencoders to compress 2D temperature fields into a lower dimension manifold. Although the autoencoder is trained on images from simulations with fixed cosmology and astrophysics, it is able to reconstruct temperature fields from simulations with different cosmological and astrophysical parameters with the same accuracy as the images it was trained on. Our autoencoder is able to identify anomalies; while visually it is able to reconstruct the CAMELS logo well, anything except the background is identified as an anomaly.
\end{itemize}

One important aspect to consider when using machine learning techniques is the trade-off between accuracy and speed-up. For the neural networks used in this paper, the computational time required for testing is negligible. The most computationally expensive part is their training. Training the autoencoder was the most demanding part of this work, as we carried out a significant tuning of the value of the hyper-parameters. Even in that case, the training only required $\sim150$ GPU hours. Running one of the CAMELS (magneto-)hydrodynamic simulations requires an average of $\sim6,000$ CPU hours. There is thus a trade-off between the computational time needed to run new and exact simulations and the time needed to train neural networks to approximate them. 

Fulfilling the main goals of the CAMELS project requires a dense exploration of the parameter space. In this case, the use of machine learning techniques will be crucial in order to learn and capture the underlying structure of the simulations and shrinking the computational requirements of such task.

We emphasize that machine learning can be used for many different purposes, from the generation of new data with some desired statistical properties to finding unknown underlying patterns. For cosmological analyses, the required accuracy will strongly depend on the considered task. For instance, generating new data to compute covariance matrices may not require percent level accuracy, while creating an emulator for a given observable may require sub-percent level accuracy. Achieving the needed accuracy will depend on different properties, from using a good architecture/model and proper value of the hyper-parameters to having enough data to train the model.

\subsection{Limitations and extensions}
\label{subsec:limitations}

We now discuss some of the limitations of the CAMEL simulations. First, the mass and spatial resolution in CAMELS do not allow us to resolve scales below $\sim 1\,h^{-1}{\rm kpc}$, while only halos with dark matter mass above $6.5\times10^9(\Omega_{\rm m}-\Omega_{\rm b})/0.251~h^{-1}M_\odot$ contain at least 100 dark matter particles. 
This implies that CAMELS cannot be used to e.g.~place constrains on the nature of dark matter using probes that rely on the distribution of matter on very small scales, such as sub-halos in the Milky Way \citep[e.g.][]{Nadler_2020}.
We refer the reader to \cite{PillepichA_16a, Pillepich_2019} and \citet{SIMBA} for resolution convergence studies of different quantities in the original IllustrisTNG and SIMBA simulations, respectively. In principle, it is possible to use deep learning techniques to increase the mass and spatial resolution of simulations \citep{Doogesh_2020}, which we plan to explore in CAMELS in future work.

Second, the volume of the simulations is relatively small: $(25~h^{-1}{\rm Mpc})^3$. This implies that long wavelength modes are not accounted for in CAMELS, which are important for the formation of large objects such as galaxy clusters and in order to set the proper normalization of the matter power spectrum on all scales. We plan to address this limitation by extending CAMELS to include larger volumes and by running separate universe (magneto-)hydrodynamic simulations \citep{Sirko05, Li_2014, Li_2018, Barreira_2019, Barreira_2020}, where an amplitude of the DC mode different than zero is taken into account. 

Third, CAMELS is limited to variations of only two cosmological parameters and four astrophysical parameters.  For example, in the (magneto-)hydrodynamic simulations we vary $\Omega_{\rm m}$ while $\Omega_{\rm b}$ is always fixed, which does not allow us to separate possible effects that may depend on $\Omega_{\rm b}/\Omega_{\rm m}$ from those of varying $\Omega_{\rm m}$ alone. 
Ideally, we would like to vary more cosmological parameters, e.g.~$h$, $n_s$, $M_\nu$, and $w$, on simulations with larger volume, to be able to perform the analysis of cosmological data on the same footing as techniques such as perturbation theory \citep{Ivanov_2019a,damico_2019, Philcox_2020}. Furthermore, it is important to vary more astrophysical parameters, as we expect that they could play an important role on different aspects of galaxy formation and evolution. We plan to expand the parameter-space volume covered by the simulations in future extensions of the CAMELS project. Machine learning can also be used to find the mapping between different models \citep[see e.g.][]{Giusarma_19}. We will investigate this route in future work.

Overall, while CAMELS represents the largest set of cosmological (magneto-)hydrodynamic simulations with full galaxy formation physics available to exploit machine learning techniques, its applicability to perform the analysis of cosmological data is still limited in different regimes.  The current simulation suite offers a unique opportunity to develop proof-of-concept techniques for the broader scientific goals of CAMELS, demonstrating e.g.~whether neural networks can learn to marginalize over baryonic effects at the field level. In future extensions of CAMELS, we will address the above limitations to create a powerful tool to extract the maximum amount of information from upcoming cosmological surveys.

\subsection{Data access}

CAMELS contains 143,922 full snapshots from a total of 4,233 simulations. For each snapshot we have stored the corresponding halo/galaxy catalogue and additional data products such as power spectra and bispectra. Details on how to access the data, the codes written to perform all analyses in this work, together with further technical details of the simulations can be found at \url{https://www.camel-simulations.org}.

\vspace{0.4cm}

\section*{ACKNOWLEDGEMENTS}
All simulations have been run in the Popeye-Simons cluster at the San Diego Supercomputer Center. This work was carried out as part of the SMAUG project. SMAUG gratefully acknowledges support from the Center for Computational Astrophysics at the Flatiron Institute, which is supported by the Simons Foundation. FVN acknowledges funding from the WFIRST program through NNG26PJ30C and NNN12AA01C. DAA was supported in part by NSF grant AST-2009687. The work of DAA, SG, DNS, RS, YL, VLT, AMD, ShH, SuH, BB, and GC has been supported by the Simons Foundation. We thank the SIMBA and IllustrisTNG collaborations for allowing us to use their respective subgrid model implementations. FVN thanks Martin White and Benjamin Wandelt for very useful discussions at the beginning of this project. We thank Miles Cranmer, Christina Kreisch, Jose Manuel Zorrilla-Matilla, Noah Kasmanoff, Elaine Cui, and Pablo Villanueva-Domingo for their help with the machine learning applications presented in this paper. We thank the anonymous referee for the constructive report, that helped us improving the clarity and readibility of the paper. We are indebted to Nick Carriero, Ian Fish, Andras Pataki, and Dylan Simon for their help with the technical difficulties we faced while running, storing, and post-processing the simulations. This work has made extensive use of the \textsc{Pylians3} libraries, publicly available at \url{https://github.com/franciscovillaescusa/Pylians3}.

\appendix

\section{A. GAN architecture}
\label{sec:GAN_architecture}

Here we provide details on the architecture used for the generator and discriminator of the GAN discussed in Section \ref{subsec:GANs}. Our model follows closely the DCGAN architecture proposed in \cite{DCGAN}. The model used for the generator is as follows:

\begin{enumerate}
\item Input: 100 channels x 1 x 1
\item ConvTranspose2d: 512 channels x 4 x 4; kernel=4, stride=1, padding=0
\item BatchNorm
\item ReLU activation (0.2)
\item ConvTranspose2d: 256 channels x 8 x 8; kernel=4, stride=1, padding=1
\item BatchNorm
\item ReLU activation (0.2)
\item ConvTranspose2d: 128 channels x 16 x 16; kernel=4, stride=1, padding=1
\item BatchNorm
\item ReLU activation (0.2)
\item ConvTranspose2d: 64 channels x 32 x 32; kernel=4, stride=1, padding=1
\item BatchNorm
\item ReLU activation (0.2)
\item ConvTranspose2d: 1 channels x 64 x 64; kernel=4, stride=1, padding=1
\item Tanh activation
\item Output: image with 1 channel x 64 x 64
\end{enumerate}

On the other hand, the architecture of the discriminator is

\begin{enumerate}
\item Input: image with 1 channel x 64 x 64
\item Conv2d: 64 channels x 32 x 32; kernel=4, stride=2, padding=1
\item LeakyReLU activation (0.2)
\item Conv2d: 128 channels x 16 x 16; kernel=4, stride=2, padding=1
\item BatchNorm
\item LeakyReLU activation (0.2)
\item Conv2d: 256 channels x 8 x 8; kernel=4, stride=2, padding=1
\item BatchNorm
\item LeakyReLU activation (0.2)
\item Conv2d: 512 channels x 4 x 4; kernel=4, stride=2, padding=1
\item BatchNorm
\item LeakyReLU activation (0.2)
\item Conv2d: 1 channels x 1 x 1; kernel=4, stride=1, padding=0
\item Sigmoid activation
\item Output: probability of image being real
\end{enumerate}

\section{B. Autoencoder architecture}
\label{sec:autoencoder_architecture}

Here we outline the architecture of the autoencoder presented in section \ref{subsec:dimensionality}. We refer the reader to \cite{Convolutional_autoencoder} and references to it for further details on convolutional autoencoders.

\begin{enumerate}
\item Input: image with 1 channel x 64 x 64
\item Conv2d: 32 channels x 32 x 32; kernel=4, stride=2, padding=1
\item LeakyReLU activation (0.2)
\item Conv2d: 64 channels x 16 x 16; kernel=4, stride=2, padding=1
\item BatchNorm
\item LeakyReLU activation (0.2)
\item Conv2d: 128 channels x 8 x 8; kernel=4, stride=2, padding=1
\item BatchNorm
\item LeakyReLU activation (0.2)
\item Conv2d: 256 channels x 4 x 4; kernel=4, stride=2, padding=1
\item BatchNorm
\item LeakyReLU activation (0.2)
\item Conv2d: 500 channels x 1 x 1; kernel=6, stride=1, padding=1
\item BatchNorm
\item LeakyReLU activation (0.2)
\item ConvTranspose2d: 256 channels x 4 x 4; kernel=4, stride=1, padding=0
\item ReLU activation (0.2)
\item ConvTranspose2d: 128 channels x 8 x 8; kernel=4, stride=2, padding=1
\item BatchNorm
\item ReLU activation (0.2)
\item ConvTranspose2d: 64 channels x 16 x 16; kernel=4, stride=2, padding=1
\item BatchNorm
\item ReLU activation (0.2)
\item ConvTranspose2d: 32 channels x 32 x 32; kernel=4, stride=2, padding=1
\item BatchNorm
\item ReLU activation (0.2)
\item ConvTranspose2d: 1 channels x 64 x 64; kernel=4, stride=2, padding=1
\item Tanh activation
\item Output: image with 1 channel x 64 x 64
\end{enumerate}

\bibliography{references}{}
\bibliographystyle{hapj}

\end{document}